\pdfoutput=1
\documentclass[final,3p,times,12pt,a4paper,sort&compress]{elsarticle}

\usepackage{amssymb,amsmath}
\usepackage{graphicx}
\usepackage{caption}
\usepackage{euscript}
\usepackage{mathrsfs}
\usepackage{isotope}
\usepackage{tikz}

\newcommand\gr[1]{\mathrm{#1}}              

\newcommand\R{\gr{R}}
\newcommand\Ham{\mathscr{H}}                
\newcommand\Lag{\mathscr{L}}                
\newcommand\PP{\EuScript{P}}                
\newcommand\AP{\EuScript{A}}                
\newcommand\DD{\mathscr{D}}                 
\newcommand\RR{\mathscr{R}}                 
\newcommand\AAA{\mathscr{A}}                

\newcommand\vek[1]{{\boldsymbol{\bf #1}}}   
\newcommand\skal[2]{\vek{#1}\cdot\vek{#2}}  
\newcommand\bra[1]{\langle#1\vert}          
\newcommand\ket[1]{\vert#1\rangle}          
\newcommand\braket[2]{\langle#1\vert#2\rangle} 
\newcommand\he[1]{#1^{\dagger}}             
\newcommand\imag{\mathrm i}                 
\DeclareRobustCommand\openone{\leavevmode\hbox{\small1\normalsize\kern-.33em1}}
\newcommand\thname{Theorem}
\newtheorem{theorem}{\thname}

\newcommand\de{\partial}
\newcommand\vp{\varphi}
\newcommand\ve{\varepsilon}
\newcommand\vr{\varrho}
\newcommand\vt{\theta}
\newcommand\op{\mathbf\Delta}               

\newcommand\nor[1]{\left\|#1\right\|}       
\newcommand\abs[1]{\left|#1\right|}         
\newcommand\Pd[2]{\partial #1/\partial #2}  
\newcommand\PD[2]{\frac{\partial #1}{\partial #2}} 
\newcommand\OD[2]{\frac{\dd #1}{\dd #2}}    
\newcommand\dd{\mathop{{\rm d}\!}\nolimits} 
\newcommand\dthree{\mathop{{\rm d}^3\!}\nolimits} 
\newcommand\dfour{\mathop{{\rm d}^4\!}\nolimits} 

\DeclareMathOperator{\re}{Re}               
\DeclareMathOperator{\im}{Im}               
\DeclareMathOperator{\Tr}{Tr}

\begin{document}

\begin{frontmatter}

\title{Spontaneous Symmetry Breaking and Nambu--Goldstone Bosons\\
in Quantum Many-Body Systems}

\author{Tom\'a\v{s} Brauner}
\ead{brauner@th.physik.uni-frankfurt.de}
\address{Institute for Theoretical Physics, Goethe University, 60438
Frankfurt am Main, Germany}
\address{Department of Theoretical Physics, Nuclear Physics
Institute ASCR, 25068 \v{R}e\v{z}, Czech Republic}

\begin{abstract}
Spontaneous symmetry breaking is a general principle, that constitutes
the underlying concept of a vast number of physical phenomena ranging from
ferromagnetism and superconductivity in condensed matter physics to the Higgs
mechanism in the standard model of elementary particles. I focus on
manifestations of spontaneously broken symmetries in systems that are not
Lorentz invariant, which include both, nonrelativistic systems as well as
relativistic systems at nonzero density, providing a self-contained review of
the properties of spontaneously broken symmetries specific to such theories.
Topics covered include: (i) Introduction to the mathematics of spontaneous
symmetry breaking and the Goldstone theorem. (ii) Minimization of Higgs-type
potentials for higher-dimensional representations. (iii) Counting rules for
Nambu--Goldstone bosons and their dispersion relations. (iv) Construction of
effective Lagrangians. Specific examples in both relativistic and
nonrelativistic physics are worked out in detail.
\end{abstract}


\end{frontmatter}


\section{Introduction}
\label{sec:introduction}

Symmetry considerations play a key role in our understanding of Nature. Ever
since the birth of science, the aesthetic beauty of symmetry has appealed to
the minds of natural philosophers, and later physicists, seeking for the origin
of the laws of Nature. With the advent of quantum mechanics in the twentieth
century, symmetry techniques based on group theory became an indispensable tool
of a theoretical physicist. While in classical physics symmetries are
straightforwardly incorporated and directly connected to physical observables
by means of the Noether theorem, the situation in the quantum theory is more
subtle. A fundamental theorem due to Wigner (see Sec.~2.2 in
\cite{Weinberg:1995v1}) states that
invariance of observables under certain transformation implies the existence of
a unitary operator on the Hilbert space of states. If the symmetry
transformation is moreover compatible with the dynamics of the system, this
unitary operator commutes with the Hamiltonian and gives rise to a
characteristic multiplet structure in its spectrum. However, there are systems
whose dynamics is invariant under a symmetry transformation, yet this symmetry
is not manifest in the spectrum or physical observables. We speak of
\emph{spontaneous symmetry breaking} (SSB). This phenomenon is ubiquitous in
condensed matter physics where it is responsible, for example, for the peculiar
behavior of superconductors, superfluids, and ferromagnets. In high energy
physics, it underlies much of our current understanding of the fundamental
interactions and the origin of masses of elementary particles.

This paper provides a pedagogical review of the properties of spontaneously
broken \emph{continuous global internal symmetries} with special emphasis on
systems lacking Lorentz invariance that is, either intrinsically nonrelativistic
systems or relativistic systems at nonzero density. (In the following, all
these systems will for simplicity be referred to as nonrelativistic, unless
explicitly indicated otherwise.) It aims to be a self-contained introduction
into the subject, requiring just moderate knowledge of quantum field theory.
Nevertheless, it will hopefully also offer some new material to the experts.

The plan of the paper is as follows. In Section~\ref{sec:basics}\ the notion of
SSB is introduced and the mathematical subtleties associated with its
implementation on the Hilbert space of a quantum system are discussed to some
depth. A specific example, a free nonrelativistic field theory, is analyzed in
detail in Section~\ref{sec:freeNR}\ illustrating explicitly the general material
of the previous section, and at the same time showing some nontrivial features
of SSB in nonrelativistic systems to be discussed in what follows. The
technically easiest way to achieve SSB in an interacting field theory is to
introduce an effective scalar field and adjust its phenomenological potential
so that it has a symmetry-breaking minimum at the tree level. While in the case
of the Higgs mechanism in the standard model of elementary particles or the
Ginzburg--Landau theory for ferromagnets this is rather trivial, it becomes a
difficult problem with the increasing dimension of the representation in which
the scalar field transforms. Section~\ref{sec:higgs}\ describes a general method
to minimize Higgs potentials; the specific example of a spin-one color
superconductor is worked out in detail. Section~\ref{sec:goldstone}\ presents,
from the theoretical point of view, the most important part of the paper. I
first formulate and prove the Goldstone theorem and point out its limitations
in case of nonrelativistic systems, essentially following the classic review
\cite{Guralnik:1968gu}. Then the issue of the counting rules for
Nambu--Goldstone (NG) bosons is introduced, starting with the seminal
contribution of Nielsen and Chadha \cite{Nielsen:1975hm}, and then discussing
the recent developments. In Section~\ref{sec:examples}\ the general results are
elucidated using several explicit examples. Finally, Section~\ref{sec:EFT}\ is
devoted to the model-independent description of SSB based on low-energy
effective field theory, worked out in the general nonrelativistic case by
Leutwyler \cite{Leutwyler:1993gf}. Section~\ref{sec:summary}\ provides a summary
and offers some comments on spontaneous breaking of spacetime symmetries, not
covered by this paper.

What I have regretfully omitted from this paper is a thorough discussion of
applications of SSB and detailed examples of its appearance in realistic
interacting quantum field theories. Even though this would certainly help the
reader close the gap between the general principles and their concrete
applications in current literature, I had several reasons for this omission.
First, the list of problems to discuss would strongly depend on personal taste,
and even with a very limited selection would increase the volume of the paper
inadequately. Second, I prefer to stay at an as general level as possible so
that the paper may be found useful by a wider audience. I thus discuss the
principles and the model-independent approach of effective field theory, and
only use specific model examples to illustrate the general results. Finally,
dealing with realistic field theories would almost inevitably mean using
suitable approximation schemes. One then has to make extra effort to show that
the obtained results are physical, and not just artifacts of the approximation.
On the contrary, most of the examples presented in the text are, sometimes in a
restricted sense, exactly solvable.

Since this is a review, most of its content is essentially compiled from
existing literature. However, some of the presented material is new. This
concerns in particular Section~\ref{sec:examples}: to the best of my knowledge
this is the first time to give an explicit, and exactly solvable, example of a
system exhibiting a NG boson whose energy is proportional to a \emph{noninteger}
power of momentum. The nonanalytic long-distance behavior of such a system
clearly precludes any description based on conventional effective field theory.
Furthermore, I have improved upon the proof of Theorem \ref{thm:schafer} on the
NG boson counting in absence of charge density. Finally, as far as I know, even
the identification of a free nonrelativistic particle as a type-II NG boson
presented in Section~\ref{sec:freeNR}\ is novel. Throughout the text I use, for
the sake of simplicity, the natural units in which $\hbar=c=1$. Also, I use the
Einstein summation convention and the Lorentz indices are contracted with the
timelike Minkowski metric, $g_{\mu\nu}=\mathrm{diag}(1,-1,-1,-1)$.

The literature dealing with various applications of the general principle of
SSB is vast, and I therefore apologize to all authors not mentioned explicitly
in the list of references for having omitted their valuable contributions. I
have, at least, tried to include papers discussing to some extent the
\emph{general} properties of spontaneously broken symmetries, in particular
those focusing on nonrelativistic systems. Besides papers covering specific
subtopics to be discussed in the following sections, I would also like to point
out several good reviews of the physics of Nambu--Goldstone bosons and their
effective field theory description
\cite{Georgi:1994qn,Kaplan:1995uv,Manohar:1996cq,Pich:1998xt,Burgess:1998ku}
as well as some specialized in the relativistic chiral perturbation theory
\cite{Scherer:2002tk,Bijnens:2006zp,Scherer:2009bt}.


\section{Basic properties of spontaneously broken symmetries}
\label{sec:basics}

Consider a quantum field theory whose action is invariant under a continuous
group of symmetry transformations. By Noether's theorem, this gives rise to the
existence of a conserved current $j^\mu(x)$, satisfying the continuity
equation, $\de_\mu j^\mu(x)=0$. In the standard, Wigner--Weyl realization of the
symmetry, the associated integral charge $Q$ serves as a generator of the
symmetry transformations, which are implemented on the Hilbert space of states
by unitary operators. The ground state is assumed to be a discrete,
nondegenerate eigenstate of the Hamiltonian. Consequently, it bears a
one-dimensional representation of the symmetry group, and therefore also is an
eigenstate of the charge $Q$. The spectrum of all eigenstates splits into
multiplets of the symmetry, corresponding to irreducible representations of the
symmetry group.

\subsection{Ground state and finite symmetry transformations}
\label{subsec:groundstate} A \emph{spontaneously broken symmetry} can be
naively characterized as such symmetry that the \emph{ground state is not an
eigenstate of its generator}. In order to give a more formal definition, let us
follow the review \cite{Guralnik:1968gu}. Given the charge density $j^0(x)$,
one introduces for an arbitrary finite space domain $\Omega$ the operator
\begin{equation}
Q_\Omega(t)=\int_\Omega\dthree\vek x\,j^0(\vek x,t)
\label{finiteVcharge}
\end{equation}
In the following I will use the same letter $\Omega$ to denote the domain and
its volume; the precise meaning of the symbol will always be clear from the
context. The \emph{symmetry breaking condition} can be restated as the
existence of a (not necessarily local) operator $\Phi$ such that
\begin{equation}
\lim_{\Omega\to\infty}\bra0[Q_\Omega(t),\Phi]\ket0\neq0
\label{SSBcondition}
\end{equation}
where $\ket0$ is a translationally invariant ground state. This expectation
value is known as the \emph{order parameter}. Clearly, this formal definition
immediately implies the previous one: if the vacuum were an eigenstate of the
charge operator, the expectation value of this commutator would have to be
zero.

It is customary to identify $Q(t)=\lim_{\Omega\to\infty}Q_\Omega(t)$ formally
with the integral charge operator. However, this operator strictly speaking does
not exist \cite{Fabri:1966fp}. Indeed, translational invariance of the vacuum
would then imply translational invariance of the state $Q(t)\ket0$, and
consequently
\begin{equation}
\bra0Q(t)Q(t)\ket0=\int\dthree\vek x\,\bra0j^0(\vek x,t)Q(t)\ket0=
\int\dthree\vek x\,\bra0j^0(\vek0,t)Q(t)\ket0
\end{equation}
which diverges unless $Q(t)\ket0=0$. This would, however, be in contradiction
with the symmetry breaking condition \eqref{SSBcondition}.

The intuitive picture of spontaneous symmetry breaking, based on the
observation that a symmetry transformation does not leave the ground state
intact, suggests high degeneracy of equivalent ground states. Indeed, since
the charge operator commutes with the Hamiltonian, so will a finite symmetry
transformation generated by this operator. It will therefore transform the
ground state into another state with the same energy. As long as the symmetry
group is continuous, we will find \emph{infinitely many degenerate ground
states}. On account of the fact that they are all connected by symmetry
transformations, they must be physically equivalent and any one of them can
serve as a starting point for the construction of the spectrum of excited
states. The issue of the choice of a ground state will be discussed in more
detail in the next subsection.

However, mathematical implementation of these ideas is subtle. The finite
volume charge operator $Q_\Omega(t)$ induces a ``finite symmetry
transformation,'' $U_\Omega(\vt,t)=\exp[\imag\vt Q_\Omega(t)]$, which in turn
gives rise to a ``rotated ground state,'' $\ket{\vt,t}_\Omega=\he
U_\Omega(\vt,t)\ket0$. However, very much like the limit
$\lim_{\Omega\to\infty}Q_\Omega(t)$ does not exist, the operator $\exp[\imag\vt
Q(t)]$ is not well defined either. In fact, it can be proved that
\begin{equation}
\lim_{\Omega\to\infty}\braket0{\vt,t}_\Omega=\lim_{\Omega\to\infty}
\bra0\exp[-\imag\vt Q_\Omega(t)]\ket0=0
\end{equation}
as will be shown later on explicit examples. It means that in the infinite
volume (thermodynamic) limit, any two ground states, formally connected by a
symmetry transformation, are actually orthogonal. The same
conclusion holds for excited states constructed above these vacua. All these
states therefore cannot be accommodated in a single separable Hilbert space,
forming rather two distinct Hilbert spaces of their own. Any of these Hilbert
spaces can, nevertheless, be taken as a basis for an equivalent description of
the system, and the choice has no observable physical consequences.

Unlike the transformations of physical states, finite symmetry transformations
of observables can be defined consistently in a theory which is sufficiently
causal. Using the Baker--Campbell--Hausdorff formula one obtains for any
operator $A$ that
\begin{equation}
\begin{split}
A_{\vt,t;\Omega}\equiv U_\Omega(\vt,t)A\he U_\Omega(\vt,t)&=
A+\imag\vt[Q_\Omega(t),A]+\frac12(\imag\vt)^2[Q_\Omega(t),[Q_\Omega(t),A]]
+\dotsb\\
\text{where}\quad
&[Q_\Omega(t),A]=\int_\Omega\dthree\vek x\,[j^0(\vek x,t),A]
\end{split}
\label{intcharge}
\end{equation}
As long as the theory satisfies the microcausality condition, that is, the
commutator of any two local operators separated by a spacelike interval
vanishes, and as long as the operator $A$ is localized in a finite domain of
spacetime, there will be a region $\Omega_0$ such that the charge density
outside this region does not contribute to the commutator,
\begin{equation}
\int_{\R^3\backslash\Omega_0}\dthree\vek x\,[j^0(\vek x,t),A]=0
\quad\text{whence}\quad
\lim_{\Omega\to\infty}[Q_\Omega(t),A]=\int_{\Omega_0}\dthree\vek x\,
[j^0(\vek x,t),A]
\end{equation}
The transformation $U_\Omega(\vt,t)A\he U_\Omega(\vt,t)$ therefore has a
well-defined limit as $\Omega\to\infty$. The expectation value of the rotated
operator $A_{\vt,t;\Omega}$ in the vacuum $\ket0$ can then be interpreted as the
expectation value of $A$ in the rotated vacuum $\ket{\vt,t}_\Omega$. One should
nevertheless keep in mind that in the limit of infinite volume the correct
formal definition proceeds as above.

Let me recall at this point that in Lorentz invariant theories the
microcausality condition is automatically guaranteed. This is, of course, not
the case for nonrelativistic field theories where interaction is often modeled
by a nonlocal instantaneous potential rather than by an exchange of a
propagating mode. However, the above argument can still be applied provided the
range of the interaction is short enough \cite{Lange:1966zz}. Typically, it is
sufficient when the correlations decay exponentially at long distance. If they
decrease just as some power of distance, the conditions for applicability of the
general arguments concerning SSB have to be inspected case by case.

\subsection{Explicit symmetry breaking and the choice of the ground state}
In the preceding subsection it was demonstrated that SSB gives rise to multiple
degenerate ground states in which the order parameter acquires different
values. I also stressed that any of these degenerate vacua can be chosen as the
physical ground state. A natural question then arises: why does the system
actually choose one of the states with a definite value of the order parameter,
and not their superposition in which the expectation value of the order
parameter could be even made to vanish? The answer to this question is tightly
connected to the presence of external perturbations and the thermodynamic limit.
A thorough discussion of this issue is given in Sec.~19.1 of
\cite{Weinberg:1996v2}.

In finite volume the ground state of a quantum system is usually (though not
necessarily always) nondegenerate. One still has a large number of states
formally connected by symmetry transformations, but their exact degeneracy is
lifted by boundary conditions. (An exception is the frequently used periodic
boundary condition that preserves translational invariance and thus inherits
many of the properties of the infinite volume system. Nevertheless, this
boundary condition seems somewhat artificial from a physical point of view.) The
unique ground state is then given by a symmetric superposition of all the
states, very much like the ground state of the quantum mechanical particle in a
potential double well. The distance of the individual energy levels is
essentially determined by the tunneling amplitude for the transition from one
state with a definite value of the order parameter to another. In the infinite
volume limit, this amplitude is exponentially suppressed and the ground states
become perfectly degenerate. In principle it is indeed possible, though
technically not so advantageous, to construct a quantum mechanical description
of a spontaneously broken symmetry using a symmetric ground state. SSB is then
manifested by long range correlations rather than nonzero vacuum expectation
values \cite{Yang:1962zz}.

In practice, the symmetry is almost never exact but is usually disturbed by
small perturbations such as external fields. A typical example is the
ferromagnet, where the role of the order parameter is played by the spontaneous
magnetization. In principle, the magnetization can take any direction, but an
arbitrarily weak external magnetic field will cause the magnetization to align
with it. This phenomenon is general. As soon as the system volume is
large enough, the energy difference induced by the external perturbation will
be much larger than the tiny intrinsic splitting of the energy levels. The
ground state will then be determined solely by the perturbation; this is called
\emph{vacuum alignment}. By choosing an appropriate perturbation one can select
the corresponding vacuum. The key step is now that there is a basis in the
space of degenerate ground states in which all observables become diagonal
operators in the limit of infinite volume (see Sec.~19.1 in
\cite{Weinberg:1996v2}). Therefore, it
is the same basis of states in which the order parameter takes definite values,
and one of which is selected by the external perturbation, regardless of what
operator is used as the perturbation. This concludes the argument that the
physical vacua are those with a definite value of the order parameter; their
superpositions do not survive the infinite volume limit. One practical means of
selecting the ground state thus is to switch on an external perturbation and
\emph{then} go to infinite volume. After this is done, the perturbation can be
adiabatically switched off without disturbing the vacuum. It is symptomatic of
SSB that the order of these two limits (infinite volume versus zero
perturbation) cannot be reversed.

Before concluding the section, let me summarize the main features of SSB. First
and most importantly, the symmetry is not realized by unitary operators on the
Hilbert space, so it does not give rise to multiplets in the spectrum. A
hallmark of SSB is the existence of an order parameter, that is, nonzero vacuum
expectation value of an operator which transforms nontrivially under the
symmetry group. There is a continuum of degenerate ground states,
labeled by different values of the order parameter. Each of these ground states
constitutes a basis of a distinct Hilbert space, all of them bearing unitarily
inequivalent representations of the broken symmetry. This is the so-called
\emph{Nambu--Goldstone realization} of symmetry. From the phenomenological
point of view, the most important consequence of SSB is the existence of soft
modes in the spectrum, the Nambu--Goldstone bosons.
Section~\ref{sec:goldstone}\ will be devoted to a detailed discussion of this
phenomenon.


\section{Example: Free nonrelativistic particle}
\label{sec:freeNR} As an illustration of the general abstract material of the
previous section, let us consider the simplest example imaginable, the free
nonrelativistic field theory (see also Chap.~2 of \cite{Miransky:1993mi}). In
order to demonstrate the variety of technical approaches to SSB, I will follow a
different path: instead of defining the charge operator by an integral of the
charge density over a finite space domain, the whole system will be quantized
in finite volume $\Omega$.

The dynamics of the system is governed by the Lagrangian density
\begin{equation}
\Lag=\imag\he\psi\PD\psi t-\frac1{2m}\nabla\he\psi\cdot\nabla\psi
\label{NRlag}
\end{equation}
Obviously, the action defined by this Lagrangian is invariant under the
following three kinds of transformations,
\begin{equation}
\psi\xrightarrow{\#1}\psi+\vt,\quad
\psi\xrightarrow{\#2}\psi+\imag\vt,\quad
\psi\xrightarrow{\#3} e^{\imag\vt}\psi
\label{NRfreetransfo}
\end{equation}
with a real parameter $\vt$. (Complex conjugated transformations for $\he\psi$
are implied.) Note that the transformation $\#3$, corresponding to the usual
conservation of particle number, leaves the Lagrangian \eqref{NRlag} itself
invariant, while the transformations $\#1,\#2$ change it by a total derivative.
Taking this into account, one infers from the Noether theorem the following
list of conserved charge and current densities,
\begin{equation}
\begin{split}
&\vr_1=\imag(\psi-\he\psi),\quad
\vr_2=\psi+\he\psi,\quad
\vr_3=\he\psi\psi\\
&\vek j_1=\frac1{2m}\nabla(\psi+\he\psi),\quad
\vek j_2=-\frac{\imag}{2m}\nabla(\psi-\he\psi),\quad
\vek j_3=\frac{1}{2\imag m}(\he\psi\nabla\psi-\psi\nabla\he\psi)
\end{split}
\end{equation}

The three independent types of transformations listed in
Eq.~\eqref{NRfreetransfo} generate the symmetry group $\gr{ISO(2)}$, that is,
the two-dimensional Euclidean group. Transformations $\#1,\#2$ correspond to
translations in the plane defined by the real and imaginary part of $\psi$,
while $\#3$ yields the rotation in this plane. Demanding as usual that the
Schr\"odinger picture fields $\psi(\vek x),\he\psi(\vek x)$ satisfy the
canonical commutation relation,
\begin{equation}
[\psi(\vek x),\he\psi(\vek y)]=\delta^3(\vek x-\vek y)
\label{NRCCR}
\end{equation}
one easily derives the algebra of the Noether charges in finite
volume, $Q_a=\int_\Omega\dthree\vek x\,\vr_a(\vek x)$,
\begin{equation}
[Q_3,Q_1]=-\imag Q_2,\quad
[Q_3,Q_2]=+\imag Q_1,\quad
[Q_1,Q_2]=2\imag\Omega
\label{NRcommutators}
\end{equation}
While the first two relations are identical to their classical counterparts,
the two translations at the classical level commute. The last commutator in
\eqref{NRcommutators} hence shows that after quantization, the algebra
$\gr{ISO(2)}$ develops a \emph{central charge}, and the representation of
finite group transformations becomes \emph{projective}.

As an aside let me remark that a projective representation of a symmetry group
of a quantum system may arise either as a result of nontrivial global topology
of the symmetry group, or as a consequence of a central charge in the Lie
algebra as above. For semi-simple Lie algebras the central charges can always be
removed by a proper redefinition of generators. In the case of the
non-semi-simple Euclidean algebras $\gr{ISO}(N)$, the central charges vanish for
all $N$ but $N=2$, which is exactly our case. Further details on the projective
representations of symmetry groups may be found in
\cite{Weinberg:1995v1} (Sec.~2.7) or \cite{Barut:1977ba} (Chap.~13).

\subsection{Hilbert space and inequivalent ground states}
In finite volume one can always define the (countable) basis of the
physical Hilbert space using the Fock construction. In the following, we will
implicitly assume periodic boundary conditions so that (discrete) translational
invariance is preserved and the basis of one-particles states may be chosen as
the eigenstates of the momentum operator. These one-particle states, labeled by
the three-momentum $\vek k$, can thus be obtained from the Fock vacuum $\ket0$
by the action of creation operators $\he a_{\vek k}$, $\ket{\vek k}\equiv\he
a_{\vek k}\ket0$. Let us suppose for the sake of generality that the
one-particle states are normalized as $\braket{\vek k'}{\vek k}=N_{\vek
k}\delta_{\vek k\vek k'}$. Then the annihilation and creation operators satisfy
the corresponding relation $[a_{\vek k},\he a_{\vek k'}]=N_{\vek k}\delta_{\vek
k\vek k'}$. The canonical commutator of the fields \eqref{NRCCR} subsequently
fixes the normalization of their expansion in the annihilation and creation
operators,
\begin{equation}
\psi(\vek x)=\sum_{\vek k}\frac{1}{\sqrt{N_{\vek k}\Omega}}
e^{\imag\skal kx}a_{\vek k}
\end{equation}
Using this expansion of the fields, one finds an explicit representation of the
symmetry generators,
\begin{equation}
Q_1=\imag\sqrt{\frac{\Omega}{N_{\vek 0}}}(a_{\vek 0}-\he a_{\vek 0}),\quad
Q_2=\sqrt{\frac{\Omega}{N_{\vek 0}}}(a_{\vek 0}+\he a_{\vek 0}),\quad
Q_3=\sum_{\vek k}\frac1{N_{\vek k}}\he a_{\vek k}a_{\vek k}
\label{NRgenerators}
\end{equation}
From here we immediately see that $\nor{Q_{1,2}\ket0}^2=\Omega\to\infty$ in
the infinite volume (thermodynamic) limit. This clearly indicates that the
operators $Q_{1,2}$ are ill-defined on the Hilbert space built above the Fock
vacuum in the limit that the space volume is sent to infinity.

The Hamiltonian of the system is in the second quantization expressed as
\begin{equation}
H=\sum_{\vek k}\frac1{N_{\vek k}}\frac{\vek k^2}{2m}\he a_{\vek k}a_{\vek k}
\label{NRHamiltonian}
\end{equation}
Obviously, the Fock vacuum $\ket0$ represents one possible ground state with
zero energy. However, we already know that other, equivalent, ground states may
be obtained from $\ket0$ by applying symmetry transformations. Let us therefore
denote
\begin{equation}
\ket z\equiv\ket{\vt_1,\vt_2}=e^{\imag(\vt_1Q_1+\vt_2Q_2)}\ket0\quad
\text{where}\quad z=\vt_1+\imag\vt_2
\end{equation}
Substituting from \eqref{NRgenerators} yields
\begin{equation}
\ket z=\exp\left[\sqrt{\frac{\Omega}{N_\vek0}}(z\he
a_{\vek0}-z^*a_{\vek0})\right]\ket0
\end{equation}
and a simple manipulation shows that all these states are in fact coherent
states, corresponding to the eigenvalue $z\sqrt{N_{\vek0}\Omega}$ of the
annihilation operator $a_{\vek0}$. Naturally, they are annihilated by all
$a_{\vek k}$ with nonzero momentum $\vek k$. They have the same energy as the
Fock vacuum since adding an arbitrary number of zero-momentum quanta does not
change the total energy of the system. The magnitude of the scalar product of
two such states is equal to
\begin{equation}
\abs{\braket{z'}{z}}=e^{-\Omega\abs{z'-z}^2}
\end{equation}
This demonstrates explicitly the general fact that in the infinite volume
limit, any two ground states connected by a broken symmetry transformation
become orthogonal.

\subsection{Spontaneous symmetry breaking and the Nambu--Goldstone boson}
Let us first analyze the pattern of symmetry breaking when the ground state is
chosen as the Fock vacuum $\ket0$. Apparently, out of the three generators $Q_a$
only the third one preserves the ground state, the other two are spontaneously
broken. In the field space this corresponds to the fact that the full Euclidean
group $\gr{ISO(2)}$ is spontaneously broken to its rotation subgroup,
$\gr{SO(2)}$. The Goldstone theorem (to be discussed in detail in Section
\ref{sec:goldstone}) now asserts that for each broken generator
there should be a state in the spectrum which couples to the associated current.
In this case there is obviously only one such state, namely the one-particle
state $\ket{\vek k}$, and the corresponding amplitudes read
\begin{equation}
\begin{split}
\bra0\vr_1(\vek x)\ket{\vek k}&=\imag
\sqrt{\frac{N_{\vek k}}{\Omega}}e^{\imag\skal kx},\quad
\bra0\vek j_1(\vek x)\ket{\vek k}=\frac{\imag\vek k}{2m}
\sqrt{\frac{N_{\vek k}}{\Omega}}e^{\imag\skal kx}\\
\bra0\vr_2(\vek x)\ket{\vek k}&=
\sqrt{\frac{N_{\vek k}}{\Omega}}e^{\imag\skal kx},\quad
\bra0\vek j_2(\vek x)\ket{\vek k}=\frac{\vek k}{2m}
\sqrt{\frac{N_{\vek k}}{\Omega}}e^{\imag\skal kx}
\end{split}
\end{equation}
All these formulas are valid in the Schr\"odinger picture where the fields are
time independent. However, transition to the Heisenberg picture is trivial in
this noninteracting field theory.

Due to the dispersion relation, $E_{\vek k}=\vek k^2/2m$, the state $\ket{\vek
k}$ represents a NG boson with momentum $\vek k$. Since its dispersion relation
is quadratic, it is the simplest possible example of a type-II NG boson,
to be discussed in depth later. Note that there is a \emph{single NG boson}
that couples to \emph{two broken generators}. As will become clear in
Section~\ref{sec:goldstone}\ this is only possible when the commutator of the
two broken generators develops nonzero vacuum expectation value. A glance at
Eq.~\eqref{NRcommutators} shows that this is indeed the case. Moreover, our
simple noninteracting field theory provides a rather nontrivial realization of
this condition. The vacuum expectation value of the commutator
$[Q_1,Q_2]$ is guaranteed by the existence of a central charge in the quantized
theory, rather than by a nonzero charge density as one might naively expect
based on the Lie algebraic structure of the symmetry.

What now if we choose as the ground state one of the coherent states $\ket z$
with nonzero $z$? In field space this means shifting the ground state off the
origin,
\begin{equation}
\bra z\psi(\vek x)\ket z=z
\end{equation}
Since the new ground state $\ket z$ is (in finite volume) connected to the Fock
vacuum $\ket0$ by a symmetry transformation, they are physically equivalent.
The corresponding symmetry breaking patterns thus also have to be the same,
just with a different basis of broken and unbroken generators. The geometric
picture is clear: any expectation value of the field always breaks the two
translations in the field space; it only preserves the rotations about the
point $\bra z\psi(\vek x)\ket z=z$.

From a physical point of view, the states $\ket z$ are very interesting
(see Sec.~2.3 in \cite{Miransky:1993mi}). Note that the expectation of the
particle number operator $Q_3$ is
\begin{equation}
\bra zQ_3\ket z=\Omega\abs z^2
\label{NRQ}
\end{equation}
The operator $Q_3$ itself is apparently not well defined in the infinite volume
limit. However, the density $\bra z\vr_3\ket z$ remains finite in this limit.
The states $\ket z$ therefore describe a finite density free nonrelativistic
many-body system with all particles occupying the lowest energy level, that is,
a Bose--Einstein condensate (this issue is discussed with much greater
mathematical rigor in \cite{Araki:1963aw}). Different choices of $z$ correspond
to condensates with a different density and/or phase.

A cautious reader may now be confused by my previous statement, based on
general considerations, that all ground states $\ket z$ are equivalent since
they are connected by symmetry transformations. Yet, systems with different
density do not look physically equivalent. The resolution of this seeming
paradox is tightly bound with measurement theory \cite{Guralnik:1968gu}. Any
device that can measure the system density must be able to distinguish states
with different density, and therefore explicitly break those symmetry
transformations that connect these states. On the other hand, if we only allow
for a measuring apparatus that preserves the symmetry of the system, then the
different ground states will really be indistinguishable.

Let us finally comment on the general technique that allows one to pick a
single ground state out of the continuum of degenerate ones. One introduces an
explicit symmetry breaking term in the Hamiltonian which splits the energy of
the ground states. Then the infinite volume limit is performed and only
afterwards the symmetry breaking term is removed. In our case such an explicit
symmetry breaking term can be proportional to the particle number operator
$Q_3$. One replaces the Hamiltonian \eqref{NRHamiltonian} with $H_\mu=H-\mu
Q_3$. The parameter $\mu$ plays the role of a chemical potential. In view of
Eq.~\eqref{NRQ} the energy of the state $\ket z$ is then $\bra zH_\mu\ket
z=-\mu\Omega\abs z^2$. In order that the system has a ground state, the
chemical potential must be non-positive. For any strictly negative value of
$\mu$ the Fock vacuum $\ket0$ becomes the single nondegenerate ground state,
while the energy of all other coherent states goes to infinity in the
thermodynamic limit. In other words, in presence of a Bose--Einstein condensate
the chemical potential must be zero, as is of course well known. Obviously, any
of the coherent states $\ket z$ can be selected as the preferred one by a proper
choice of the explicit symmetry breaking term.

Even though this section was devoted to a free nonrelativistic particle, it is
amusing to compare to the free \emph{relativistic} massless complex scalar
field theory. There, the action possesses the same symmetry group
$\gr{ISO(2)}$, while the spectrum contains two NG bosons with energy
proportional to momentum: the quantum of the field and its antiparticle. The
spectrum is therefore dramatically different even though the symmetry breaking
pattern is the same. The reason is that in the relativistic case the shift
generators $Q_1$ and $Q_2$ commute even after quantization since the
contributions to the commutator from the particle and antiparticle sectors
cancel each other. The algebra $\gr{ISO(2)}$ thus does not develop a
central charge and the implementation of broken symmetry as well as the
spectrum of NG bosons is usual.


\section{Achieving spontaneous symmetry breaking: Minimization of Higgs
potentials}
\label{sec:higgs}
Demonstrating that a given physical system exhibits SSB requires two
key steps. First, one has to identify a suitable order parameter. This is
usually very difficult to do from first principles. The reason is that SSB is a
nonperturbative phenomenon, so it cannot be achieved at any finite order of
perturbation theory based on a ground state which preserves the symmetry. One
then often relies on observations (such as in the case of ferromagnets
where the order parameter, the spontaneous magnetization, is obvious) or
physical insight (such as in the case of superconductors, where pair
correlations are crucial).

The second step is the actual calculation of the order parameter. From the
qualitative point of view, there are essentially two approaches. The first,
physically more satisfactory but technically much more difficult, is to find a
self-consistent (thus nonperturbative) symmetry-breaking solution to the
equations of motion, starting from whatever degrees of freedom are present in
the theory. The NG bosons then typically appear as collective modes of these
elementary degrees of freedom. This is the case, for instance, of the
Heisenberg model of a ferromagnet to be discussed in Section~\ref{sec:examples},
the celebrated Nambu--Jona-Lasinio model \cite{Nambu:1961tp,Nambu:1961fr}, or
the technicolor theories in high energy physics. The most serious technical
drawback of this approach is that one usually has to introduce an Ansatz for the
symmetry-breaking solution and show in turn that it is consistent with the
equation of motion. Thus one only gets such solutions that are put in by hand
from the beginning.

The second, more phenomenological approach is to introduce an effective field
with suitable transformation properties under the symmetry in question, which
creates the NG bosons and whose ground state expectation value at the same time
serves as the order parameter. This is done, for example, in the Higgs mechanism
in the standard model of elementary particles or in the Ginzburg--Landau theory
of phase transitions. While the minimization of a quartic potential with one
complex doublet scalar field as in the standard model is straightforward, it
can become a complicated task provided the order parameter transforms in a
higher-dimensional representation of the symmetry group such as in some models
of grand unification. It is therefore worthwhile to ask what can be said about
the minima of a potential just from symmetry considerations. The investigation
of this issue was initiated by Michel \cite{Michel:1971th,Michel:1980pc}, and
in the next subsection I will review some basic results. A heuristic method to
minimize general Higgs-type potentials, developed by Kim \cite{Kim:1981xu},
will then be presented and illustrated on an explicit nontrivial example.

\subsection{Group action on the order parameter space}
\label{subsec:Michel} Let me introduce, in a non-rigorous manner following
Sec.~6.2 of \cite{Vollhardt:1990vw}, some basic mathematical terms needed for
the discussion of invariant potentials and their minima. Let $\gr G$ be a
compact Lie group, acting smoothly on an infinitely differentiable manifold $M$.
The points on this manifold may be thought of as values of the order parameter,
and the action of a group element, $g\in\gr G$, on a point $\phi\in M$ will be
denoted simply as $g\phi$. For each point we also denote as $\gr H_\phi$ the
set of all group elements that leave $\phi$ invariant, that is, $\gr
H_\phi=\{g\in\gr G\,|\,g\phi=\phi\}$. This set forms a subgroup of $\gr G$ and
is called the \emph{little (or isotropy) group} of $\phi$. In physical terms,
it consists exactly of those transformations left unbroken when the order
parameter takes the value $\phi$.

A very important notion that we will further use is that of the \emph{orbit}
$G(\phi)$ of a given point $\phi$. It consists of all points of $M$ which can
be reached from $\phi$ by a (naturally, broken) symmetry transformation,
$G(\phi)=\{g\phi\,|\,g\in\gr G\}$. The relation defined by the condition that
two points be connected by a group transformation is an equivalence, and
the group orbits then define a decomposition of the manifold $M$ into
equivalence classes. Any potential $V$ on the manifold which is invariant under
the group action, $V(\phi)=V(g\phi)$ for all $\phi\in M$ and $g\in\gr G$, may
be thought of as a function on the orbits. The minimization problem for a given
potential on $M$ can therefore be reformulated as a minimization of a function
on the space of orbits.

It is clear from the definition of the orbit that two points on the same orbit
have isomorphic little groups. This is in fact an immediate consequence of the
stronger statement that the two little groups are \emph{conjugate}, $\gr
H_{g\phi}=g\gr H_\phi g^{-1}$. In addition to the orbit $G(\phi)$ there may be
other points on the manifold whose little group is conjugate to $\gr H_\phi$.
For example, when $M$ is a linear space such as in the Higgs mechanism,
multiplying $\phi$ by any (nonzero) number we obviously get a point with the
same little group as $\phi$. The set of all points with little groups conjugate
to $\gr H_\phi$ is called a \emph{stratum}, $S(\phi)$. Intuitively, a stratum
consists of all points of the same symmetry ``class'': they have the same
unbroken subgroup and the same symmetry-breaking pattern. In the phase
diagram, a stratum would be associated with a particular phase. At a phase
transition, the order parameter moves from one stratum to another, and the
symmetry class changes.

We are now ready to formulate a fundamental theorem which constrains possible
stationary points of group-invariant potentials \cite{Michel:1971th}:
\begin{theorem}[Michel]
\label{thm:michel1}
Let $\gr G$ be a compact Lie group acting smoothly on the real manifold $M$ and
let $\phi\in M$. Then the orbit $G(\phi)$ is critical, that is, \emph{every
smooth real $\gr G$-invariant function on $M$ is stationary on $G(\phi)$ if and
only if $G(\phi)$ is isolated in its stratum}, that is, there is a neighborhood
$U_\phi$ of $\phi$ such that $U_\phi\cap S(\phi)=G(\phi)$.
\end{theorem}
It is worth emphasizing that the theorem makes no particular assumption about
the form of the invariant function, so it may be a Higgs-type quartic
potential as well as the full quantum effective potential whose power expansion
would contain terms of arbitrarily high orders in $\phi$. In case $M$ is a
linear space Theorem \ref{thm:michel1} is actually not very useful. For every
point $\phi\in M$ its stratum $S(\phi)$ also contains all points $\lambda\phi$
with nonzero $\lambda$, and the orbit $G(\phi)$ is therefore never isolated in
its stratum unless $\phi=0$. So in this case, the theorem simply asserts a
rather obvious fact that the origin $\phi=0$ is a stationary point of every $\gr
G$-invariant function on the space $M$.

On the other hand when the manifold $M$ is not a linear space, Theorem
\ref{thm:michel1} may have remarkably strong consequences. Let us consider a
manifold $M$ made of the values of the order parameter of fixed ``length'' and
arbitrary ``direction'', that is, constrain the linear space of the previous
paragraph by the requirement $\nor\phi=1$, where $\nor\cdot$ is a suitably
defined norm. This is a situation one deals with in the low energy effective
descriptions of spontaneously broken symmetries (such as the so-called nonlinear
sigma model). As I will show in the next subsection, also the minimization of a
quartic Higgs potential can under certain circumstances be reduced to
minimization over directions of the order parameter, its norm being fixed. When
classifying possible shapes of the order parameter by their symmetry properties,
one often encounters the situation that the form of the order parameter is fixed
up to a symmetry transformation (and trivial rescaling). These states are called
\emph{inert} in condensed matter physics (see Sec.~6.2.3 in
\cite{Vollhardt:1990vw}). By definition, their stratum consists of a single
(hence isolated) orbit. The inert states are therefore found among stationary
states of any $\gr G$-invariant function on the manifold of order parameters of
fixed norm. Explicit examples will be shown later.

Before concluding the general discussion let me stress that an invariant
function can, of course, have other stationary states than those guaranteed by
Theorem \ref{thm:michel1}. Indeed, \emph{any} state can be realized as a
minimum of a suitably defined potential. To see this, note that a given group
orbit can be uniquely specified by the values $c_\alpha$ of a set of group
invariants, say $\PP_\alpha(\phi)$. The desired potential then reads
$f(\phi)=\sum_\alpha[\PP_\alpha(\phi)-c_\alpha]^2$. Such potentials will,
however, typically involve terms of high order. If we restrict ourselves to
Higgs-type polynomials of fourth order, the list of allowed stationary states
may become more narrow. Another point to emphasize is that Michel's theorem
\ref{thm:michel1} makes a claim about stationary states of a given function,
while in physical applications one would usually like to know the (absolute)
minimum. To that end, Michel conjectured \cite{Michel:1980pc} that if the
representation of the symmetry group $\gr G$ of the Higgs-type fourth order
potential is irreducible on the real, its absolute minima are realized by orbits
with maximal little groups. Even though this rule turned out not be hold in
general (see \cite{Slansky:1981yr,Meljanac:1985br} and references therein), it
provides a useful guide for locating the minimum of Higgs potentials. A more
rigorous and complete analysis of this problem may be found in
\cite{Abud:1983id}.

\subsection{Minimization of Higgs potentials}
An elegant general method to minimize Higgs potentials was developed by Kim
\cite{Kim:1981xu}. I will restrict to the simplest case where the order
parameter transforms in an irreducible representation of the symmetry group.
Then there is a single invariant quadratic term in the potential, proportional
to $\sum_i\phi_i^2\equiv\nor\phi^2$. (We can without lack of generality assume
that the components $\phi_i$ of the order parameter $\phi$ are real.) Assuming
further reflection symmetry or any other constraint which rules out a cubic
term, the most general potential up to fourth order in the field reads
\begin{equation}
V(\phi)=-\frac12m^2\nor\phi^2+\frac14\sum_\alpha\lambda_\alpha
\PP^{(4)}_\alpha(\phi)
\label{GLpot}
\end{equation}
where $\PP_\alpha^{(4)}$ is a set of algebraically independent fourth-order
invariants in $\phi$. (The mass term $m^2$ was written suggestively with a
minus sign to ensure that a nontrivial minimum exists.) One of these invariants
is always $\PP^{(4)}_1(\phi)=(\nor\phi^2)^2$. Factoring it out, one gets
\begin{equation}
V(\phi)=-\frac12m^2\nor\phi^2+\frac14\left(\nor\phi^2\right)^2
\biggl[\lambda_1+\sum_{\alpha\neq1}\lambda_\alpha\AP_\alpha(\phi)\biggr]
\end{equation}
where $\AP_\alpha(\phi)=\PP^{(4)}_\alpha(\phi)/\PP^{(4)}_1(\phi)$. These
quantities are dimensionless and represent the ``angles'', that is, the
orientation of the condensate in field space. The main observation now is that
minimization of the potential with respect to $\phi$ is equivalent to
successive minimization with respect to the norm $\nor\phi$ and the angles
$\AP_\alpha$. For fixed values of the angles, the norm of the field and the
value of the potential in the minimum are given by
\begin{equation}
\nor\phi^2_{\text{min}}=\frac{m^2}{\lambda_1+\sum_{\alpha\neq1}
\lambda_\alpha\AP_\alpha},\quad
V_{\text{min}}(\phi)=-\frac14m^2\nor\phi^2_{\text{min}}
\label{kimaux}
\end{equation}

\begin{figure}
\begin{center}
\usepgflibrary{arrows}
\begin{tikzpicture}
  \draw (-0.5,0) to (5,0);
  \draw (0,-0.5) to (0,5);
  \draw (4.8,-0.3) node {$\AP_2$};
  \draw (-0.3,4.8) node {$\AP_3$};
  \filldraw[color=gray] (2,1.5) to [out=60,in=200] (4,3) to [out=120,in=270]
                        (3.5,5) to [out=170,in=40] (1.5,4.5) to [out=275,in=60]
                        (1,2.5) to [out=-20,in=110] (2,1.5);
  \draw[color=red] (-1,0.5) to (5,2.5);
  \draw[color=red,style=dashed] (-1,-0.2) to (5,1.8);
  \filldraw[color=red] (2,1.5) circle (2pt);
  \draw (2,1.2) node {$K$};
  \draw (4.2,3) node {$L$};
  \draw (3.8,5.1) node {$M$};
  \draw (1.3,4.6) node {$N$};
  \draw (0.8,2.5) node {$O$};
  \draw[->,>=stealth',thick] (-0.5,2.0/3.0) to (-5.0/6.0,5.0/3.0);
  \draw (-1.5,1.2) node {$(\lambda_2,\lambda_3)$};
\end{tikzpicture}
\end{center}
\caption{Geometric minimization of the quantity
$\Xi(\AP_\alpha)\equiv\sum_{\alpha\neq1}\lambda_\alpha\AP_\alpha$. The region
of allowed values of $\AP_\alpha$ is indicated by gray shading. The lines of
constant $\Xi(\AP_\alpha)$, having the common normal vector
$(\lambda_2,\lambda_3)$, are in red. The red dot denotes the position of the
absolute minimum of the potential.}
\label{fig:Kim}
\end{figure}
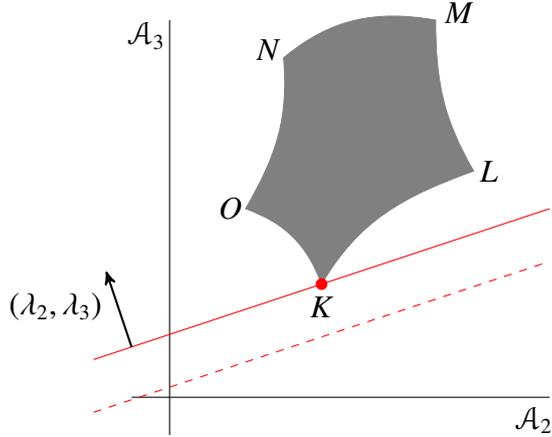

In order to find the absolute minimum of the potential, one has to maximize
$\nor\phi_{\text{min}}$, and hence minimize the expression
$\Xi(\AP_\alpha)\equiv\sum_{\alpha\neq1}\lambda_\alpha\AP_\alpha$ for given
couplings $\lambda_\alpha$. This may be done using an appealing geometric
picture. Consider for simplicity the case of three independent quartic
invariants $\PP_\alpha^{(4)}(\phi)$, that is, two angles, $\AP_2$ and $\AP_3$.
These cannot acquire arbitrary values, but rather span some domain in a
two-dimensional plane which is indicated by the gray shaded area in
Fig.~\ref{fig:Kim}; it will be referred to as the \emph{target space}. The shape
of the target space is independent of the couplings $\lambda_\alpha$,
and is a sole characteristics of the algebraic structure of the symmetry group
and its particular representation.

The set of constant $\Xi(\AP_\alpha)$ is in the $(\AP_2,\AP_3)$ plane
represented by a straight line with the normal vector $(\lambda_2,\lambda_3)$.
If we choose too low a value of $\Xi$, the line will not intersect the target
space, that is, there is no $\phi$ that would yield the desired value of
$\Xi$. This corresponds to the dashed red line. As we increase $\Xi$, the line
shifts parallel until it touches for the first time the target space (solid
line, point $K$). This point determines the absolute minimum of
$\Xi(\AP_\alpha)$ and thus also the absolute minimum of the potential
$V(\phi)$. It is then a matter of mere algebra do find out which shapes of the
order parameter map onto the point $K$ in the target space.

This geometric picture can be conveniently used to scan the whole phase diagram
as the couplings $\lambda_\alpha$ are varied. First, it is obvious that the
shape of the order parameter in the minimum does not depend on $\lambda_1$;
this merely affects its overall magnitude, and at the same time must be large
enough to ensure the boundedness of the potential from below. Coming back to
the example with three quartic invariants, the phase diagram can then be
plotted in the $(\lambda_2,\lambda_3)$ plane. For any values of the couplings,
the ground state will always be represented by a point on the boundary of the
target space. If the boundary is concave, as is the case of its edges adjacent
to the point $K$, then by tuning the couplings continuously the ground state
will reside at $K$, and then change abruptly to $L$ or $O$. In other words, the
system will undergo a first-order phase transition. If, on the other hand, the
boundary has a convex segment, such as the edge between the $M,N$ vertices,
then tuning the couplings smoothly will result in a continuous change of the
ground state. In the phase diagram, one will then observe an $M$-phase, an
$MN$-phase (where the ground state travels from $M$ to $N$ and typically has a
different symmetry structure), and finally an $N$-phase, the three of them
separated by a sequence of two second-order transitions. An explicit example
will be given in the next subsection.

When the number of quartic invariants differs from three, the situation is very
similar. First of all, in the case of two invariants only the solution is
trivial. The target space is one-dimensional and there are just two phases,
corresponding to the minimum and maximum allowed values of $\AP_2(\phi)$. This
happens, for instance, in the Color--Flavor Locked phase of dense three-flavor
quark matter \cite{Iida:2000ha}. For more quartic invariants the target space
is multidimensional, and keeping track of all faces and edges of its boundary,
which determine the phases appearing in the phase diagram, may become
increasingly difficult. Yet the presented technique seems to be the most
elegant way to accomplish this task. The same method can also be used, with
necessary complications, when the representation of the order parameter is
reducible, or when other (cubic or higher-order) terms are present in the
potential.

\subsection{Example: spin-one color superconductor}
\label{subsec:spin1CSC}
As an explicit nontrivial example, I will now discuss the phases of a spin-one
color superconductor. The details of all derivations may be found in
\cite{Brauner:2008ma}, see also \cite{Bailin:1983bm,Schmitt:2004et}
for a partial discussion of the problem. A similar analysis for the case of
$d$-wave pairing, that is, a theory of a traceless complex symmetric matrix with
$\gr{SO(3)\times U(1)}$ symmetry, was performed in \cite{Mermin:1974me}.

\begin{table}
\setlength{\tabcolsep}{2pt}
\begin{center}
\begin{tabular}{||c|c|c|c||}
\hline\hline
Oblate & Cylindrical & $\ve$ & A\\
\hline $\gr{SO(2)_V}$ & $\gr{SO(2)_V\times U(1)_L}$ & $\gr{SO(2)_V\times
U(1)_L}$ &
$\gr{SU(2)_L\times SO(2)_V\times U(1)_L}$\\
\hline $\begin{pmatrix}
\Delta_1 & +\imag a & 0\\
-\imag a & \Delta_1 & 0\\
0 & 0 & \Delta_2
\end{pmatrix}$
& $\begin{pmatrix}
\Delta & +\imag a & 0\\
-\imag a & \Delta & 0\\
0 & 0 & 0
\end{pmatrix}$
& $\begin{pmatrix}
\Delta_1 & +\imag\Delta_1 & 0\\
-\imag\Delta_1 & \Delta_1 & 0\\
0 & 0 & \Delta_2
\end{pmatrix}$
& $\begin{pmatrix}
1 & +\imag & 0\\
-\imag & 1 & 0\\
0 & 0 & 0
\end{pmatrix}$\\
\hline\hline
CSL & Polar & $\gr{N_1}$ & $\gr{N_1}$\\
\hline $\gr{SO(3)_V}$ & $\gr{SU(2)_L\times SO(2)_R\times U(1)_L}$ &
$\gr{U(1)_L}$ &
$\gr{SU(2)_L\times U(1)_L}$\\
\hline $\begin{pmatrix}
1 & 0 & 0\\
0 & 1 & 0\\
0 & 0 & 1
\end{pmatrix}$
& $\begin{pmatrix}
0 & 0 & 0\\
0 & 0 & 0\\
0 & 0 & 1
\end{pmatrix}$
& $\begin{pmatrix}
0 & 0 & 0\\
z_1 & z_2 & z_3\\
z_4 & z_5 & z_6
\end{pmatrix}$
& $\begin{pmatrix}
0 & 0 & 0\\
0 & 0 & 0\\
z_1 & z_2 & z_3
\end{pmatrix}$\\
\hline\hline
\end{tabular}
\end{center}
\caption{Classification of different ground states of a spin-one color
superconductor, based on the pattern of spontaneous breaking of
\emph{continuous} symmetries. First line: name of the phase; second line:
unbroken continuous symmetry; third line: representative element of the
stratum. Lower indices ${}_{\text{L,R}}$ denote subgroups of $\gr{U(3)_L}$ and
$\gr{SO(3)_R}$, while ${}_{\text V}$ stands for a ``diagonal'' subgroup,
mixing transformations from the two.}
\label{tab:spin1clas}
\end{table}
Spin-one color superconductivity is a viable candidate phase for the ground
state of cold dense quark matter at moderate densities (see
\cite{Alford:2007xm} for a review). The most favored pairing pattern involves
quarks of a single flavor, pairing in the color $\gr{SU(3)}$ antitriplet and
spin $\gr{SO(3)}$ triplet channel. All we need to know for the purposes of the
present paper is that the order parameter $\op$ is a complex $3\times3$ matrix,
transforming as $\op\to U\op R$ where $U\in\gr{U(3)_L}$ is unitary and
$R\in\gr{SO(3)_R}$ is orthogonal. (The overall phase transformations,
complementing the color $\gr{SU(3)}$, stem from conservation of baryon number.)
Given the fact that a complex $3\times3$ matrix involves eighteen degrees of
freedom, twelve of which can be transformed away, the order parameter is fully
specified by a set of six independent real numbers. It can always be cast in the
form
\begin{equation}
\op=\begin{pmatrix}
\Delta_1 & \imag a_3 & -\imag a_2\\
-\imag a_3 & \Delta_2 & \imag a_1\\
\imag a_2 & -\imag a_1 & \Delta_3
\end{pmatrix}
\end{equation}
This expression is very convenient for the analysis of the possible
inequivalent structures of the order parameter, that is, the group orbits and
strata. Focusing for simplicity on continuous symmetry transformations only,
there are altogether eight nontrivial strata, summarized in
Tab.~\ref{tab:spin1clas}. Three of them correspond to inert states: A, polar,
and CSL (abbreviated from Color--Spin Locking). It is worth emphasizing that the
isomorphy of little groups of two different values of the order parameter does
not necessarily imply that they lie in the same stratum. Indeed, the cylindrical
and $\ve$, as well as the polar and A, phases do have isomorphic little groups.

\begin{figure}
\parbox{0.5\textwidth}{\includegraphics[width=0.48\textwidth]{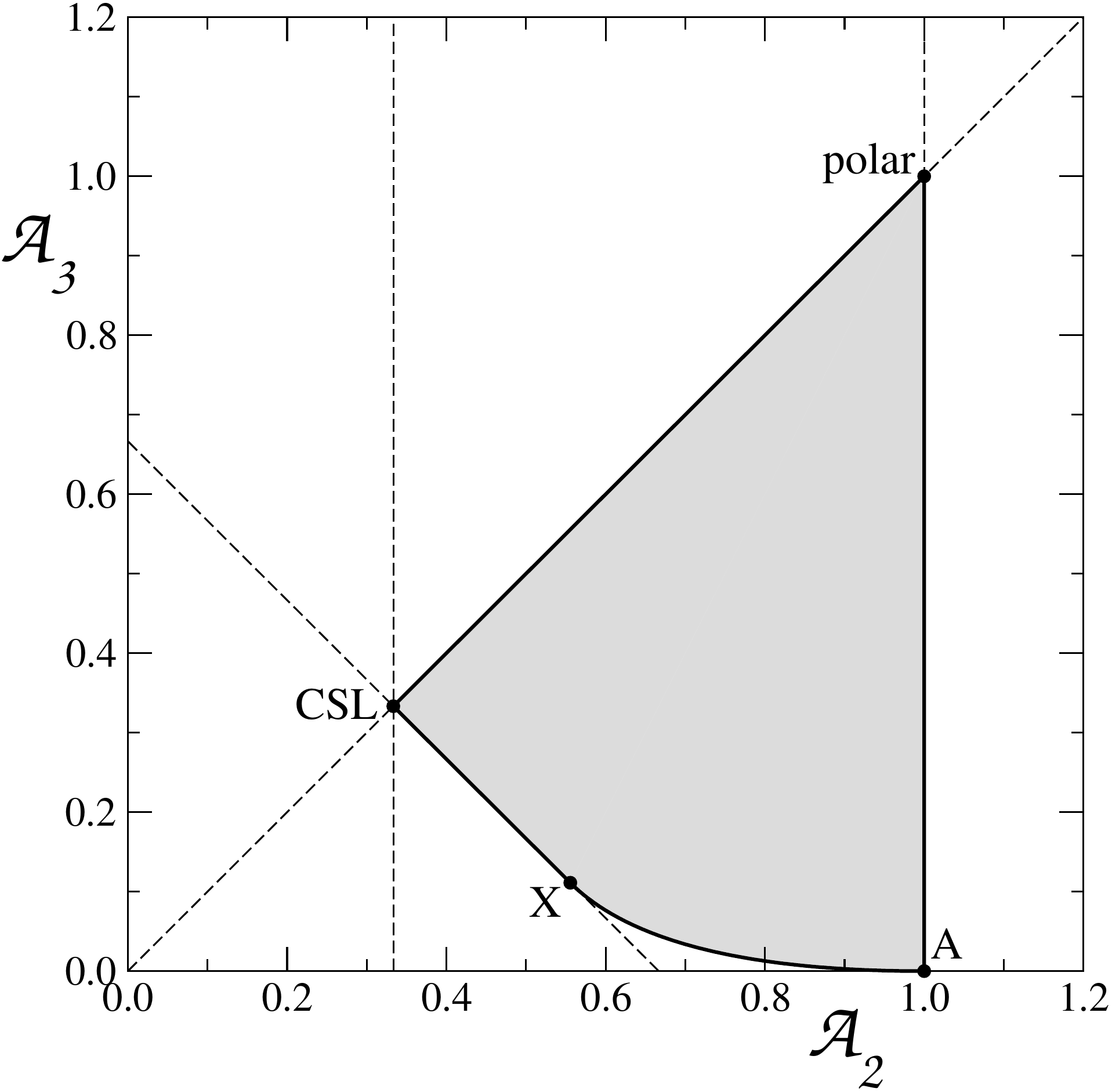}}\hfill
\parbox{0.45\textwidth}{\includegraphics[width=0.45\textwidth]{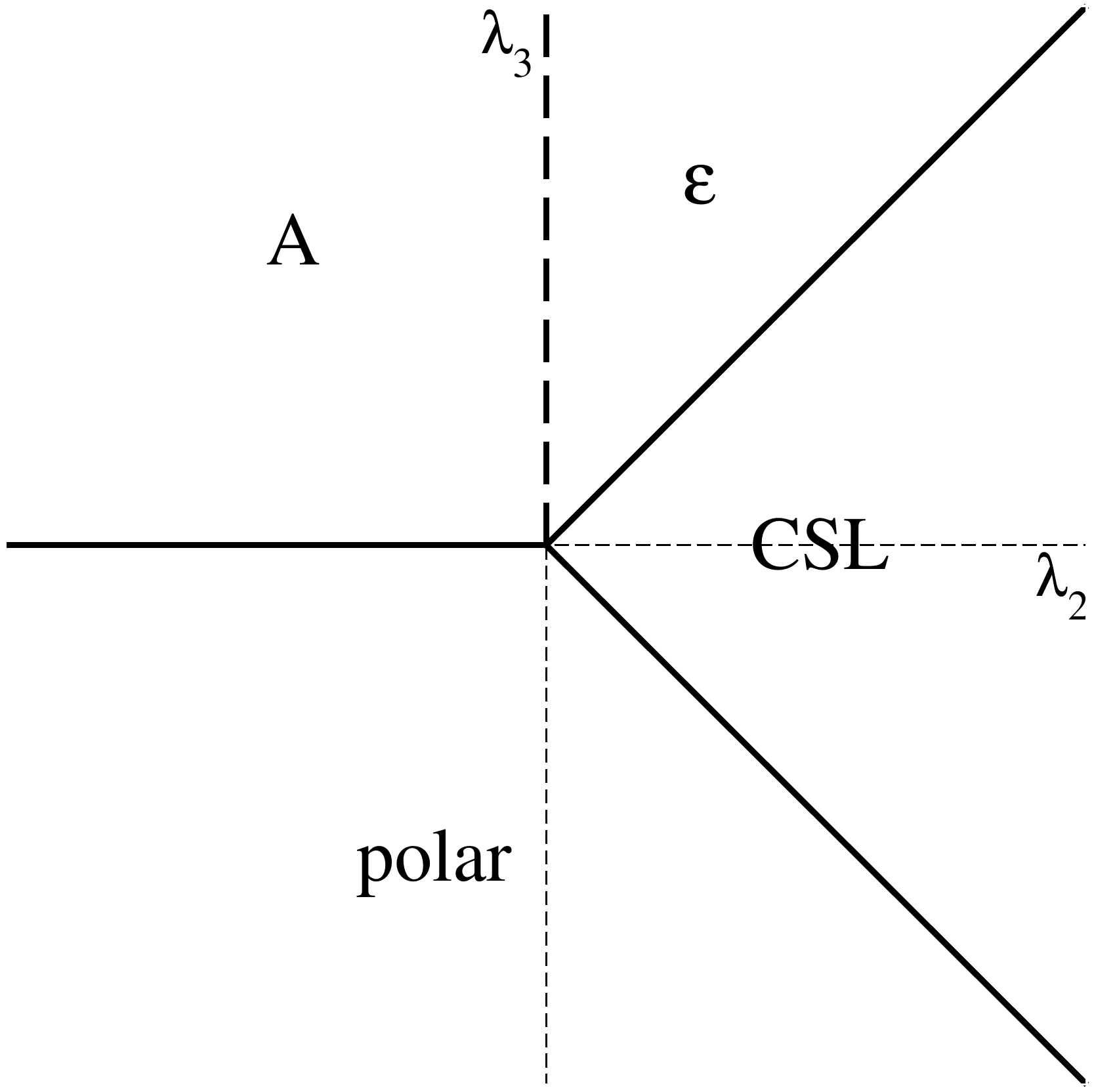}}
\caption{Left panel: target space of a spin-one color superconductor. The
dashed lines are given by various bounding inequalities as described in the
text. The $AX$ segment of the boundary is occupied by the noninert $\ve$ phase.
Right panel: phase diagram of the spin-one color superconductor as a function
of the quartic couplings. Thick solid and dashed lines stand for first and
second order phase transitions, respectively.}
\label{fig:spin1PD}
\end{figure}
To decide which of these phases occupy a part of the phase diagram one needs to
write down the most general $\gr{U(3)_L\times SO(3)_R}$ invariant potential of
the Higgs type \eqref{GLpot},
\begin{equation}
V(\op)=-\frac12m^2\sqrt{\PP^{(4)}_1(\op)}+\frac14\sum_{\alpha=1}^3
\lambda_\alpha\PP^{(4)}_\alpha(\op)
\end{equation}
where
\begin{equation}
\PP^{(4)}_1(\op)=[\Tr(\op\he\op)]^2,\quad
\PP^{(4)}_2(\op)=\Tr(\op\he\op\op\he\op),\quad
\PP^{(4)}_3(\op)=\Tr[\op\op^T\he{(\op\op^T)}]
\end{equation}
The three independent quartic invariants satisfy the following inequalities
which establish the shape of the target space:
\begin{equation}
\frac{\PP^{(4)}_1}3\leq\PP^{(4)}_2\leq\PP^{(4)}_1,\quad
0\leq\PP^{(4)}_3\leq\PP^{(4)}_2,\quad
\frac23\PP^{(4)}_1\leq\PP^{(4)}_2+\PP^{(4)}_3
\end{equation}
and finally
\begin{equation}
\sqrt{\PP^{(4)}_1}\leq\sqrt{\PP^{(4)}_3}+\sqrt{\PP^{(4)}_2-\PP^{(4)}_3}
\end{equation}
which holds only for those values of $\op$ satisfying
$\PP^{(4)}_3\leq\PP^{(4)}_1/9$. The conditions for the saturation of these
inequalities determine which strata appear on the boundary of the target space,
that is, what are the candidate ground states.

The target space of the spin-one color superconductor is plotted in
the left panel of Fig.~\ref{fig:spin1PD}. The three inert states are represented
by single points, forming the corners of the target space. This is in
agreement with the general discussion since their strata consist of a single
orbit each. Accordingly, they occupy the majority of the phase diagram, shown in
the right panel of Fig.~\ref{fig:spin1PD}. However, a part of the phase diagram
belongs to the noninert $\ve$ phase, which corresponds to the $AX$ segment of
the boundary of the target space. This phase was missed in a previous analysis
\cite{Bailin:1983bm}. In contrast to the sketch in Fig.~\ref{fig:Kim}, the
remaining segments of the target space boundary are not strictly convex, but
straight. Consequently, right at the phase transition between two adjacent
phases, many more states can actually coexist than just the two. Thus, the
A--polar line is occupied by all matrices of rank one, the polar--CSL
line by all real matrices (up to a symmetry transformation), and the CSL--X
segment by all matrices of the oblate type with
$\Delta_2=\sqrt{\Delta_1^2+a^2}$.

\begin{figure}
\usepgflibrary{arrows}
\begin{center}
\begin{tikzpicture}[->,auto,>=triangle 45]
  \node (csl) at (0,0) {\fbox{CSL}};
  \node (a) at (3,0) {\fbox{A}};
  \node (pol) at (6,0) {\fbox{Polar}};
  \node (cyl) at (0,-3) {\fbox{Cylindrical}};
  \node (eps) at (3,-3) {\fbox{$\ve$}};
  \node (n2) at (6,-3) {\fbox{$\gr{N_2}$}};
  \node (obl) at (1.5,-6) {\fbox{Oblate}};
  \node (n1) at (4.5,-6) {\fbox{$\gr{N_1}$}};
  \draw (a) to (cyl);
  \draw (cyl) to (obl);
  \draw (a) to (eps);
  \draw (eps) to (n1);
  \draw (pol) to (eps);
  \draw (a) to (n2);
  \draw (eps) to (obl);
  \draw (pol) to (n2);
  \draw (n2) to (n1);
  \draw (cyl) to (n1);
  \draw (csl) to [out=225,in=180] (obl);
\end{tikzpicture}
\end{center}
\caption{Hierarchy of the little groups in a spin-one color superconductor.
Arrows indicate successive breaking into smaller and smaller subgroups.
The relation $\gr{G_1\to G_2}$ means that the group $\gr{G_2}$ is conjugate to
a subgroup of $\gr{G_1}$, or in other words, the orientation of the order
parameter in the two phases can be chosen so that $\gr{G_2}\subset\gr{G_1}$.}
\label{fig:spin1scheme}
\end{figure}
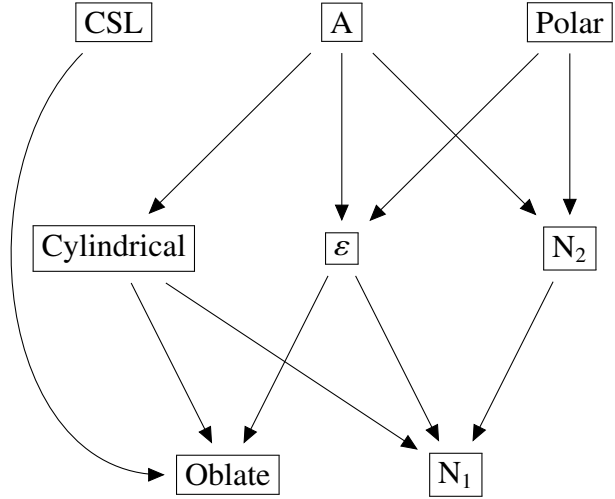
In view of Michel's conjecture mentioned at the end of
Section~\ref{subsec:Michel}\ it is useful to investigate the little groups of
the eight strata listed in Tab.~\ref{tab:spin1clas}. Their hierarchy is
displayed in Fig.~\ref{fig:spin1scheme}. The three inert states all have
maximal little groups, which is actually a general feature of the inert states.
On the other hand, the little group of the $\ve$ phase is non-maximal, and its
presence in the phase diagram thus provides a counterexample to Michel's
conjecture. Note that this does not follow from the mere fact that the $\ve$
state is noninert, as even such states can in principle have maximal little
groups (see Sec.~6.2.4 in \cite{Vollhardt:1990vw}). The $\ve$ and A phases are
connected by a second-order phase transition; the fact that the little group of
the $\ve$ phase is a subgroup of that of the A phase is in accord with the
general Landau theory of phase transitions. At the other end of the hierarchy of
symmetry breaking stand the oblate and $\text N_1$ states which have no
nontrivial (continuous) subgroups, hence representing the most general states
preserving a part of the symmetry.


\section{Goldstone theorem and the counting of Nambu--Goldstone bosons}
\label{sec:goldstone}

\subsection{Goldstone theorem}
One of the most striking consequences of SSB is the existence of soft modes in
the spectrum whose energy vanishes in the long-wavelength limit. This is the
celebrated Goldstone theorem \cite{Goldstone:1961eq,Goldstone:1962es} and the
soft modes are usually referred to as the Nambu--Goldstone bosons. Since the
aim of the paper is to discuss SSB in nonrelativistic systems, which involves a
number of subtleties, I will now briefly review the standard proof of the
Goldstone theorem in a version applicable to these systems.

From a physical point of view, the most important ingredient responsible for
the presence of NG bosons is, apart from the symmetry breaking itself, the
existence of a conserved charge. In Section~\ref{subsec:groundstate}\ the
integral charge $Q_\Omega(t)$ was defined in finite volume and it was shown
that it does not exist in the limit $\Omega\to\infty$. Since SSB is based on
the existence of the order parameter \eqref{SSBcondition} which, being a
commutator of the broken charge, is well defined even in infinite volume, let
us inspect its time dependence. To that end, one takes the commutator
$[\de_\mu j^\mu(x),\Phi]$ and integrates it over the domain $\Omega$. Using the
continuity equation and the Gauss theorem one arrives at
\begin{equation}
\de_0[Q_\Omega(t),\Phi]+\int_{\de\Omega}[\dd\vek\sigma\cdot\vek j,\Phi]=0
\label{surface_int}
\end{equation}
The key technical assumption of the Goldstone theorem, apart from translational
invariance of the vacuum, is the vanishing of the surface integral in the
infinite volume limit. This, upon taking the vacuum expectation value,
guarantees that the \emph{order parameter is time independent}. The same remark
as at the end of Section~\ref{subsec:groundstate}\ applies here: in causal
theories the vanishing of the surface term is guaranteed as long as the
operator $\Phi$ is localized to a finite domain of spacetime. (In practice, it
is often even strictly local.) In acausal theories such as some nonrelativistic
models with instantaneous interaction, the surface integral tends to zero in
the infinite volume limit when the interaction is of finite range or decreases
exponentially with distance. In case of long-range interactions, however, the
disappearance of the surface term must be checked case by case.

With these remarks in mind, I will from now on simply assume that the surface
term in Eq.~\eqref{surface_int} vanishes. This makes sure that the order
parameter is time independent as $\Omega\to\infty$. It is given by the spatial
integral of $\bra0[j^0(\vek x,t),\Phi]\ket0$. Inserting partition of unity in
terms of eigenstates of the Hamiltonian and using translational invariance of
the vacuum, this becomes
\begin{equation}
\bra0[j^0(\vek x,t),\Phi]\ket0=\sum_n\int\frac{\dthree\vek k}{(2\pi)^3}
\left[e^{-\imag k\cdot x}\bra0j^0(0)\ket{n_{\vek k}}\bra{n_{\vek k}}\Phi\ket0-
e^{+\imag k\cdot x}\bra0\Phi\ket{n_{-\vek k}}\bra{n_{-\vek k}}j^0(0)\ket0
\right]
\label{Goldstone_proof}
\end{equation}
The summation runs over discrete labels distinguishing different excitation
branches in the spectrum as well as continuous internal variables of the
multiparticle states. The states are normalized by $\braket{n_{\vek k}}{m_{\vek
q}}=(2\pi)^3\delta_{mn}\delta^3(\vek k-\vek q)$. Integral over total
three-momentum of the modes is indicated explicitly. Integrating now over the
domain $\Omega$ yields
\begin{multline}
\bra0[Q_\Omega(t),\Phi]\ket0=\sum_n\int\frac{\dthree\vek k}{(2\pi)^3}
\left[e^{-\imag E_{n,\vek k}t}\vp_\Omega(\vek k)\bra0j^0(0)
\ket{n_{\vek k}}\bra{n_{\vek k}}\Phi\ket0-\right.\\
\left.-e^{+\imag E_{n,-\vek k}t}\vp_\Omega(-\vek k)
\bra0\Phi\ket{n_{-\vek k}}\bra{n_{-\vek k}}j^0(0)\ket0 \right]
\label{Goldstone_proof2}
\end{multline}
where $\vp_\Omega(\vek k)=\int_\Omega\dthree\vek x\,e^{\imag\skal kx}$. In a
large volume, this function is strongly peaked around $\vek k=\vek0$ and thus
only states with low momentum will contribute to the right hand side. At the
same time we know that as $\Omega\to\infty$ the left hand side becomes time
independent. This is only possible when the energy of the contributing states
vanishes in the limit of zero momentum. Finally, we use the broken symmetry
assumption which states that the left hand side actually is nonzero, so that
there must be at least one state that couples to both the broken current
$j^0(x)$ and the interpolating field $\Phi$. The most general formulation of
the Goldstone theorem therefore is:
\begin{theorem}[Goldstone]
Spontaneous breaking of a continuous global internal symmetry implies the
existence of a mode in the spectrum such that $\lim\limits_{\vek
k\to\vek0}E_{\vek k}=0$.
\end{theorem}
Several remarks to the theorem, its technical assumptions and derivation are in
order. First of all, the theorem guarantees existence of $\emph{a}$ NG mode in
the spectrum. In the most general formulation it does not tell us \emph{how
many} NG bosons there are. In Lorentz invariant theories there turns out to be
exactly one NG boson for each broken generator. (Also, due to the strong
constraints on the form of the dispersion relation, a NG boson is then simply a
massless particle.) Perhaps this is the reason why, somewhat unfortunately,
many textbooks focused on relativistic quantum field theory and its
applications to particle physics overlook the issue of NG boson counting. Here,
it will be discussed in detail in the following subsection for the case of
internal symmetries. It is also quite nontrivial when spacetime symmetries are
spontaneously broken. For more details, the reader is referred to
\cite{Low:2001bw}.

Second, the Goldstone theorem provides information about the low energy
behavior of the NG boson. Since it constitutes a whole excitation branch in the
spectrum, it will (but does not necessarily need to) presumably also exist at
high momentum. However, in this range its behavior is determined by details of
short distance physics, hence it is nonuniversal and cannot be predicted solely
from the broken symmetry. Examples are the acoustic phonon in crystalline
solids whose dispersion relation is linear at low momentum, but gradually
flattens until the group velocity becomes zero at the edge of the Brillouin
zone, or the Bogolyubov--Anderson mode in Bose--Einstein condensates whose
dispersion at high momentum becomes that of a free boson, essentially
insensitive to the presence of the condensate.

Third, as was repeatedly stressed above, a sufficient technical condition for
the integral charge $Q$ to be time independent, and therefore for the Goldstone
theorem to hold, is the causality which is inherent in all Lorentz invariant
theories. In such a case the consequences of the Goldstone theorem cannot be
escaped. Yet it does not mean that the predicted gapless NG boson can actually
be observed, for it may appear in the unphysical sector of the Hilbert space.
A distinguished class of systems where this may happen are gauge theories.
In continuum, they are ill-defined unless the gauge is fixed appropriately.
The remaining global symmetry can then be broken spontaneously. In
covariant gauges that preserve Lorentz invariance and hence causality, a NG
boson appears, but it is in the unphysical part of the Hilbert space, as
mentioned above \cite{Boulware:1962zz,Englert:1964et,Guralnik:1964eu}. On the
other hand, in noncovariant (such as Coulomb) gauges, the presence of
a long-range interaction invalidates the Goldstone theorem. This is well known
to happen, for instance, in electric superconductors where the soft mode
acquires nonzero energy even in the long-wavelength limit. In the lattice
formulation where there is no need for gauge fixing, spontaneous breaking of
the gauge symmetry is ruled out by Elitzur's theorem \cite{Elitzur:1975im}.

Fourth, the assumption of unbroken translational invariance of the vacuum is
natural. Without translational invariance, there is no conserved momentum to
label the excitations so the whole quasiparticle picture loses its meaning. Note
that for the sake of the low-energy physics that the Goldstone theorem is
concerned with, only discrete translational invariance like in crystalline
solids is in principle sufficient. The values of momentum are then restricted to
a finite range, but this still allows for well-defined low-momentum excitations.

Finally, it should be made clear that the Goldstone theorem speaks of a
\emph{limit} of the energy as momentum goes to zero. This is ensured by the
derivation above where integration is first performed over a finite space
domain $\Omega$. The function $\vp_\Omega(\vek k)$ is thus smeared around the
origin and only in the limit $\Omega\to\infty$ goes to $(2\pi)^3\delta^3(\vek
k)$. This distinction is very important since in principle there can be
isolated states with zero momentum and energy which do not represent NG bosons,
but still give a spurious contribution to the right hand side of
Eq.~\eqref{Goldstone_proof2}. Thanks to the fact that the integrand is smeared
by the factor $\vp_\Omega(\vek k)$, these states fortunately do not affect the
value of the momentum integral. In finite volume, such states may be, for
example, the other, degenerate ground states. One can then show explicitly
\cite{Lange:1966zz} that their contribution is suppressed when the limit of
infinite volume is taken.

Concerning the observable consequences of SSB, I have focused exclusively on
the presence of massless particles in the spectrum so far. However, the
Goldstone theorem has in fact much broader consequences. The reason is that we
can consider not just the (quasi-)local field $\Phi$, but in general an
$n$-point Green's function constructed from such fields as the interpolating
field that gives rise to an order parameter. An immediate corollary of the
Goldstone theorem then is that if two Green's functions, connected by a
symmetry transformation, are different then there must be a NG mode in the
spectrum. For example, if the masses of two particles, created by fields that
lie in the same multiplet of the symmetry, are different then the spectrum
exhibits a NG state \cite{Frishman:1966fk}. This makes it clear that a
spontaneously broken symmetry by no means resembles symmetry that is broken
explicitly. Indeed, the former still gives rise to \emph{exact} constraints on
the Green's functions of the theory, conveniently summarized in terms of a set
of Ward identities.

\subsection{Goldstone boson counting: Dispersion relations}
\label{subsec:GBdispersions} As already suggested in the previous subsection,
it is in general a complicated task to determine the precise number of NG
bosons and its connection to the number of broken symmetry generators. In order
to appreciate how frequently one deals with nontrivial realizations of broken
symmetry, let me list several examples from nonrelativistic as well as
relativistic physics. The most profound example is that of ferromagnets versus
antiferromagnets. In both the $\gr{SU(2)}$ symmetry of spin rotations is
spontaneously broken by mutual spin alignment to the $\gr{U(1)}$ subgroup of
rotations about the direction of total magnetic moment (staggered, in the case
of antiferromagnets). This means that two generators are spontaneously broken.
While an antiferromagnet possesses two NG modes (spin waves or \emph{magnons}),
both with a linear dispersion relation at low momentum, there is just one NG
mode in a ferromagnet, its dispersion relation being quadratic. Similarly, in
the so-called canted phase of ferromagnets, the $\gr{SU(2)}$ symmetry is
completely broken, and there is one NG mode of either of the above mentioned
types \cite{Sachdev:1996sa}.

The same phenomenon was predicted to occur in multicomponent Bose--Einstein
condensates of alkalic atomic gases \cite{Ho:1998ho,Ohmi:1998om}. In particular
with three spin polarizations the global symmetry in question is
$\gr{SO(3)\times U(1)}$, corresponding to rotational invariance and conservation
of particle number. There are two phases, roughly analogous to the polar and A
phases of superfluid \isotope[3]{He} or the spin-one color superconductor
analyzed in Section~\ref{subsec:spin1CSC}, leaving unbroken $\gr{SO(2)}$ and
$\gr{U(1)}$ subgroups which are isomorphic, but realized differently due to a
different structure of the ground state. (The respective values of the order
parameter lie in different strata.) The three broken generators give rise to
three NG bosons with linear dispersions in the polar phase, and to one linear
and one quadratic NG mode in the A phase. Similar conclusions have recently
been reported for spin-two Bose--Einstein condensates \cite{Uchino:2009ya}.

In a three-component Fermi gas
\cite{Honerkamp:2003hh,Honerkamp:2004ho,He:2006ne}, the global $\gr{SU(3)\times
U(1)}$ symmetry is spontaneously broken in presence of Cooper pairs built from
fundamental fermions. Due to Pauli principle, these pairs transform as an
antisymmetric tensor (antitriplet) of $\gr{SU(3)}$, leaving an $\gr{SU(2)\times
U(1)}'$ symmetry unbroken. The five broken generators lead to one linear and two
quadratic NG bosons. One should perhaps add that while in (anti)ferromagnets the
dispersion relation of magnons can be probed directly by means of neutron
scattering, similar direct measurements for superfluid atomic gases are beyond
the reach of current experiments.

In relativistic physics, the issue of NG boson counting has been discussed
primarily in the context of dense nuclear and quark matter. For instance, a
condensate of neutral kaons can accompany the Color--Flavor Locked (CFL) phase
of three-flavor quark matter at very high densities. Such a system exhibits a
global $\gr{SU(2)\times U(1)}$ symmetry stemming from isospin and strangeness
conservation. The condensate leaves just its $\gr{U(1)}'$ subgroup intact,
thus breaking spontaneously three generators. These give rise to one linear and
one quadratic NG boson
\cite{Schafer:2001bq,Miransky:2001tw,Andersen:2006ys}. The same behavior
as in condensed matter ferromagnets can be observed in nuclear matter
\cite{Beraudo:2004zr}. Yet other examples can be found among the various color
superconducting phases of dense quark matter. In a two-flavor color
superconductor, known as the 2SC phase, the symmetry as well as its breaking
pattern are the same as in the above mentioned three-component Fermi gas
\cite{Blaschke:2004cs}. However, in this case the symmetry is gauged, and
as will become clear later, this result may thus be just an artifact of the
model treatment based on a global color invariance. On the other hand, in a
spin-one color superconductor, one can find spin waves and the ``abnormal'' NG
bosons with quadratic dispersion relation are then physical
\cite{Buballa:2002wy}.

All the preceding examples suggest that there is a deeper connection between
the number of NG bosons and their dispersion relations. Quite generally, when
there are some modes with quadratic dispersion relation, the total number of NG
bosons seems less than the number of broken generators, in contrast to the
naive expectation based on experience with Lorentz invariant systems.
Incidentally, a great deal of understanding of this problem was achieved long
before many of the above listed examples were even known. More than thirty
years ago, Nielsen and Chadha \cite{Nielsen:1975hm} formulated the following
theorem:
\begin{theorem}[Nielsen and Chadha]
\label{thm:NC}
Assume that translational invariance is not completely broken spontaneously and
that there are no long-range interactions. Then the energy of a NG boson is
analytic in momentum. Denoting NG modes whose energy is proportional to an odd
power of momentum as type-I, and those whose energy is proportional to an even
power of momentum as type-II, \emph{the number of type-I NG bosons plus twice
the number of type-II NG bosons is greater than or equal to the number of broken
symmetry generators.}
\end{theorem}
The requirement on the absence of long-range forces was technically formulated
as a commutativity condition: for any two local operators $A(x)$ and $B(0)$
there is a real positive number $\tau$ such that
\begin{equation}
\abs{\bra0[A(\vek x,t),B(0)]\ket0}\to e^{-\tau\abs{\vek x}}\quad
\text{as}\quad\abs{\vek x}\to\infty
\end{equation}
It is this assumption which is responsible for the analyticity of the Fourier
transform of the commutator \eqref{Goldstone_proof}, and in turn for the
analyticity of the dispersion relation. Obviously it is a stronger assumption
than is necessary for the Goldstone theorem itself. Indeed, in
Section~\ref{sec:examples}\ I will show examples of systems with long-range
forces where the Goldstone theorem is still valid, but the NG dispersion
relation is not analytic.

The actual proof of the claim about the number of NG bosons is algebraic and
somewhat involved, so I will not repeat details and instead refer the reader to
the original paper \cite{Nielsen:1975hm}. It is notable that the statement
of the Nielsen--Chadha theorem is rather general. First, it does not specify the
power of momentum to which energy is proportional. Second, it only gives an
inequality for the number of NG bosons. On the other hand, I am not aware of
any example of a NG boson whose energy would be proportional to a higher power
of momentum than two. Also, the question of the possible general saturation of
the inequality for the number of NG bosons is yet to be understood.
Nevertheless, note that the counting may be obscured by the presence of gapless
non-NG modes, as recognized recently in the context of spin-two Bose--Einstein
condensates \cite{Uchino:2009ya}.

\subsection{Goldstone boson counting: Charge densities}
\label{subsec:charge_densities}
What I have not mentioned before in the list of examples exhibiting type-II NG
bosons was that they are all accompanied by nonvanishing density of some of the
conserved charges. Indeed, this is exactly the sought property that
distinguishes ferromagnets and antiferromagnets: the ferromagnetic ground state
features nonzero net spin density. The connection of the NG boson counting and
the presence of charge density in the ground state is rather general and a
number of (partial) results exists in this respect. I will start with a theorem
due to Sch\"afer \emph{et al.} \cite{Schafer:2001bq}:
\begin{theorem}[Sch\"afer \emph{et al.}]
\label{thm:schafer}
If $\bra0[Q_a,Q_b]\ket0=0$ for all pairs of broken generators $Q_a,Q_b$, then
the number of NG bosons is at least equal to the number of broken generators.
\end{theorem}
Note that in \cite{Schafer:2001bq} the theorem was formulated as a strict
equality. However, this does not follow from the proof presented there. In
general it is difficult to place an upper bound on the number of NG modes; see
also the remark at the end of the previous subsection in this respect. The
original version of the proof was based on the identification of vectors
$Q_a\ket0$ with the NG states. This may be troublesome because as explained in
Section~\ref{sec:basics}\ these vectors are not well defined in the infinite
volume limit. Also, it is not clear \emph{a priori} that the number of linearly
independent vectors $Q_a\ket0$ is equal to the number of NG states in the
spectrum. Below I present a slightly more rigorous modification of the proof
which avoids this step.

Take $j^0_a(x)$ as the broken charge density in Eq.~\eqref{Goldstone_proof} and
$j^0_b(0)$ as the interpolating field for the NG mode. Integrating over the
space in the infinite volume limit then yields
\begin{equation}
\bra0[Q_a,j^0_b(0)]\ket0=\sum_n\left[
\bra0j^0_a(0)\ket{n_{\vek0}}\bra{n_{\vek0}}j^0_b(0)\ket0-
\bra0j^0_b(0)\ket{n_{\vek0}}\bra{n_{\vek0}}j^0_a(0)\ket0
\right]
\label{currentOP}
\end{equation}
Denoting $M_{an}=\bra0j^0_a(0)\ket{n_{\vek0}}$, this equals $(M\he M)_{ab}-(M\he
M)_{ba}$. The assumption of the theorem is equivalent to the requirement that
$M\he M$ be symmetric, and hence real (it is automatically Hermitian). It may
therefore be diagonalized by a real orthogonal transformation, $M\he M\to RM\he
MR^T$, which is equivalent to $M\to RM$, that is, to a change of basis of the
broken generators. (It is essential that the transformation matrix $R$ is real,
for otherwise $R_{ab}Q_b$ would not necessarily be a generator of the symmetry.)
Assume now that the number of NG states, $n_{\text{NG}}$, is smaller than the
number of broken currents, $n_{\text{BC}}$. The rank of the
$n_{\text{BC}}\times n_{\text{NG}}$ matrix $M$ then cannot be larger than
$n_{\text{NG}}$, and so cannot the rank of the $n_{\text{BC}}\times
n_{\text{BC}}$ diagonal matrix $M\he M$. The basis of broken generators may be
ordered so that the first $n_{\text{NG}}$ diagonal elements of $M\he M$ are
nonzero while the remaining $n_{\text{BC}}-n_{\text{NG}}$ ones are zero. Since
$M\he M$ is just the matrix of scalar products of rows of $M$, this means that
the last $n_{\text{BC}}-n_{\text{NG}}$ rows are zero. Consequently, the
generators $Q_{n_{\text{NG}}+1},\dotsc, Q_{n_{\text{BC}}}$ have zero matrix
elements with all NG states $\ket{n_{\vek0}}$, which is in contradiction with
the assumption that they are broken. The theorem is thus proved.

A short precaution is appropriate regarding the interpretation of Theorem
\ref{thm:schafer}. Since the symmetry generators furnish a Lie algebra, it is
tempting to conclude that nonzero density of some of the conserved charges is a
necessary condition for the number of NG bosons to be smaller than the number
of broken generators \cite{Schafer:2001bq,Brauner:2005di}. This is indeed the
case in all systems listed in the previous subsection. However, the
otherwise trivial example of a free nonrelativistic particle (see
Section~\ref{sec:freeNR}) teaches us that it is possible to achieve nonzero
vacuum expectation value of a commutator of two generators even if the
expectations of all generators themselves vanish. The resolution is that the Lie
algebra of symmetry generators picks a central charge upon quantization of the
theory. One therefore cannot generally conclude that nonzero charge density is
necessary in order to have fewer NG bosons than broken generators, unless
additional assumptions are made such as that the Lie algebra of symmetry
generators is semi-simple, in which case there are no nontrivial central charges
(see Sec.~2.7 in \cite{Weinberg:1995v1}).

Looking back at Eq.~\eqref{currentOP}, one observes that in case that the left
hand side is actually nonzero, there must be a NG mode that couples to both
broken generators $Q_a,Q_b$ \cite{Brauner:2005di}. This indicates at a very
elementary level that the opposite to Theorem \ref{thm:schafer} should also
hold: nonzero vacuum expectation value of a commutator of two broken generators
implies that the number of NG bosons is smaller than the number of broken
generators. However, it would be desirable to prove this statement at a similar
level of rigor as Theorem \ref{thm:schafer} itself.

\begin{figure}
\usepgflibrary{arrows}
\begin{center}
\begin{tikzpicture}[<->,auto,>=triangle 45]
  \node (disp) at (0,0) {\fbox{\parbox{1.7cm}{dispersion\\ relations}}};
  \node (num) at (2,-3.464) {\fbox{\parbox{1.5cm}{number\\ of NGBs}}};
  \node (dens) at (4,0) {\fbox{\parbox{1.5cm}{charge\\ densities}}};
  \draw (disp) to node [swap] {\cite{Nielsen:1975hm}} (num);
  \draw (disp) to [out=45,in=135] node [swap] {\cite{Leutwyler:1993gf}} (dens);
  \draw[style=dashed] (dens) to [out=225,in=-45] node [swap]
{\cite{Brauner:2007uw}} (disp);
  \draw[style=dashed] (num) to node [swap] {\cite{Schafer:2001bq}} (dens);
\end{tikzpicture}
\end{center}
\caption{Characteristics of NG spectrum and their interconnections. Nielsen
and Chadha \cite{Nielsen:1975hm} provided a general relation between the number
of NG bosons and their dispersion relations. The connection of NG dispersion
relations and the presence of charge densities was clarified by Leutwyler
\cite{Leutwyler:1993gf} using low-energy effective field theory. Partial results
are indicated by dashed lines.}
\label{fig:NGBchar}
\end{figure}
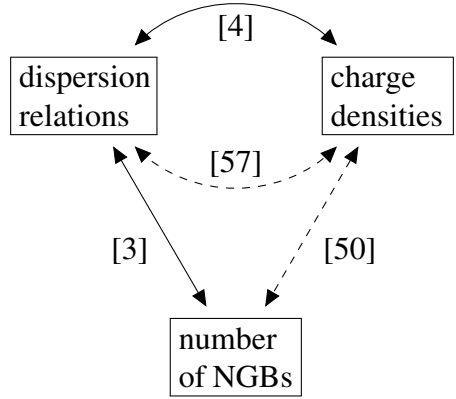
While the Nielsen--Chadha theorem \ref{thm:NC} relates the number of NG bosons
to their dispersion relations, Theorem \ref{thm:schafer} of Sch\"afer
\emph{et al.} provides a (partial) connection of the number of NG bosons to
charge density. (Keeping in mind the example of Section~\ref{sec:freeNR}\ I use
the term ``charge density'' here as a synonym for ``expectation value of a
commutator of two broken generators.'') However, there is also a rather
straightforward connection between the latter two, that is, NG dispersion
relations and charge density. This was shown by Leutwyler
\cite{Leutwyler:1993gf} in the framework of low-energy effective field theory.
A general argument based just on basic symmetry assumptions and analyticity is
presented below. For a summary of all these relations, see
Fig.~\ref{fig:NGBchar}.

Consider now two broken generators, $Q_a$ and $Q_b$, such that their commutator
has a nonzero vacuum expectation value. I will demonstrate that, quite
generally, this implies existence of one NG boson with nonlinear (typically
quadratic) dispersion relation at low momentum. For the sake of simplicity I
will assume rotational invariance. This may seem rather restrictive, note
however that even the low-energy behavior of many solid state systems such as
ferromagnets or superconductors is, to leading order, described by a
rotationally invariant effective field theory. Keeping this in mind, the
transition amplitude for the annihilation of the NG boson by the current
operator can be parameterized as
\begin{equation}
\bra0j^\mu_a(0)\ket{n_{\vek k}}=\imag k^\mu_{\text{on}} F_{an}(\abs{\vek
k})+\imag\delta^{\mu0}G_{an}(\abs{\vek k})
\label{current_amplitude}
\end{equation}
where $k^\mu_{\text{on}}$ is the on-shell wave vector of the one-particle
state, that is, $k^0_{\text{on}}$ is given by the dispersion relation,
$k^0_{\text{on}}=E_{n,\vek k}$. Using current conservation, this leads to the
following equation for the NG dispersion relation,
\begin{equation}
(E^2_{n,\vek k}-\vek k^2)F_{an}+E_{n,\vek k}G_{an}=0
\label{typeIIGBdisp}
\end{equation}
A few special cases  are worth mentioning. When $G_{an}=0$ the standard Lorentz
invariant dispersion relation is recovered. When the ratio $G_{an}/F_{an}$ is
small of order $\mathcal O(\abs{\vek k})$ in the limit $\vek k\to\vek0$, the
dispersion relation comes out linear with phase velocity differing from one.
Finally, when $G_{an}/F_{an}$ has a nonzero limit as $\vek k\to\vek0$, the
dispersion relation takes the low-momentum form
\begin{equation}
E_{n,\vek k}=\vek k^2\frac{F_{an}}{G_{an}}
\label{typeIIGBdisplowk}
\end{equation}
This is the case of most interest since it turns out to be implied by the
presence of nonzero charge density.

In order to decide which of the possibilities outlined above is actually
realized, one needs more information about the dynamics of the system. To that
end, consider the time-ordered current--current correlation function,
$D^{\mu\nu}_{ab}(x-y)=-\imag\bra0T\{j^\mu_a(x)j^\nu_b(y)\ket0$. Let us also
assume the following form of the equal-time commutator of two charge density
operators, $[j_a^0(\vek x,t),j_b^0(\vek y,t)]=\imag\delta^3(\vek x-\vek
y)C_{ab}(\vek x,t)$. By this general notation the possibility of a nontrivial
central charge is taken into account. Taking the divergence of the correlation
function and employing the K\"all\'en--Lehmann spectral representation, one
arrives at the following Ward identity in momentum space,
\begin{multline}
\imag\bra0C_{ab}(0)\ket0=k_{\mu}D^{\mu0}_{ab}(k^0,\vek k)=\\
=\sum_n\left[\frac{(k_{\mu}k_{\text{on}}^{\mu}F_{an}+k^0G_{an})
(k_{\text{on}}^0F^*_{bn}+G^*_{bn})}
{k^0-E_{n,\vek k}+\imag\ve}
-\frac{(k_{\text{on}}^0F_{bn}+G_{bn})
(k_{\mu}\widetilde k_{\text{on}}^{\mu}F^*_{an}+k^0G^*_{an})}
{k^0+E_{n,\vek k}-\imag\ve}\right]
\label{WI}
\end{multline}
where $\widetilde k^\mu=(k^0,-\vek k)$. The index $\nu$ was set to zero since
only then the left-hand side can have nonzero vacuum expectation value due to
the assumed rotational invariance. In order that the particle and antiparticle
poles on the right-hand side are canceled, they must appear at energies
satisfying Eq.~\eqref{typeIIGBdisp}, which is just another derivation of the NG
dispersion relation. Upon cancelation of the poles, the right-hand side of
Eq.~\eqref{WI} takes the form
\begin{equation}
2\imag\im\left[E_{n,\vek k}^2F_{an}F^*_{bn}+E_{n,\vek
k}(F_{an}G^*_{bn}+F^*_{bn}G_{an})+G_{an}G^*_{bn}\right]
\end{equation}
Taking now the limit of zero momentum, the broken symmetry condition expressed
in terms of the nonvanishing expectation value of $C_{ab}(0)$ implies a
``density rule,''
\begin{equation}
\im G_{an}G^*_{bn}=\frac12\bra0C_{ab}(0)\ket0
\label{density_rule}
\end{equation}
In this expression, $G_{an}$ stands for the zero-momentum limit of the
amplitude function defined by Eq.~\eqref{current_amplitude}.

This is the desired result. It means that when the commutator of $Q_a$ and $Q_b$
has nonzero vacuum expectation value, the amplitudes $G_a,G_b$ have finite
limits as $\vek k\to\vek0$. According to Eq.~\eqref{typeIIGBdisplowk} the
dispersion relation of the NG boson that couples to both broken generators is
then nonlinear. In fact, in all cases I am aware of it is quadratic. However,
this does not strictly speaking follow from the general argument presented here.

\subsection{Linear sigma model}
\label{subsec:lsm}
The above arguments connecting the number of NG bosons, their dispersion
relations and charge densities are general, yet not as conclusive as one might
desire. This last part of the section will therefore be devoted to a discussion
that is less general, but allows much stronger statements. For many purposes it
is often sufficient to introduce an effective scalar field whose vacuum
expectation value plays the role of an order parameter. This was actually the
underlying motivation behind Section~\ref{sec:higgs}\ where the general problem
of minimization of potential for such a field was considered. Let me first
recall one of the standard proofs of the Goldstone theorem that applies to
Lorentz invariant theories \cite{Goldstone:1962es}.

Consider a set of real scalar fields $\phi_i$, transforming linearly in a real
representation of a symmetry group with generators $(T_a)_{ij}$, supposed to be
pure imaginary and Hermitian. This means in practice that the symmetry group is
realized by orthogonal transformations, which can always be ensured as long as
the group is compact (see Chap.~7 of \cite{Barut:1977ba}). The restriction to
real fields is not essential for every complex field can be decomposed into two
real ones, and I do not make any assumptions about (ir)reducibility of the group
representation.

For linearly realized symmetry, the quantum effective potential
$V_{\text{eff}}(\phi)$ of the theory is invariant under the same transformations
as the classical Lagrangian (see Sec.~16.4 in \cite{Weinberg:1996v2}). The
condition of invariance therefore reads
\begin{equation}
\PD{V_{\text{eff}}}{\phi_i}(T_a)_{ij}\phi_j=0
\end{equation}
for all generators $T_a$. Differentiating with respect to $\phi$ and setting it
to its vacuum expectation value $\vp$, satisfying
$\Pd{V_{\text{eff}}(\vp)}{\phi_i}=0$, one obtains
\begin{equation}
\frac{\de^2V_{\text{eff}}(\vp)}{\de\phi_i\de\phi_k}(T_a)_{ij}\vp_j=0
\end{equation}
The matrix of second partial derivatives is the mass matrix of the theory and
as long as Lorentz invariance is preserved, it completely determines the
spectrum of the system. This is the Goldstone theorem at work: spontaneous
breaking of the generator $T_a$ by definition means that it does not annihilate
the vacuum, that is, $(T_a)_{ij}\vp_j\neq0$. But then the real vector
$\chi_{ai}=\imag(T_a)_{ij}\vp_j$ is a zero mode of the mass matrix, hence
corresponding to a NG boson.

How many NG bosons are there? It seems almost obvious that there is one for
each broken generator, but let us be a bit more precise at this point. The
vectors $\chi_a$ give rise to a matrix of scalar products,
$h_{ab}=\chi^T_a\chi_b=\vp^TT_aT_b\vp$. This matrix is real, symmetric and
positive semi-definite, and may be diagonalized by an orthogonal
transformation, $h\to RhR^T$. This in turn corresponds to the change of basis
of the Lie algebra of generators, $T_a\to R_{ab}T_b$. After the transformation,
the nonzero vectors $\chi_a$, equal in number to the number of broken
generators, form an orthogonal system. Since they are all zero modes of
the mass matrix, it is now obvious that there is exactly one NG mode for each
broken generator.

In order to see what difference lack of Lorentz invariance brings in, I will
now switch to a class of theories with ``minimal'' breaking of Lorentz
symmetry, induced by chemical potential. These describe many-body systems with
relativistic dynamics. Lorentz invariance is broken explicitly by medium
effects. The main message will be that \emph{the number of zero modes of the
mass matrix is still equal to the number of broken generators, but in general no
longer coincides with the number of NG modes}.

Consider a general model for a \emph{complex} field $\phi$, defined by the
Lagrangian \cite{Brauner:2005di}
\begin{equation}
\Lag=\he{D_\mu\phi}D^\mu\phi-V(\phi)
\end{equation}
where $V(\phi)$ is a renormalizable potential with terms up to fourth order in
the field $\phi$. The covariant derivative reads $D_\mu\phi=(\de_\mu-\imag
A_\mu)\phi$ and makes the Lagrangian formally gauge invariant
(see Sec.~2.4 in \cite{Kapusta:2006kg}). In the end the background gauge field
is set to $A_\mu=(\sum_a\mu_aT_a,\vek0)$, where $\mu_a$ are chemical potentials
associated with a set of mutually commuting conserved charges represented by
matrices $T_a$. Upon expanding the covariant derivatives, the Lagrangian becomes
\begin{equation}
\Lag=\de_\mu\he\phi\de^\mu\phi-2\im\he\phi A^\mu\de_\mu\phi-\tilde V(\phi)
\end{equation}
with the modified potential $\tilde V(\phi)=V(\phi)-\he\phi A^\mu A_\mu\phi$.
The chemical potential thus gives a negative contribution to the mass matrix,
and when it exceeds certain critical value, the perturbative vacuum $\vp=0$
will no longer be stable and the field will condense. As soon as the field
develops nonzero expectation value, some of the symmetry generators
become spontaneously broken. Keeping in mind that the zero modes of the mass
matrix are $\imag T_a\vp$, one reparameterizes the field as
\begin{equation}
\phi(x)=\vp+H(x)+\imag\Pi(x)\vp
\end{equation}
Here $\Pi(x)$ is a linear combination of broken generators and represents
the NG fields, while $H(x)$ stands for the massive, ``Higgs'' modes. After some
manipulation, the bilinear part of the Lagrangian which governs the spectrum at
tree level, becomes
\begin{multline}
\Lag_{\text{bilin}}=\de_\mu\he H\de^\mu H-\tilde V_{\text{bilin}}(H)-
2\im\he HA^\mu\de_\mu H+\\
+\he\vp\de_\mu\Pi\de^\mu\Pi\vp-4\re\he HA^\mu\de_\mu\Pi\vp
-\im\he\vp A^\mu[\Pi,\de_\mu\Pi]\vp
\label{Lbilin}
\end{multline}
where $\tilde V_{\text{bilin}}$ is the bilinear part of the modified potential,
which only depends on the Higgs field $H$. This is the manifestation of the NG
nature of the field $\Pi$: it drops from the mass part of the Lagrangian.

The Noether currents in the model are $j^\mu_a=-\imag(D^\mu\he\phi T_a\phi-
\he\phi T_aD^\mu\phi)$, and the charge densities in the ground state read
$j^0_a=2\he\vp A^0T_a\vp$. Consequently, the last term in the Lagrangian
\eqref{Lbilin} is proportional to the density of a commutator of two
generators. Not surprisingly, it plays a crucial role in the structure of
the NG spectrum, in accordance with Theorem \ref{thm:schafer}. In order to
understand the spectrum of the system governed by the Lagrangian \eqref{Lbilin},
it is desirable to simplify the problem as much as possible; it looks like one
has to deal with a very complicated mixing of all the modes. Fortunately, it
turns out that when the fields $H,\Pi$ are properly decomposed into irreducible
representations of the unbroken subgroup, the whole Lagrangian splits into
sectors in which just two modes can mix with each other at a time (see
\cite{Brauner:2005di} for details). A key role in the argument is played by the
statement, whose validity is not restricted to the framework of the
linear sigma model, formulated here as a simple theorem:
\begin{theorem}
It is possible to choose a basis of the symmetry generators in such a way that
only mutually commuting charges have nonzero vacuum expectation value.
\end{theorem}
Unfortunately, I am not aware of a completely general proof of this claim.
However, for the wide class of systems whose symmetry group is the special
unitary group and its products, the proof is simple. Denote the expectations of
the charge operators as $q_a=\bra0Q_a\ket0$. Forming a linear combination with
the generator matrices $T_a$ one gets a Hermitian matrix $q_aT_a$. Under the
unitary group transformation $U$ on the Hilbert space, the charge
\emph{operators} transform in the adjoint representation, $Q_a\to UQ_a\he U$.
This in turn translates into the transformation rule for the charge expectations
$q_a$. However, under the adjoint action of the unitary group, the \emph{matrix}
$q_aT_a$ can always be diagonalized. In such a diagonal representation, only
mutually commuting charges have nonzero expectation values, as was to be proved.

The search for charges with nonzero expectation value is greatly facilitated by
the fact that they must transform as singlets under the unbroken symmetry. The
candidates can therefore be found by a decomposition of the Lie algebra of
generators into irreducible representations of the unbroken subgroup.
Once all charges that have nonzero vacuum expectation value have been
identified, they can be completed to form the Cartan subalgebra of the Lie
algebra. The standard root decomposition of Lie algebras
(see Chaps.~6 and 8 of \cite{Georgi:1982jb}) then guarantees that the remaining
generators split up into pairs whose commutators lie in the Cartan subalgebra.
Only such pairs can be mixed by the last term in the Lagrangian \eqref{Lbilin}.
The task to determine the NG spectrum thus boils down to the problem of mixing
of two fields, captured by the Lagrangian
\begin{equation}
\Lag_{\text{bilin}}=\frac12(\de_\mu\pi)^2+\frac12(\de_\mu
H)^2-\frac12f^2(\mu)H^2-g(\mu)H\de_0\pi
\end{equation}
As the notation suggests, $\pi$ is the NG field while $H$ is a Higgs field as
long as its mass function $f(\mu)$ is nonzero. The coefficient $g(\mu)$ accounts
for the last two terms in Eq.~\eqref{Lbilin}, which have just a single time
derivative. The spectrum of this Lagrangian consists of two modes with squared
energies at low momentum given by the following formulas,
\begin{equation}
E_{1,\vek k}^2=f^2(\mu)+g^2(\mu)+\dotsb,\quad
E_{2,\vek k}^2=\frac{f^2(\mu)}{f^2(\mu)+g^2(\mu)}\vek k^2+
\frac{g^4(\mu)}{[f^2(\mu)+g^2(\mu)]^3}\vek k^4+\dotsb
\end{equation}
If $f(\mu)\neq0$, the Lagrangian indeed describes a massive state and a type-I
NG boson with linear dispersion relation at low momentum. On the other hand, if
$f(\mu)=0$, that is, if both $\pi$ \emph{and} $H$ have no mass term, the
spectrum contains a massive mode and only one NG boson of type-II with
quadratic dispersion relation at low momentum, $E_{2,\vek k}=\vek
k^2/\abs{g(\mu)}$. This is a sheer consequence of the term in the Lagrangian
with a single time derivative, and in turn of the nonzero expectation value of
a commutator of two broken generators. One can thus conclude with the strong
statement that \emph{there is one type-II NG boson with quadratic dispersion
relation for each pair of generators whose commutator has nonzero vacuum
expectation value}. Explicit examples, further elucidating this result,
will be analyzed in the next section.


\section{Further examples}
\label{sec:examples}
In order to shed some light on the general arguments given in the preceding
section, I will now work out in detail some examples of systems featuring
type-II NG bosons. Interestingly enough, these examples turn out to be
nontrivial and thus hopefully contribute to deeper understanding of the
underlying physics. I will start with a nonrelativistic, completely solvable
toy model.

\subsection{Nonrelativistic Boulware--Gilbert model}
\label{subsec:boulware}
Following \cite{Guralnik:1968gu}, consider a nonrelativistic version of the
model introduced in \cite{Boulware:1962zz}, defined in finite volume $\Omega$
by the Hamiltonian
\begin{equation}
\Ham=\frac12\int_\Omega\dthree\vek x\,\left[\pi^2+(\nabla\phi)^2\right]+
\frac1{2}\int_\Omega\dthree\vek x\dthree\vek y\,\pi(\vek x)V(\vek x-\vek
y)\pi(\vek y)
\end{equation}
The canonical coordinate and momentum satisfy the usual commutation relation,
$[\phi(\vek x),\pi(\vek y)]=\imag\delta^3(\vek x-\vek y)$. Since the
Hamiltonian is bilinear, this is essentially a free field theory.

The theory can be solved by switching to Fourier transformed variables,
$\phi_{\vek k}=\int_\Omega\dthree\vek x\,\phi(\vek x)e^{-\imag\skal kx}$, and
analogously for $\pi_{\vek k}$ and $V_{\vek k}$. The Hamiltonian thus becomes
\begin{equation}
\Ham=\frac1{2\Omega}\sum_{\vek k}\left[\vek k^2\phi_{\vek k}\phi_{-\vek k}+
(1+V_{\vek k})\pi_{\vek k}\pi_{-\vek k}\right]
\label{HamFT}
\end{equation}
From here it is already straightforward to derive the Lagrangian,
\begin{equation}
\Lag=\frac1{2\Omega}\sum_{\vek k}\left(\frac{\dot\phi_{\vek k}\dot\phi_{-\vek
k}}{1+V_{\vek k}}-\vek k^2\phi_{\vek k}\phi_{-\vek k}\right)
\label{Boulware_Lagrangian}
\end{equation}
Inverting the kernel of this bilinear Lagrangian one obtains the
time-ordered Green's function (propagator) of the theory,
\begin{equation}
D(k_0,\vek k)=\frac{1+V_{\vek k}}{k_0^2-E_{\vek k}^2+\imag\ve}
\end{equation}
where $E_{\vek k}^2=\vek k^2(1+V_{\vek k})$ is the dispersion relation of the
one-particle excitation of the model.

Apparently, when the potential $V(\vek x)$ falls off sufficiently fast at large
distance so that a finite limit $\lim_{\vek k\to\vek0}V_{\vek k}$ exists, the
theory has a gapless excitation. This is a NG boson associated with the
invariance of the Lagrangian \eqref{Boulware_Lagrangian} under the constant
shift,
$\phi(\vek x)\to\phi(\vek x)+\vt$. The associated current is $\vek
j=-\nabla\phi$ and charge density $\vr=\pi$. The integral charge then is
$Q_\Omega=\int_\Omega\dthree\vek x\,\vr(\vek x)=\pi_{\vek0}$, which indeed
commutes with the Hamiltonian.

Let us now have a closer look at the NG dispersion relation in the case of a
long-range interaction. Assume that the potential is radially symmetric,
regular at the origin and drops as $1/r^\alpha$ at large distance $r$,
considering the following approximation,
\begin{equation}
V(r)\approx\begin{cases}
V_0 &\text{for}\quad r\ll r_0\\
\gamma/{r^\alpha} &\text{for}\quad r\gg r_0
\end{cases}
\label{longdistpot}
\end{equation}
The momentum integration in
\begin{equation}
V_{\vek k}=\frac{4\pi}k\int_0^\infty\dd r\, rV(r)\sin kr=
\frac{4\pi}{k^3}\int_0^\infty\dd x\,xV(x/k)\sin x
\label{FTpot}
\end{equation}
where $k=\abs{\vek k}$, can then be split into three regions, $0<x<kr_0$,
$kr_0<x<1$, and $1<x<\infty$, which are well separated provided the momentum is
small enough. In the integration over the first region we approximate the
potential by $V_0$ and the sine by its argument, getting $\frac43\pi V_0r_0^3$.
In the second region we approximate the potential by $\gamma/r^\alpha$ and the
sine by its argument, which yields
$4\pi\gamma\left(k^{\alpha-3}-r_0^{3-\alpha}\right)/(3-\alpha)$. In the last
region we approximate just the potential, leading to the integral
$4\pi\gamma k^{\alpha-3}\int_1^\infty\dd x\,\sin x/x^{\alpha-1}$. We do not
need to evaluate this integral explicitly, it is enough to know that it
converges as long as $\alpha>1$. Putting all the pieces together, we arrive at
the asymptotic scaling of the energy at low momentum,
\begin{equation}
E_{\vek k}\sim
\begin{cases}
k & \text{for}\quad3\leq\alpha\\
k^{(\alpha-1)/2} & \text{for}\quad1<\alpha\leq 3
\end{cases}
\end{equation}
For $\alpha\leq1$ the Fourier transform of the potential does not exist and the
theory is ill-defined in three spatial dimensions.

The conclusion is that for $\alpha>3$ the range of the interaction is short
enough to ensure applicability of the Nielsen--Chadha theorem \ref{thm:NC}; the
energy is analytic in momentum. On the other hand, for $1<\alpha\leq3$ the
interaction is long-ranged, but there is still a NG boson. It is, however, a
very nontrivial one. The nonanalytic structure of the dispersion relation
clearly precludes a low-energy description in terms of an effective Lagrangian.
Expansion in powers of momentum is impossible in such a case.

Another interesting special case is the screened Coulomb interaction,
$V(r)=\gamma e^{-\mu r}/4\pi r$ \cite{Guralnik:1968gu}. Recalling that $V_{\vek
k}=\gamma/(\vek k^2+\mu^2)$, one observes that the system can be stable even
with an attractive interaction, provided it is not too strong so that
$V_{\vek0}\geq-1$. In the limiting case, $\gamma=-\mu^2$, we find $E_{\vek
k}=\vek k^2/\sqrt{\vek k^2+\mu^2}$. Energy is quadratic in momentum and the NG
boson is type-II! Does it mean that we have found an explicit example of a
system exhibiting a strict inequality in the Nielsen--Chadha theorem
\ref{thm:NC}? Not really, because there are in fact \emph{two} broken conserved
charges. The reason is that for $V_{\vek0}=-1$, the operator $\phi_{\vek0}$
also commutes with the Hamiltonian \eqref{HamFT}. There are thus two conserved
charges, $\pi_{\vek0}$ and $\phi_{\vek0}$, both being spontaneously broken.
Moreover, their commutator is proportional to the unit operator, and hence
obviously has nonzero vacuum expectation value in accord with the general
arguments of the previous section. Yet another nontrivial fact is that the new
conserved charge is not of the Noether type; it is not associated with a
symmetry of the Lagrangian. This emphasizes the fact that the Goldstone theorem
assumes just the existence of a conserved charge, and it is not important where
this charge comes from.

\subsection{Heisenberg ferromagnet}
\label{subsec:heisenberg}
The next example of SSB is already quite realistic and represents a model for
isotropic ferromagnets \cite{Lange:1966zz}. The degrees of freedom are
spin-$\frac12$ particles fixed to nodes of a crystal lattice, represented by
operators $\vek s_i$. (The Roman subscript labels the lattice sites.) The
Hamiltonian of the model reads
\begin{equation}
\Ham=-\frac12\sum_{ij}J_{ij}\vek s_i\cdot\vek s_j
\label{Heisenberg}
\end{equation}
It is invariant under simultaneous rotations of all the spins, forming the
group $\gr{SU(2)}$. As long as all the couplings $J_{ij}$ are positive, the
ground state of the Hamiltonian is obviously formed by all spins aligned in the
same direction. Formally, one rewrites $\vek s_i\cdot\vek s_j=\frac12(\vek
s_i+\vek s_j)^2-\frac34$. With all spins aligned, each pair resides in a state
with total spin one so that the ground state energy is
$-\frac18\sum_{ij}J_{ij}$.

Any state of the $i$th particle may be labeled by a direction $\vek n$ as
$\ket{i,\vek n}$, meaning that it is an eigenstate of the projection $\skal
ns_i$ with the eigenvalue $1/2$. The degenerate ground states of the
Hamiltonian \eqref{Heisenberg} are then $\ket{0_{\vek
n}}=\prod_{i=1}^N\ket{i,\vek n}$, where $N$ is the total number of lattice
sites. Using an explicit expression for the eigenvectors of spin-$\frac12$
operators, one finds the scalar product of two states associated with different
directions, $\abs{\braket{\vek n_1}{\vek n_2}}=\cos(\vt_{\vek n_1,\vek n_2}/2)$,
where $\vt_{\vek n_1,\vek n_2}$ is the angle between the two directions. The
scalar product of two ground states of the Heisenberg model therefore is
\begin{equation}
\abs{\braket{0_{\vek n_1}}{0_{\vek n_2}}}=\left(\cos\frac{\vt_{\vek n_1,\vek
n_2}}2\right)^N
\end{equation}
and it apparently goes to zero as $N\to\infty$ unless $\vek n_1$ and $\vek n_2$
are identical. This is in agreement with the discussion in
Section~\ref{sec:basics}\ where it was pointed out as one of the characteristic
features of SSB.

To construct the whole Hilbert space above the ground state $\ket{0_{\vek n}}$,
one can conveniently use the formalism of creation and annihilation operators.
To that end, recall that the two-dimensional space of spin $\frac12$ may be
viewed as the Fock space of the fermionic oscillator. One defines annihilation
operators $a_i(\vek n)$ and creation operators $\he a_i(\vek n)$ so that
$a_i(\vek n)\ket{i,\vek n}=0$ and $\{a(\vek n),\he a(\vek n)\}=1$. These are
nothing else than the lowering and raising operators familiar from the theory
of angular momentum. In addition to their defining relations above, they
satisfy $[a_i(\vek n),\he a_i(\vek n)]=2\skal ns_i$. Note that, in this
setting, annihilation and creation operators at different lattice sites
\emph{commute} rather than anticommute as usual. The change of sign induced by
the interchange of two distinguishable fermions is, however, merely a
convention.

The Hilbert space of the Heisenberg ferromagnet is set up as a Fock space above
the vacuum $\ket{0_{\vek n}}$. In the ground state all spins point in the
direction $\vek n$, while the excited states are obtained by the action of the
creation operators $\he a_i(\vek n)$ that flip the spin at the $i$th lattice
site to the opposite direction. The countable basis of the space Hilbert space
contains all vectors of the form $\he a_{i_1}(\vek n)\he a_{i_2}(\vek
n)\dotsb\ket{0_{\vek n}}$ where a \emph{finite} number of spins are flipped. It
is now obvious that in the infinite volume limit, all basis vectors from the
Hilbert space corresponding to a direction $\vek n_1$ are orthogonal to all
basis vectors from the Hilbert space characterized by a different direction
$\vek n_2$. This means that these two spaces are completely orthogonal.

The symmetry group $\gr{SU(2)}$ is generated by the operator of the total spin,
$\vek S=\sum_i\vek s_i$. Only rotations about the direction $\vek n$ of the
ground state, that is, the spontaneous magnetization, are unbroken. The
unbroken group $\gr{U(1)}$ is generated by the spin projection $\skal nS$. The
two remaining generators of the symmetry group are spontaneously broken. They
are obviously ill-defined operators, as are the induced finite transformations,
for they take any state out of the above constructed separable Hilbert space.

In order to identify the NG boson state(s), one rewrites the Hamiltonian
\eqref{Heisenberg} in terms of the annihilation and creation operators. Since
the orientation of the ground state is now fixed, the argument $(\vek n)$ of
these operators will be for simplicity omitted. First observe that
\begin{equation}
\vek s_i\cdot\vek s_j=-\frac12(\he a_i-\he a_j)(a_i-a_j)+\he a_ia_i\he
a_ja_j+\frac14
\label{Heis_Ham_second_quantization}
\end{equation}
The Hamiltonian preserves the ``particle number,'' that is, the number of
flipped spins generated by the operator $\sum_i\he a_ia_i$. This is of course,
up to irrelevant constants, nothing but the $\vek n$-projected component of the
total spin, which is not spontaneously broken and thus can be used to label
physical states. Let us restrict our attention to the ``one-particle'' space,
spanned on the basis $\ket i=\he a_i\ket{0_{\vek n}}$. The physical reason
behind this restriction is that the sought NG boson turns out to be the spin
wave, a traveling perturbation induced by flipping a single spin.

On the one-particle space, the second term on the right hand side of
Eq.~\eqref{Heis_Ham_second_quantization} gives zero unless $i=j$. The
one-particle Hamiltonian thus reads, up to an irrelevant constant,
\begin{equation}
H_{1\text P}=\frac14\sum_{ij}J_{ij}(\he a_i-\he a_j)(a_i-a_j)
\end{equation}
and acts on the basis states as
\begin{equation}
H_{1\text P}\ket i=\frac12\sum_jJ_{ij}(\ket i-\ket j)
\label{spin_wave_ham}
\end{equation}
It is worth emphasizing that so far we have nowhere used the discrete
translational invariance implied by the symmetries of the crystal lattice. In
fact, the indices $i,j$ have been just labels distinguishing different degrees
of freedom. However, as we already know, translational invariance is needed in
order to have well defined quasiparticle excitations carrying conserved
momentum. The stationary states can then be sought as the common eigenstates of
the discrete translation operators, that is, the plane waves, $\ket{\vek
k}=\sum_i e^{\imag\skal kx_i}\ket i$. Assuming finally that the couplings
depend just on the distance of the lattice sites, $J_{ij}=J(\abs{\vek x_i-\vek
x_j})$, one finds by direct substitution that the plane wave $\ket{\vek k}$
indeed is an eigenstate of the one-particle Hamiltonian \eqref{spin_wave_ham},
with the energy given by
\begin{equation}
E_{\vek k}=\frac12(J_{\vek0}-J_{\vek k})
\label{E_magnon}
\end{equation}
where $J_{\vek k}$ is the discrete Fourier transform, $J_{\vek k}=\sum_i J(\vek
x_i)e^{-\imag\skal kx_i}$. As a concrete example, the nearest neighbor
interaction of strength $J$ on a square lattice of spacing $\ell$ would give
the dispersion $E_{\vek k}=J(3-\cos k_x\ell-\cos k_y\ell-\cos k_z\ell)=
2J\left(\sin^2\frac{k_x\ell}2+\sin^2\frac{k_y\ell}2+\sin^2\frac{k_z\ell}2\right)$.
The dispersion relation of the NG boson is quadratic at low momentum, hence it
is of type-II. In fact, this follows directly from the Nielsen--Chadha theorem
\ref{thm:NC} since there are two broken generators and only one NG state. In
this case one can easily see that acting with the two broken generators on the
ground state $\ket{0_{\vek n}}$ formally produces states that differ just by a
phase factor.

At the end, let us again look more closely at the NG dispersion relation
\eqref{E_magnon} and assume a long-range potential as in
Eq.~\eqref{longdistpot}. Since we are only interested in the low-momentum
behavior, the discrete Fourier transform of the interaction can be replaced
with the continuous radial integral \eqref{FTpot} with an additional prefactor
$1/\ell^3$ taking into account the volume of the elementary lattice cell. In
order that the spectrum be well defined at all, the Fourier transform at zero
momentum must exist, which is true only for $\alpha>3$. By an analysis similar
to that for the Boulware--Gilbert model one finds that
\begin{equation}
E_{\vek k}\sim
\begin{cases}
k^2 & \text{for}\quad5\leq\alpha\\
k^{\alpha-3} & \text{for}\quad3<\alpha\leq 5
\end{cases}
\end{equation}
In conclusion, for $\alpha\geq5$ the interaction drops sufficiently rapidly so
that the Nielsen--Chadha theorem applies and the dispersion relation is
analytic, concretely quadratic since the two broken generators demand a
type-II NG boson. For $3<\alpha\leq5$ the NG boson still exists, but its energy
is once again nonanalytic in momentum. Remember that even though we have not
solved the Heisenberg model \eqref{Heisenberg} completely, Eq.~\eqref{E_magnon}
still represents the \emph{exact} energy of exact eigenstates of the
Hamiltonian, and therefore this conclusion about the NG energy is not just an
artifact of some approximation. Finally, note that for $\alpha\geq5$ the
Fourier transform at low momentum may be evaluated by Taylor expansion up to
order $\vek k^2$, resulting in the dispersion relation
\begin{equation}
E_{\vek k}=\frac{\pi\vek k^2}{3\ell^3}\int_0^\infty\dd r\,r^4J(r)
\end{equation}

\subsection{Linear sigma model}
For a final example let me get back to the linear sigma model. Unlike the
previous two examples where the NG state and its dispersion relation were
constructed exactly, here just the classical approximation will be
used. As a trade-off one is able to gain deeper insight in the nature of type-II
NG bosons.

For illustration purpose, I will restrict to the model with $\gr{SU(2)\times
U(1)}$ symmetry, defined by the Lagrangian
\cite{Schafer:2001bq,Miransky:2001tw}
\begin{equation}
\Lag=D_\mu\he\phi D^\mu\phi-M^2\he\phi\phi-\lambda(\he\phi\phi)^2
\label{Lag_Miransky}
\end{equation}
Here $\phi$ is a complex doublet of scalar fields and the covariant derivative
includes chemical potential $\mu$ associated with the $\gr{U(1)}$ factor of the
symmetry group, $D_\nu\phi=(\de_\nu-\imag\delta_{\nu0}\mu)\phi$. For $\mu>M$
the perturbative Fock vacuum becomes unstable with respect to fluctuations of
the field $\phi$, and the field develops nonzero vacuum expectation value,
$\vp$. At tree level, $\he\vp\vp\equiv v^2=(\mu^2-M^2)/(2\lambda)$. The four
Noether currents of the theory have the form
\begin{equation}
j^a_\nu=-2\im\he\phi\tau^a\de_\nu\phi+2\mu\delta_{\nu0}\he\phi\tau^a\phi,\quad
j_\nu=-2\im\he\phi\de_\nu\phi+2\mu\delta_{\nu0}\he\phi\phi
\label{lsmcurrents}
\end{equation}
where $\tau^a$ are the Pauli matrices. Choosing the orientation of the ground
state so that $v$ is real and positive and resides solely in the lower
component of $\varphi$, the only unbroken generator will be
$\tau^+=\frac12(\openone+\tau^3)$. The broken generators can be conveniently
chosen as $\tau^-=\frac12(\openone-\tau^3)$ and $\tau^{1,2}$. Then, only the
generator $\tau^-$ has nonzero expectation value in the ground state.
Therefore, the commutator $[\tau^1,\tau^2]$ also has nonzero expectation value
and we expect one type-II NG boson which couples to both $j^1_\nu$ and
$j^2_\nu$. In addition, there should be one type-I NG boson coupled to
$\tau^-$.

Remember from Section~\ref{subsec:lsm}\ that the NG fields can be identified as
$\imag T_a\vp$, using the broken generators $T_a$. This means that the upper
component of $\phi$, $\phi_1$, represents the type-II NG boson while the
imaginary part of the lower component, $\phi_2$, represents the type-I NG boson.
The real part of $\phi_2$ corresponds to fluctuations of the magnitude of the
condensate, and is expected to be massive. Moreover, $\phi_1$ carries unit
charge of the unbroken $\gr{U(1)}'$ symmetry generated by $\tau^+$ while
$\phi_2$ does not. Therefore, these two components do not mix and their bilinear
Lagrangians and propagators can be evaluated separately.

One thus finds that $\phi_1$ excites modes with dispersion relations
$E_{\mp,\vek k}=\sqrt{\vek k^2+\mu^2}\mp\mu$. The upper sign corresponds to the
type-II NG boson; indeed, at low momentum the energy is approximately $E_{-,\vek
k}\approx\vek k^2/2\mu$. The lower sign describes a mode with mass $2\mu$. As
predicted in Section~\ref{subsec:lsm}\ the presence of nonzero vacuum
expectation value of the commutator of two broken generators turns the two
associated modes into one massive state and one NG boson of type-II. Similarly,
the dispersion relations of the two modes in $\phi_2$ are found to be
\begin{equation}
E^2_{\mp,\vek k}=\vek k^2+3\mu^2-M^2\mp\sqrt{4\mu^2\vek k^2+(3\mu^2-M^2)^2}
\end{equation}
Again, the upper sign refers to the NG boson, whose dispersion relation at low
momentum this time is $E_{-,\vek k}\approx\abs{\vek
k}\sqrt{(\mu^2-M^2)/(3\mu^2-M^2)}$.

Apart from the dispersion relations, the knowledge of the propagators allows us
to determine the couplings of the modes to the fields from the residua of the
respective poles using the K\"all\'en--Lehmann spectral representation
\cite{Brauner:2006xm}. Denoting the type-I and type-II boson states as
$\ket{\pi(\vek k)}$ and $\ket{G(\vek k)}$, one obtains
\begin{equation}
\begin{split}
\bra0\phi_2(0)\ket{\pi(\vek k)}&=\frac1{\sqrt{2E_{-,\vek k}}}
\frac1{\sqrt{E_{+,\vek k}^2-E_{-,\vek k}^2}}
\left[-(E_{-,\vek k}-\mu)^2+\vek k^2+2\mu^2-M^2\right]\\
\bra0\he\phi_2(0)\ket{\pi(\vek k)}&=-\frac1{\sqrt{2E_{-,\vek k}}}
\frac1{\sqrt{E_{+,\vek k}^2-E_{-,\vek k}^2}}
\left[-(E_{-,\vek k}+\mu)^2+\vek k^2+2\mu^2-M^2\right]
\end{split}
\end{equation}
and
\begin{equation}
\bra0\phi_1(0)\ket{G(\vek k)}=\frac1{\sqrt{2\sqrt{\vek k^2+\mu^2}}}
\end{equation}
and finally $\bra0\he\phi_1(0)\ket{G(\vek k)}=0$ thanks to the unbroken
$\gr{U(1)}'$ charge. These expressions allow one to evaluate, with the
help of Eq.~\eqref{lsmcurrents}, the couplings of the NG bosons to the broken
currents \cite{Brauner:2007uw},
\begin{equation}
\begin{split}
\bra0j_1^\nu(0)\ket{G(\vek k)}&=-\imag\bra0j_2^\nu(0)\ket{G(\vek k)}=
v(k^\nu_{\text{on}}+2\mu\delta^{\nu0})\bra0\phi_1(0)\ket{G(\vek k)}\\
\bra0j_-^\nu(0)\ket{\pi(\vek k)}&=v\left[(k^\nu_{\text{on}}+2\mu\delta^{\nu0})
\bra0\phi_2(0)\ket{\pi(\vek k)}-(k^\nu_{\text{on}}-2\mu\delta^{\nu0})
\bra0\he\phi_2(0)\ket{\pi(\vek k)}\right]
\end{split}
\end{equation}
With all these formulas at hand one readily checks all general formulas derived
in Section~\ref{subsec:charge_densities}\ such as the density rule
\eqref{density_rule}.

In order to gain further insight into the nature of the type-II NG boson, let
us investigate the solution of the classical equation of motion. The bilinear
part of the Lagrangian \eqref{Lag_Miransky} containing just the field $\phi_1$
is
\begin{equation}
\Lag_{\text{bilin},1}
=2\imag\mu\he\phi_1\de_0\phi_1+\de_\mu\he\phi_1\de^\mu\phi_1
\end{equation}
up to a total derivative. This has plane wave solutions of the type
$\phi_1(x)=ae^{-\imag k\cdot x}$ with $k_0=\sqrt{\vek k^2+\mu^2}-\mu$, which is
exactly the dispersion relation of the type-II NG boson in the model.
Substituting into Eq.~\eqref{lsmcurrents} one arrives at the expressions for
the currents carried by this plane wave,
\begin{equation}
\begin{split}
j_1^\nu=-2v(k^\nu_{\text{on}}+2\delta^{\nu0}\mu)\im\phi_1&,\quad
j_2^\nu=-2v(k^\nu_{\text{on}}+2\delta^{\nu0}\mu)\re\phi_1\\
j_+^\nu=2(k^\nu_{\text{on}}+\delta^{\nu0}\mu)\abs{\phi_1}^2&,\quad
j_-^\nu=2\delta^{\nu0}\mu v^2
\end{split}
\label{planewave}
\end{equation}
The second line shows that the plane wave is associated with a uniform current
of the unbroken charge, $\tau^+$. This corresponds to the observation made
above that the field $\phi_1$, and hence its quantum $\ket{G(\vek k)}$,
carries unit charge of the unbroken group $\gr{U(1)}'$. Also, there is uniform
background density of the $\tau_-$ charge, induced by the condensate $v$. It is
independent of the plane wave since the associated current does not couple to
$\ket{G(\vek k)}$.

The first line of Eq.~\eqref{planewave} reveals that the isospin vector of
charge densities rotates in the $(1,2)$ plane perpendicular to the direction of
the condensate. In other words, the type-II NG boson is a circularly polarized
isospin wave. This is, of course, also in accord with the fact that it carries
the unbroken charge. One may then naturally ask where the plane wave with
opposite polarization is. In this model, this ``antiparticle'' of the type-II
NG boson is heavy; its mass is $2\mu$.

Interestingly, the answer to this question is quite different in a
nonrelativistic ferromagnet, analyzed in the previous subsection. There, too,
the magnon is a spin wave which carries (minus) one unit of the unbroken
charge, that is, projection of spin into the direction of total magnetization.
However, in this case, the spin wave with opposite circular polarization does
not even exist. The explanation is twofold. First, quantum mechanically, the
ground state is the state with the highest possible spin, and the magnon
corresponds to a propagating perturbation caused by flipping of one of the
spins. The spin wave polarized in the opposite way would have to have total
spin one unit higher than the ground state, which is not possible. Classically,
imagine the magnon as a single test spin propagating on a background of other
spins. The Heisenberg Hamiltonian \eqref{Heisenberg} shows that this background
acts on the test spin as a homogeneous magnetic field. The evolution of a spin
in a magnetic field is well known as the Larmor precession. Its sense is fixed
by the magnetic moment of the spin; rotation of the spin in the opposite
direction is not possible.


\section{Low-energy effective field theory for NG bosons}
\label{sec:EFT}
Effective field theory (EFT) is a powerful approach to problems involving
several energy scales that was developed in the full generality in particle
physics \cite{Weinberg:1978kz}, but has grown into an indispensable tool in
essentially all branches of physics that use the methods of quantum field
theory. The basic assumption is that the physical degrees of freedom split into
groups well separated in energy. In case one is interested only in
low-energy, or long-distance, observables, EFT may be used to describe the
system in a way which involves just the low-energy degrees of freedom. The
effects of the microscopic, short-distance dynamics are incorporated in the
values of coupling constants of the effective theory. For more details see the
review papers \cite{Georgi:1994qn,Kaplan:1995uv,Manohar:1996cq,Pich:1998xt}.

There are fortunate cases where the low-energy EFT can be derived directly from
the underlying microscopic theory by ``integrating out'' the heavy, high-energy
degrees of freedom. A profound example is the Euler--Heisenberg Lagrangian for
the electromagnetic field, which can be inferred from quantum electrodynamics
by eliminating electrons or any other massive charged particles present. In such
cases EFT still provides a very convenient tool which dispenses with the
degrees of freedom irrelevant for the description of low-energy physics.
However, the true power of EFT lies in its successful application to systems
where the high-energy degrees of freedom cannot be simply integrated out of the
microscopic theory, either because of its nonperturbative nature, or because
the low-energy modes are not even present in the microscopic theory (such as
the hadrons that are not the fundamental degrees of freedom in quantum
chromodynamics). The utility of EFT then relies on the observation that field
theory is merely a convenient way to incorporate the general principles such as
unitarity and cluster decomposition but contains no further dynamical
assumptions, and it can therefore reproduce the predictions at low energy of
\emph{any} microscopic theory as long as it involves the appropriate low-energy
degrees of freedom and symmetries, see e.g.~\cite{Burgess:1998ku} or Sec.~19.5
in \cite{Weinberg:1996v2}.

A distinguished class of systems featuring separation of scales is represented
by those with SSB. The NG bosons, guaranteed by the Goldstone theorem, are then
the low-energy degrees of freedom. In the most frequent case that there are no
other gapless states in the spectrum, they are in fact the only degrees of
freedom in the low-energy EFT. A systematic approximation scheme for
calculations of low-energy observables is then provided by expansion in powers
of momentum, or derivatives. (This requires absence of long-range interactions,
as shown in Sections~\ref{subsec:boulware}\ and \ref{subsec:heisenberg}\ on
explicit examples with non-analytic low-energy behavior that makes momentum
expansion impossible.) For a detailed explanation how to set up a
consistent power counting scheme, see \cite{Manohar:1996cq,Burgess:1998ku}.
The standard procedure to construct invariant Lagrangians for NG bosons
\cite{Coleman:1969sm,Callan:1969sn} is briefly reviewed in the following
subsection. After then I will point out some specific issues concerning its
application to nonrelativistic systems.

\subsection{Coset construction of effective Lagrangians}
As was shown already in Section~\ref{sec:goldstone}\ NG states are created
from the vacuum by the action of the broken symmetry generators. Formally, a
\emph{global} broken symmetry transformation, which merely moves the vacuum
into an equivalent ground state, may be viewed as a zero-momentum NG boson. The
finite-momentum NG state can then be excited by a \emph{local} infinitesimal
broken symmetry transformation. This corresponds to the intuitive picture of NG
modes as small local fluctuations of the order parameter. Denoting the vacuum
expectation value of the order parameter field $\phi(x)$ as $\vp$, all possible
values of the order parameter that can be reached from $\vp$ by a symmetry
transformation span a manifold, known as the \emph{coset space} of the
broken symmetry, $\gr{G/H}$. (Strictly speaking, the coset space may in general
consist of several orbits of the symmetry group $\gr G$. For the construction
of the effective Lagrangian, only the orbit containing the vacuum $\vp$ is
important.) The elements of the coset space may be thought of as the sets $g\gr
H=\{gh\,|\,h\in\gr H\}$ with arbitrary $g\in\gr G$. Thanks to the group
properties, two sets $g_1\gr H$ and $g_2\gr H$ are either identical or disjoint,
and $\gr{G/H}$ is the space of these classes.

While the ground state is represented by a single point in the coset space, a
general NG field configuration can be viewed as a map from the spacetime to the
coset space, $\phi(x):\gr{R^4}\to\gr{G/H}$. The task to construct the most
general $\gr G$-invariant Lagrangian for the NG bosons is equivalent to the
geometric problem of constructing a $\gr G$-invariant function on $\gr{G/H}$,
given the action of the group on the order parameter.

Let the symmetry group $\gr G$ be compact and semi-simple. Then one can choose
a basis $T_a$ of its Lie algebra such that $\Tr(T_aT_b)=\delta_{ab}$
(see Chap.~1 of \cite{Barut:1977ba}). Splitting the generators $T_a$ into the
unbroken ones, $V_\alpha$, generating the unbroken subgroup $\gr H$, and the
broken ones, $A_i$, it follows that $\Tr(V_\alpha
V_\beta)=\delta_{\alpha\beta}$, $\Tr(A_iA_j)=\delta_{ij}$, and $\Tr(V_\alpha
A_i)=0$. Furthermore, the structure constants $f_{abc}$ of the Lie algebra of
$\gr G$ are fully antisymmetric and the broken generators span a representation
of the unbroken group, hence the commutation relations
\begin{equation}
[V_\alpha,V_\beta]=\imag f_{\alpha\beta\gamma}V_\gamma,\quad
[V_\alpha,A_i]=\imag f_{\alpha ij}A_j,\quad
[A_i,A_j]=\imag f_{ij\alpha}V_\alpha+\imag f_{ijk}A_k
\end{equation}
Any element $g\in\gr G$ can be, at least in the neighborhood of unity,
decomposed as $g=e^{\imag\pi^iA_i}e^{\imag v^\alpha V_\alpha}$. Since $e^{\imag
v^\alpha V_\alpha}$ leaves the ground state intact, the points of the coset
space can be parameterized as $\phi(\pi)=U(\pi)\vp$, where
$U(\pi)=e^{\imag\pi^iA_i}$. Thus, the NG fields $\pi^i$ serve as coordinates on
$\gr{G/H}$. The action of the symmetry group on $\gr{G/H}$ is defined by left
multiplication, $\phi(\pi)\xrightarrow{g}g\phi(\pi)=\phi(\tilde\pi)$.
Decomposing $gU(\pi)$ as $U(\tilde\pi)h(g,\pi)$, where $h(g,\pi)=e^{\imag
v^\alpha(g,\pi)V_\alpha}$ belongs to the unbroken subgroup $\gr H$, one arrives
at the transformation rule
\begin{equation}
U(\pi)\xrightarrow{g}U(\tilde\pi(g,\pi))=gU(\pi)\he h(g,\pi)
\label{coset_transfo}
\end{equation}
This defines a nonlinear realization of the group $\gr G$ on $\gr{G/H}$. In
order that the group structure be preserved, the function $h:\gr
G\times\gr{G/H}\to\gr H$ must satisfy basic constraints such as
$h(\openone,\pi)=\openone$ and the associativity,
$h(g'g,\pi)=h(g',\tilde\pi(g,\pi))h(g,\pi)$. Also, for transformations from the
unbroken subgroup it is independent of $\pi$, $h(h,\pi)=h$, so that \emph{the
unbroken group acts on $\gr{G/H}$ linearly}, $\tilde\pi^iA_i=h(\pi^iA_i)\he h$.

The effective Lagrangian is most easily constructed using the
\emph{Maurer--Cartan form},
\begin{equation}
\omega(\pi)=-\imag\he U(\pi)\dd U(\pi)
\end{equation}
Employing the very useful formula,
\begin{equation}
e^{-A}\dd e^A=\int_0^1\dd t\,e^{-tA}(\dd A)e^{tA}
\label{Baker_fla}
\end{equation}
one easily realizes that the Maurer--Cartan form lies in the Lie algebra of
$\gr G$. Under the group action \eqref{coset_transfo} it transforms as
\begin{equation}
\omega\xrightarrow{g}\tilde\omega(g,\omega)=h\omega\he h-\imag h\dd\he h
\label{Maurer_transfo}
\end{equation}
Since $h\dd\he h$ lies in the Lie algebra of $\gr H$, it is useful to decompose
the Maurer--Cartan form into projections on the spaces of unbroken and broken
generators, $\omega=\omega_\parallel+\omega_\perp$. The ``longitudinal'' part
transforms as a gauge field,
$\omega_\parallel\xrightarrow{g}h\omega_\parallel\he h-\imag h\dd\he h$, while
the ``transverse'' part transforms covariantly $\omega_\perp\to
h\omega_\perp\he h$.

Should it be necessary to include other, non-NG modes in the effective theory,
one proceeds as follows. Taking advantage of the fact that the unbroken
subgroup $\gr H$ is realized linearly even on the NG bosons, consider a set of
``matter'' fields $\psi$ that transform in the representation $\RR$ of $\gr H$,
$\psi\xrightarrow{h}h_\RR\psi$. This is promoted to a nonlinear realization of
the full group $\gr G$, defined by
\begin{equation}
\psi\xrightarrow{g}h_\RR(g,\pi)\psi
\end{equation}
Any Lagrangian invariant under $\gr H$ can be made invariant under the full
group $\gr G$ using this procedure. Note that since the transformation matrix
$h_\RR(g,\pi)$ depends on the NG fields $\pi$ and thus implicitly on the
spacetime coordinates, the transformation becomes local. In order that terms
containing derivatives of $\psi$ be still $\gr G$-invariant, one uses the
longitudinal component of the Maurer--Cartan form,
$\omega_\parallel=\omega_{\parallel\mu}\dd x^\mu$, to promote derivatives to
covariant ones, $\de_\mu\psi\to(\de_\mu+\imag\omega_{\parallel\mu})\psi$. The
most general $\gr G$-invariant Lagrangian can then be constructed
\cite{Burgess:1998ku} from the objects $\omega_\perp,\psi$ and their covariant
derivatives. For instance, the expression
\begin{equation}
\Lag_{\text{bilin}}=\Tr(f_\pi^2\omega_\perp\omega_\perp)
\label{coset_kinterm}
\end{equation}
obviously includes the kinetic term for the NG bosons, as can be seen with the
help of Eq.~\eqref{Baker_fla}. The $f_\pi^2$ is a Hermitian, positive definite
matrix that must commute with all matrices from $\gr H$ in order that
$\Lag_{\text{bilin}}$ be $\gr G$-invariant. If the broken generators transform
irreducibly under $\gr H$, it is proportional to the unit matrix by Schur's
lemma, hence providing merely an overall normalization of the Lagrangian. In
general it contains one arbitrary parameter per each irreducible
representation, thus giving an independent normalization of the kinetic term for
each multiplet of NG bosons.

One often meets the situation that the symmetry in question is local rather
than global. This can happen both, when the system interacts with a dynamical
gauge (such as electromagnetic) field and when external sources are coupled to
the conserved currents of the theory \cite{Leutwyler:1993iq}. Consider a gauge
field $\AAA$ transforming under local $\gr G$ transformations as usual as
$\AAA\xrightarrow{g}g\AAA\he g+\imag g\dd\he g$. It is straightforward to see
that the Maurer--Cartan form, defined now by $\imag\omega=\he U(\pi)\DD
U(\pi)=\he U(\pi)(\dd-\imag\AAA)U(\pi)$, satisfies the same rule
\eqref{Maurer_transfo} as in the case of a global symmetry.

This concludes the brief review of the coset construction of effective
Lagrangians. It is useful to note that while the choice of the coset
parameterization as $U(\pi)=e^{\imag\pi^iA_i}$ is very convenient in particular
because it renders the action of the unbroken subgroup linear, the construction
is not limited to this choice. In fact, the key elements are only the geometry
of the coset space and its symmetry. The choice of coordinates $\pi^i$ has no
observable effects; one speaks of \emph{reparameterization invariance} of
physical observables \cite{Coleman:1969sm}.

\subsection{Geometric interpretation}
The kinetic term \eqref{coset_kinterm} can be evaluated more or less
explicitly. Using the orthogonality of the generators, the Maurer--Cartan form
can be expanded in $V_\alpha,A_i$. Denoting $\omega_\perp=\omega_{\perp i}A_i$,
one finds
\begin{equation}
\omega_{\perp i}=\Sigma_{ij}\dd\pi^j,\quad
\Sigma_{ij}=\int_0^1\dd t\,\Tr\left(
A_ie^{-\imag t\pi^kA_k}A_je^{+\imag t\pi^kA_k}
\right)
\end{equation}
The kinetic term acquires the form $g_{ij}\dd\pi^i\dd\pi^j$ with
$g_{ij}=\Sigma_{ki}\Sigma_{lj}\Tr(f_\pi^2A_kA_l)$. (I deliberately use
differentials instead of spacetime derivatives in order to emphasize the
geometric nature of the quantities, and to avoid having to treat space and
time derivatives separately.) The matrix $g_{ij}$ can be interpreted as a
\emph{metric} on the coset space. The construction of the most general term in
the effective Lagrangian with two derivatives therefore reduces to the
geometric problem of finding all $\gr G$-invariant metrics on $\gr{G/H}$
\cite{Leutwyler:1993gf,Leutwyler:1993iq}.

In order to see what are the coordinate transformations that leave the metric
$g_{ij}$ invariant, one must explicitly evaluate the change of the fields
$\pi^i$ under a $\gr G$ transformation. Writing Eq.~\eqref{coset_transfo} as
\begin{equation}
e^{-\imag\pi^iA_i}e^{\imag\tilde\pi^iA_i}e^{\imag v^\alpha V_\alpha}=
e^{-\imag\pi^iA_i}ge^{\imag\pi^iA_i}
\label{auxtransfo}
\end{equation}
one sets $g=e^{\imag\vt^aT_a}$ and expands both sides to first order in the
infinitesimal transformation parameters $\vt^a$. (No assumption on the
smallness of $\pi^i$ is made though.) Denoting further
\begin{equation}
\R_{\alpha i}=\int_0^1\dd t\,\Tr\left(V_\alpha e^{-\imag t\pi^kA_k}A_i
e^{+\imag t\pi^kA_k}\right)
\end{equation}
$e^{-\imag\pi^kA_k}V_\alpha e^{\imag\pi^kA_k}=V_\beta P^{(VV)}_{\beta\alpha}+
A_iP^{(AV)}_{i\alpha}$, and $e^{-\imag\pi^kA_k}A_i
e^{\imag\pi^kA_k}=V_\alpha P^{(VA)}_{\alpha i}+A_jP^{(AA)}_{ji}$,
Eq.~\eqref{auxtransfo} reduces to
\begin{equation}
(V_\alpha R_{\alpha j}+A_i\Sigma_{ij})\delta\pi^j+V_\alpha v^\alpha=
\left[V_\alpha P^{(VV)}_{\alpha\beta}+A_iP^{(AV)}_{i\beta}\right]\vt^\beta+
\left[V_\alpha P^{(VA)}_{\alpha j}+A_iP^{(AA)}_{ij}\right]\vt^j
\label{master_pi}
\end{equation}
Comparing coefficients at $A_i$, a formula for the shift of NG fields,
$\delta\pi^i$, follows,
\begin{equation}
\delta\pi^i=(\Sigma^{-1})^{ij}\left[P^{(AV)}_{j\beta}\vt^\beta+P^{(AA)}_{jk}\vt^k\right]
\equiv\xi^i_{a}\vt^a
\label{killing}
\end{equation}
The set of objects $\xi^i_a$ that depend just on the coset geometry define the
\emph{Killing vectors} on $\gr{G/H}$. They represent infinitesimal motions on
$\gr{G/H}$ induced by the symmetry group transformations. A similar
expression for the parameters $v^\alpha$ of the compensating transformation
$h(g,\pi)$ is implied.

The above formulas were derived for infinitesimal transformation parameters
$\vt^a$, but hold for arbitrary values of the NG fields $\pi^i$. In practice,
the vacuum is usually set at $\pi^i=0$ and for calculation of low-energy
observables such as scattering amplitudes, one gets by with a power expansion
in $\pi^i$. All expressions of this subsection can then be evaluated explicitly
in terms of the structure constants of the Lie algebra of symmetry generators.
To first order in the NG fields one finds, for instance,
\begin{equation}
\Sigma_{ij}=\delta_{ij}-\frac12f_{ijk}\pi^k,\quad
\xi^i_\alpha=f^i_{\ j\alpha}\pi^j,\quad
\xi^i_j=\delta^i_j-\frac12f^i_{\ jk}\pi^k
\end{equation}
This essentially confirms the picture that underlies all discussions of
effective Lagrangians for NG bosons: unbroken symmetry is realized linearly on
the NG fields whereas broken transformations are equivalent, to lowest order,
to a mere shift. This makes sure that the NG fields correspond to gapless modes
in the spectrum.

It is worthwhile to inquire how the Maurer--Cartan form changes in presence
of a gauge field. According to our discussion above, it picks an additional
contribution,
\begin{equation}
-\he U(\pi)\AAA U(\pi)= -\left[V_\beta
P^{(VV)}_{\beta\alpha}+A_iP^{(AV)}_{i\alpha}\right]\AAA^\alpha -\left[V_\alpha
P^{(VA)}_{\alpha i}+A_jP^{(AA)}_{ji}\right]\AAA^i
\label{gauged_Maurer}
\end{equation}
One then immediately finds that the transverse component of the Maurer--Cartan
form that determines the kinetic term of the NG bosons, modifies according to
$\omega_{\perp i}=\Sigma_{ij}\dd\pi^j\to\Sigma_{ij}\DD\pi^j$, where
$\DD\pi^i=\dd\pi^i-\xi^i_a\AAA^a$. This shows that even in this very general
setting where the symmetry group is realized nonlinearly, the infinitesimal
shift of the fields under group motion serves to construct their
gauge-covariant derivative.

In the above derivation, the metric $g_{ij}$ was derived from the
Maurer--Cartan form and its $\gr G$-invariance followed from the $\gr
G$-invariance of the Lagrangian. It is instructive to reverse the argument and
use an invariant metric on the coset space to construct
Eq.~\eqref{coset_kinterm} in a purely geometric way. To this end, recall that
we deal with compact symmetry groups that can be represented by unitary
matrices. The space of complex matrices possesses a natural distance function,
invariant under both left and right multiplication by unitary matrices, defined
by $d(x,y)=\nor{x-y}$, where $\nor x=\sqrt{\Tr(\he xx)}$. (I use the term
\emph{distance} instead of metric to distinguish this object from the metric
\emph{tensor}, which roughly speaking measures the distance on the tangent
space at unity.) This induces a distance function on the coset $\gr{G/H}$,
manifestly invariant under the action \eqref{coset_transfo} of $\gr G$,
\begin{equation}
d_{\gr{G/H}}(x,y)=\min_{h\in\gr H}d(xh,y)
\label{coset_distance}
\end{equation}
It is easy to show that this prescription indeed yields a well defined
distance. First, the unbroken subgroup $\gr H$ is compact so that the
minimum in Eq.~\eqref{coset_distance} exists. Furthermore, $d_{\gr{G/H}}$ is
independent of the choice of the representative elements $x,y$ for the cosets
$x\gr H,y\gr H$ and it is zero if and only if $x$ and $y$ lie in the same coset.
Finally, it is straightforward to prove the triangle inequality as a
consequence of that for the distance $d$.

To determine the metric \emph{tensor} following from
Eq.~\eqref{coset_distance}, one takes the points $x,y$ infinitesimally far
apart, $x=U(\pi)$ and $y=U(\pi+\dd\pi)$, whence
\begin{equation}
d_{\gr{G/H}}(x,y)=\min_{h\in\gr H}\nor{U(\pi+\dd\pi)-U(\pi)h}=
\min_{h\in\gr H}\nor{\he U(\pi)U(\pi+\dd\pi)-h}=
\min_{h\in\gr H}\nor{\imag\omega(\pi)+\openone-h}
\end{equation}
The minimum will be realized for $h$ infinitesimally close to unity, hence
$\openone-h$ will lie in the Lie algebra of $\gr H$, and affect only the
longitudinal part of the Maurer--Cartan form, $\omega_\parallel(\pi)$. Clearly,
the minimum is achieved when $\openone-h$ exactly cancels
$\omega_\parallel$ so that
\begin{equation}
d_{\gr{G/H}}(x,y)=\nor{\omega_\perp}=\sqrt{\Tr(\omega_\perp\omega_\perp)}
\label{coset_metric_intr}
\end{equation}
Consequently, the bilinear Lagrangian \eqref{coset_kinterm} is directly related
to a $\gr G$-invariant metric on the coset space $\gr{G/H}$. The only
difference is the factor $f_\pi^2$ present in Eq.~\eqref{coset_kinterm}.
Mathematically, this comes from the fact that Eq.~\eqref{coset_metric_intr}
represents an intrinsic metric of the coset space, invariant under both left
and right $\gr G$ transformations, while Eq.~\eqref{coset_kinterm} is by
construction invariant under left $\gr G$ multiplication, but only under $\gr
H$ acting from the right \cite{Leutwyler:1993iq}.

\subsection{Nonrelativistic effective Lagrangians}
In the preceding two subsections a general method how to build effective
Lagrangians was presented. The construction is based on an implicit
\emph{assumption} that the Lagrangian is invariant under the symmetry group of
the system. Global symmetry of the theory can be conveniently formulated in
terms of invariance of the generating functional of Green's functions of the
conserved currents, $j^\mu_a(x)$, with respect to \emph{local} transformations
of external sources, $\AAA^a_\mu(x)$, coupled to these currents
\cite{Gasser:1983yg,Gasser:1984gg}. This implies a set of Ward identities that
the theory must satisfy in order to preserve the symmetry. In Lorentz
invariant theories without quantum anomalies, the most general solution of the
Ward identities can indeed always be obtained from a Lagrangian invariant under
the symmetry group \cite{Leutwyler:1993iq}. On the other hand, in systems that
lack Lorentz invariance (``nonrelativistic'' systems) this conclusion is no
longer valid. The most general solution of the Ward identities to the lowest
order of the derivative expansion was given by Leutwyler \cite{Leutwyler:1993gf}
who showed that in general it is just the action, not the Lagrangian, what is
invariant.

The Lagrangian of a low-energy EFT in principle contains an infinite number of
terms. The predictive power of EFT lies in the fact that at any desired
accuracy of an observable to be calculated, only a finite number of terms
contribute. The Lagrangian is systematically expanded in powers of space ($s$)
and time ($t$) derivatives, $\Lag=\sum_{s,t}\Lag^{(s,t)}$. In order to achieve
a consistent power counting scheme, it is convenient to assign the external
source $\AAA^a_\mu$ the same order as the derivative $\de_\mu$. However, the
scaling of spatial and temporal derivatives may in general differ. For the sake
of simplicity, I will assume that the system is invariant under the continuous
rotation group $\gr{SO(3)}$. This is certainly plausible for fluids, and even
for some crystalline (such the plain cubic) structures, where the anisotropy
induced by the lattice only appears at higher orders in the derivative
expansion. The first few terms in the effective Lagrangian then read,
\begin{equation}
\Lag^{(0,1)}=C_i(\pi)\dot\pi^i+\vr_a(\pi)\AAA^a_0,\quad
\Lag^{(0,2)}=\frac12\bar g_{ij}(\pi)\DD_0\pi^i\DD_0\pi^j,\quad
\Lag^{(2,0)}=-\frac12g_{ij}(\pi)\DD_k\pi^i\DD_k\pi^j
\end{equation}
where the dot denotes a time derivative, and the covariant derivative,
$\DD_\mu\pi^i=\de_\mu\pi^i-\xi^i_a\AAA^a_\mu$, was introduced below
Eq.~\eqref{gauged_Maurer}. Both $g_{ij}(\pi)$ and $\bar
g_{ij}(\pi)$ are $\gr G$-invariant metrics on the coset space. It follows from
the above discussion that if the NG fields transform irreducibly under the
unbroken symmetry, they must be identical functions on $\gr{G/H}$ up to a
trivial scale factor. Both $\Lag^{(0,2)}$ and $\Lag^{(2,0)}$ are obviously
invariant under simultaneous gauge transformations of the NG fields and the
external sources.

It would be tempting to conclude that the coefficient functions $C_i(\pi)$ and
$\vr_a(\pi)$ are related by $\vr_a=-C_i\xi^i_a$ so that
$\Lag^{(0,1)}=C_i(\pi)\DD_0\pi^i$. However, this does not follow from general
symmetry considerations. Instead, the Ward identities encoded in the gauge
invariance of the generating functional imply a set of constraints for all
coefficients $C_i(\pi),\vr_a(\pi),g_{ij}(\pi),\bar g_{ij}(\pi),\xi^i_a(\pi)$
\cite{Leutwyler:1993gf}. The functions $\xi^i_a(\pi)$, defined by
Eq.~\eqref{killing}, are Killing vectors of \emph{both} invariant metrics,
$g_{ij}(\pi)$ and $\bar g_{ij}(\pi)$, that is, they obey the Killing
equation $\nabla_i\xi_{ja}+\nabla_j\xi_{ia}=0$. The Riemannian covariant
derivative $\nabla_i$ is defined with the help of the Christoffel symbol,
$\nabla_i\xi_{ja}=\de_i\xi_{ja}-\Gamma^k_{ij}\xi_{ka}$. (Following
\cite{Leutwyler:1993gf} I use the notation $\de_i\equiv\Pd{}{\pi^i}$, not to be
mixed up with a derivative with respect to the space coordinate.) Note
that using the two metrics to lower the index in $\xi^i_a$, one obtains two
\emph{different} objects, $\xi_{ia}$ and $\bar\xi_{ia}$, that satisfy Killing
equations with respective Christoffel symbols.

Since the Killing vectors $\xi^i_a$ generate the symmetry transformation
\eqref{killing} on the coset space, they must reproduce the structure of the
Lie algebra of the symmetry generators, $\dd{}_a\xi^i_b-\dd{}_b\xi^i_a=f^c_{\
ab}\xi^i_c$, where $\dd{}_a\equiv\xi^i_a\de_i$. Finally, the coefficients
$C_i(\pi)$ and $\vr_a(\pi)$ satisfy the identities
\begin{equation}
\dd{}_a\vr_b=f^c_{\ ab}\vr_c,\quad \xi^i_a(\de_iC_j-\de_jC_i)=\de_j\vr_a
\label{constr_leut}
\end{equation}
While the former asserts that $\vr_a(\pi)$ transforms in the adjoint
representation of the symmetry group and fixes its value at any point of
$\gr{G/H}$ in terms of that at the origin ($\pi=0$), the latter determines the
function $C_i(\pi)$ up to a gradient. This ambiguity is intrinsic to EFT and has
no observable consequences. Indeed, changing $C_i(\pi)$ by a gradient merely
modifies the Lagrangian by a total time derivative, hence leaving the action
intact.

Several remarks are in order here. First, $\Lag^{(0,1)}$ is in general not
gauge invariant: under the transformation \eqref{killing} it changes by a total
derivative, $\delta\Lag^{(0,1)}=\OD{}{t}\left[\vt^a(C_i\xi^i_a+\vr_a)\right]$.
Second, evaluating the action at its extremum in order to establish the leading
contribution to the generating functional, one finds by differentiation with
respect to $\AAA^a_0$ that $\vr_a(\pi)$ at the origin determines the density of
the conserved charge, $\bra0j^0_a(x)\ket0=\vr_a(0)$. When all charge
densities are zero, it follows from Eq.~\eqref{constr_leut} that both
$\vr_a(\pi)$ and $C_i(\pi)$ vanish identically and the lowest order Lagrangian
is essentially the same as in Lorentz invariant theories, up to a possible
redefinition of the phase velocity of the NG bosons. The energy of the NG
bosons is linear in momentum so that one assigns space and time derivatives the
same degree in the power expansion, and the leading order Lagrangian reads
$\Lag=\Lag^{(0,2)}+\Lag^{(2,0)}$. Also, when the symmetry is Abelian
Eq.~\eqref{constr_leut} guarantees that $\vr_a(\pi)=\vr_a(0)$ on the whole
coset space, and in turn $C_i(\pi)=0$. Since constant $\vr_a(\pi)$ does not
contribute to the dynamics of NG bosons, merely generating background charge
density, the equation of motion and dispersion relation of the NG bosons will
be the same as above.

On the other hand, nonzero density of a non-Abelian charge leads to nonzero
$C_i(\pi)$. The Lagrangian is gauge invariant only up to a total derivative, and
the term with the single time derivative leads to a quadratic dispersion
relation of the NG boson. One must then count a time derivative as two space
derivatives so that the lowest order Lagrangian is
$\Lag=\Lag^{(0,1)}+\Lag^{(2,0)}$. This confirms the observation that density of
a non-Abelian charge implies a type-II NG boson in the spectrum, made in
Section~\ref{sec:goldstone}\ on a general ground.

It is instructive to illustrate the construction of effective Lagrangians on an
explicit example. The most convenient one is provided by the (anti)ferromagnet.
Both these systems possess an $\gr{SU(2)}$ symmetry group of spin rotations,
which is spontaneously broken by an ordered ground state to its $\gr{U(1)}$
subgroup. As already mentioned in Section~\ref{subsec:GBdispersions}\ in a
ferromagnet all spins sitting on the crystal lattice are aligned in the same
direction and there is a net total magnetization. Accordingly, the two broken
generators give rise to one type-II NG boson with quadratic dispersion
relation. On the contrary, in an antiferromagnet antiparallel alignment of
neighboring spins is favored and the ground state, though ordered, does not
develop nonzero spin density. There are two NG bosons with linear dispersion
relations, corresponding to two linearly polarized spin waves.

The coset space is $\gr{SU(2)/U(1)\simeq SO(3)/SO(2)}$ in both cases and it is
equivalent to a two-sphere, $S^2$. It can therefore be most conveniently
represented by a unit vector, $\vek n$. The symmetry group $\gr{SU(2)}$ acts on
$\vek n$ through rotations, which can be parameterized by a vector $\vek g$
whose direction and magnitude represent the axis and angle of the rotation,
respectively. An infinitesimal symmetry transformation therefore reads
$\delta\vek n=\vek n\times\vek g$, and the covariant derivative in presence of
an external source $\vek\AAA_\mu$ consequently $\DD_\mu\vek n=\de_\mu\vek n+
\vek\AAA_\mu\times\vek n$. Since all charge densities are zero in the
antiferromagnet, the term $\Lag^{(0,1)}$ is missing in the Lagrangian. The
leading order Lagrangian thus is
\begin{equation}
\Lag_{\text{antiferro}}=\frac12F_t^2\DD_0\vek n\cdot\DD_0\vek n
-\frac12F_s^2\DD_k\vek n\cdot\DD_k\vek n
\end{equation}
Obviously, for an arbitrary set of coordinates $\pi^1,\pi^2$ on the coset, the
invariant metrics will be $g_{ij}=F_s^2\de_i\vek n\cdot\de_j\vek n$ and
$\bar g_{ij}=F_t^2\de_i\vek n\cdot\de_j\vek n$. From the Lagrangian one can
also extract the Killing vectors, $\vek h_i=-F_s^2\vek n\times\de_i\vek n$.
Indices are raised with the inverse metric,
$g^{ij}=\ve^{ik}\ve^{j\ell}\de_k\vek n\cdot\de_\ell\vek n/F_s^2\nor{\de_1\vek
n\times\de_2\vek n}^2$.

The case of the ferromagnet is more subtle. The ground state has nonzero charge
density and accordingly the term $C_i(\pi)$ appears in the Lagrangian. First
observe that the zero component of the external field, $\AAA_0^a$, couples to
the Noether charge density, which is proportional to the total magnetization.
Hence $\AAA_0^a$ is interpreted as an external magnetic field. The direction of
the magnetization in the ground state is aligned with it so that the
vector of couplings $\vr_a(\pi)$ is simply proportional to the order parameter
$\vek n(\pi)$ up to a constant, $\vek\vr=\alpha\vek n$. By explicitly solving
the constraint \eqref{constr_leut}, the function $C_i(\pi)$ can be shown to
have a topological nature, related to the Brouwer degree of the map $\vek
n(\pi)$ \cite{Leutwyler:1993gf}. Constructing a path $\tilde\pi^i(\pi,\lambda)$,
$0\leq\lambda\leq1$ such that $\tilde\pi^i(\pi,0)=0$ and
$\tilde\pi^i(\pi,1)=\pi^i$, and setting $\vek n=\vek
n\bigl(\tilde\pi(\pi,\lambda)\bigr)$, one has
\begin{equation}
C_i(\pi)=\alpha\int_0^1\dd\lambda\,(\de_i\vek n\times\de_\lambda\vek n)
\cdot\vek n
\end{equation}
The integral is independent of the choice of the path, connecting the origin
with the point $\pi^i$.  This concludes the construction of the leading order
effective Lagrangian for the ferromagnet. Details can be found in
\cite{Leutwyler:1993gf}.

One comment concerning the difference between the antiferromagnet and the
ferromagnet is worthwhile here. Due to the presence of a term with a single
time derivative, the effective Lagrangian of the ferromagnet is sometimes
claimed to break time reversal invariance. However, this is somewhat
misleading. What breaks the time reversal is the ferromagnetic \emph{ground
state}, thanks to the nonzero charge density, that is, a zero component of a
conserved current. On the other hand, any microscopic description of a
ferromagnet will be time reversal invariant. Indeed, for example the
Hamiltonian \eqref{Heisenberg} can describe both the ferromagnetic and the
antiferromagnetic state, depending on the sign of the coupling constants
$J_{ij}$. Both systems therefore must have the same symmetry. Since one of the
basic requirements on the low-energy EFT is that it should reproduce the
symmetry of the underlying theory, the effective Lagrangian of the ferromagnet
must actually inherit this time reversal invariance. What really distinguishes
ferromagnets from antiferromagnets is the way time reversal is implemented on
the coset space, which is natural because this implementation does depend on
the structure of the ground state. The transformation of the NG fields under
time reversal in the antiferromagnet is such that it prohibits the term with a
single time derivative, whereas in the ferromagnet it is allowed
\cite{Roman:1999ro}.

\subsection{Applications of effective field theory}
Effective field theory has become an invaluable tool of a theoretical physicist
in all branches of physics where the methods of quantum field theory are used.
It often facilitates calculations that would otherwise be very complicated, or
hard to carry out at all. It would be hopeless to even try to give a
representative list of all its applications. I will therefore just mentioned
briefly a few explicit examples that are related to other material presented
in this paper. The application of EFT to ferromagnets was worked out in
\cite{Leutwyler:1993gf,Roman:1999ro,Hofmann:1998pp,Hofmann:2001ck}, and
in a similar spirit also for antiferromagnets
\cite{Hofmann:1997qm,Kampfer:2005ba,Brugger:2006dz,Hofmann:2009ru}. The utility
of EFT of course goes beyond the mere derivation of the equation of motion and
dispersion relation of spin waves. One can calculate other observables as well
such as the cross section for scattering of magnons, or of slow neutrons off the
magnons \cite{Burgess:1998ku}. The use of EFT as well as other techniques for
the nonrelativistic weakly interacting Bose gas and Bose--Einstein condensation
is reviewed in \cite{Andersen:2003qj}. See also \cite{Andersen:2002nd} for the
calculation of the damping rate of the corresponding NG boson and the
construction of the effective Lagrangian based on Galilean invariance. Subtle
features associated with a consistent definition of thermodynamics in various
approximation schemes for Bose condensed systems are discussed in detail in
\cite{Yukalov:2008yu}. Extension of Galilean invariance in nonrelativistic
physics to general coordinate invariance was worked out in \cite{Son:2005rv}
and applied to strongly attractive Fermi gas at unitarity.

Among the varied applications of EFT, I would like to discuss in a little
more detail the technique to construct the effective action for relativistic
superfluids from the equation of state, developed by Son \cite{Son:2002zn}. It
is based on the observation that in a relativistic many-body system at zero
temperature, the chemical potential is the sole source of Lorentz violation.
Lorentz invariance can therefore be restored when the chemical potential is
treated as a zero component of a background gauge field, very much like in
Section~\ref{subsec:lsm}\ where the linear sigma model was investigated.

Consider a relativistic system with Abelian $\gr{U(1)}$ symmetry that acts on
the order parameter $\phi$ as a phase transformation: $\phi\to\phi
e^{\imag\vt}$. Assigning a chemical potential $\mu$ to the Noether charge of
this symmetry and denoting as in Section~\ref{subsec:lsm}\ $A_\mu=(\mu,\vek0)$,
the quantum effective action of the system, $\Gamma_{\text{eff}}[A_\mu,\phi]$,
will be invariant under a simultaneous \emph{local} transformation of $\phi$ and
$A_\mu$, $A_\mu\to A_\mu+\de_\mu\vt$. As soon as the order parameter develops
nonzero vacuum expectation value, $\vp$, it can be conveniently parameterized
as $\phi(x)=e^{\imag\pi(x)}[\vp+H(x)]$. The phase transformation of $\phi$ is
equivalent to a shift of the NG field, $\pi\to\pi+\vt$.

At low energy the modes created by the ``Higgs'' field cannot be excited. In
order to obtain an effective action for the NG boson only, one ``integrates
out'' the Higgs mode by minimizing the effective action with respect to the
modulus $\abs\phi$. The leading term in the derivative expansion of the
resulting effective action, $\Gamma_{\text{eff}}[A_\mu,\pi]$, depends just on
$A_\mu$ and $\de_\mu\pi$. The underlying gauge invariance makes sure that they
appear in the effective Lagrangian only in the combination
$\DD_\mu\pi=\de_\mu\pi-A_\mu$. At the minimum of the effective action, the
effective Lagrangian is equal to minus the energy density of the ground state
in the background field $A_\mu$ (see Sec.~16.3 in \cite{Weinberg:1996v2}).
Recalling that the energy density is equal to minus the thermodynamic pressure,
$p(\mu)$, and observing that $\mu=\sqrt{A^\mu
A_\mu}=\sqrt{\DD_\mu\pi\DD^\mu\pi}$ for a constant field $\pi$, the full
dependence on $\DD_\mu\pi$ can be restored with the knowledge of the function
$p(\mu)$, that is, the equation of state,
\begin{equation}
\Gamma_{\text{eff}}[\mu,\pi]=
\int\dfour x\,p(\sqrt{\DD_\mu\pi\DD^\mu\pi})=
\int\dfour x\,p\Bigl(\sqrt{(\de_0\pi-\mu)^2-(\nabla\pi)^2}\Bigr)
\end{equation}
One therefore arrives at the conclusion that to leading order in the derivative
expansion, the full quantum effective action of the superfluid NG boson is
completely determined by the equation of state. Since the equation of state is
usually straightforward to calculate perturbatively, this is a rare example of
a system where the low-energy EFT can be derived explicitly from the underlying
microscopic theory. The effective Lagrangian can be used to derive transport
coefficients and other observables associated with low-energy processes in
which only the NG bosons can be excited \cite{Manuel:2004iv}.


\section{Conclusions}
\label{sec:summary} In the present paper, I have provided a self-contained
introduction to spontaneous symmetry breaking and the properties of the
associated NG bosons. I discussed to some detail the mathematical subtleties,
connected with the implementation of a broken symmetry on the Hilbert space of
states. The focus of the paper was on nontrivial features of SSB specific to
nonrelativistic systems. These include unconventional dispersion relations of
NG bosons as well as improved rules for their counting. The general results
were demonstrated on several, mostly exactly solvable, examples.

Due to limited space, several issues were touched only superficially, or
not at all. The first of them is the possibility of explicit symmetry breaking.
If weak enough, spontaneous breaking of such symmetry will give rise to a
pseudo-NG boson. Its energy at zero momentum is not strictly zero, but it is
still small compared to the characteristic energy scale of the system. However,
the counting of NG bosons then becomes blurred, since it may not be possible to
distinguish an ``approximately linear'' dispersion relation from an
``approximately quadratic'' one. Only when the explicit symmetry breaking is
weak, one can still distinguish type-I and type-II NG bosons
\cite{Brauner:2007uw}.

A special case of explicit symmetry breaking is represented by a wide class of
theories where a part of the Lagrangian, responsible for the NG boson spectrum
at the tree level or in a similar approximation, has a higher symmetry than the
rest of it. This typically happens in relativistic scalar theories coupled to
gauge fields \cite{Weinberg:1972fn,Coleman:1973jx}, but it can also occur in
some condensed matter systems such as superfluid \isotope[3]{He}
\cite{Volovik:1982vo,Volovik:1983vo}. In the appropriate approximation, the
system then exhibits more gapless states than would correspond to the symmetry
of the full theory, some of them stemming from the extended symmetry of the
part of the Lagrangian. These spurious NG bosons receive a gap once quantum
corrections are taken into account. This mechanism can be responsible for the
presence of naturally light states in the spectrum.

Second, in this paper I have considered exclusively continuous internal
symmetries. While spontaneous breaking of a discrete symmetry does not give
rise to NG bosons, spontaneous breaking of spacetime symmetries is subtle.
First of all, the NG field configurations can be imagined as small local
fluctuations of the order parameter. Since the local versions of different
spacetime transformations may coincide, there are typically fewer NG bosons
than broken generators \cite{Low:2001bw}. Examples of spontaneous breaking of
spacetime symmetries include homogeneous but anisotropic states
such as in superfluid \isotope[3]{He} (see \cite{Vollhardt:1990vw} for an
extensive review), spin-one color superconductors \cite{Schmitt:2004et}, or even
imbalanced spin-zero color superconductors \cite{Muther:2002ej}, and in
Bose--Einstein condensates of relativistic vector fields
\cite{Sannino:2001fd,Gusynin:2003yu}, as well as inhomogeneous states such as
crystalline solids \cite{Leutwyler:1996er} or superconductors with inhomogeneous
pairing \cite{Casalbuoni:2003wh}. Accordingly, the behavior of NG bosons may be
highly nontrivial. For example, in helical ferromagnets
\cite{Dzyaloshinsky:1958dz,Moriya:1960mo} the local magnetization field forms a
spiral structure. Along the axis of the helix, the average magnetization is zero
and the NG boson is type-I. In the transverse directions, the NG boson feels the
uniform magnetization and is type-II like in ordinary ferromagnets.

My main motivation for writing this paper was to provide a review on
spontaneous symmetry breaking that would be general enough to cover both,
relativistic many-body systems as well as intrinsically nonrelativistic systems,
and thus to bridge the gap between the communities. If a reader with expertise
in high energy physics discovers that there is more to SSB and NG bosons than
usually presented in the textbooks on particle physics, and if a reader with
background in condensed matter or atomic physics finds here a general framework
for the variety of fascinating phenomena he or she is familiar with, my goal
will be achieved.


\section*{Acknowledgments}
I am indebted to Ji\v{r}\'{\i} Ho\v{s}ek for introducing me to the subtleties
of spontaneous symmetry breaking in nonrelativistic systems, and for continuous
support of my research on this topic. Many thanks belong to Jens~O.~Andersen for
reading the manuscript carefully and suggesting numerous improvements. I am
also grateful to Xu-guang Huang for pointing out several typos in the
manuscript. This work was supported by the ExtreMe Matter Institute EMMI in the
framework of the Helmholtz Alliance Program of the Helmholtz Association
(HA216/EMMI).


\bibliography{NGrefs}

\begin{thebibliography}{10}
\expandafter\ifx\csname url\endcsname\relax
  \def\url#1{\texttt{#1}}\fi
\expandafter\ifx\csname urlprefix\endcsname\relax\def\urlprefix{URL }\fi
\expandafter\ifx\csname href\endcsname\relax
  \def\href#1#2{#2} \def\path#1{#1}\fi

\bibitem{Weinberg:1995v1}
S.~Weinberg, The Quantum Theory of Fields, Vol.~I, Cambridge University Press,
  Cambridge, UK, 1995.

\bibitem{Guralnik:1968gu}
G.~S. Guralnik, C.~R. Hagen, T.~W.~B. Kibble, Broken symmetries and the
  goldstone theorem, in: R.~L. Cool, R.~E. Marshak (Eds.), Advances in Particle
  Physics, Vol.~II, Wiley, New York, NY, USA, 1968, pp. 567--708.

\bibitem{Nielsen:1975hm}
H.~B. Nielsen, S.~Chadha, On how to count goldstone bosons, Nucl. Phys. B105
  (1976) 445--453.
\newblock \href {http://dx.doi.org/10.1016/0550-3213(76)90025-0}
  {\path{doi:10.1016/0550-3213(76)90025-0}}.

\bibitem{Leutwyler:1993gf}
H.~Leutwyler, {Nonrelativistic effective Lagrangians}, Phys. Rev. D49 (1994)
  3033--3043.
\newblock \href {http://arxiv.org/abs/hep-ph/9311264}
  {\path{arXiv:hep-ph/9311264}}, \href
  {http://dx.doi.org/10.1103/PhysRevD.49.3033}
  {\path{doi:10.1103/PhysRevD.49.3033}}.

\bibitem{Georgi:1994qn}
H.~Georgi, {Effective Field Theory}, Annu. Rev. Nucl. Part. Sci. 43 (1993)
  209--252.

\bibitem{Kaplan:1995uv}
D.~B. Kaplan, {Effective Field Theories}Lectures given at 7th Summer School in
  Nuclear Physics: ``Symmetries'', Seattle, WA, USA, June 19-30, 1995.
  arXiv:nucl-th/9506035.
\newblock \href {http://arxiv.org/abs/nucl-th/9506035}
  {\path{arXiv:nucl-th/9506035}}.

\bibitem{Manohar:1996cq}
A.~V. Manohar, {Effective Field Theories}, in: H.~Latal, W.~Schweiger (Eds.),
  {Perturbative and Nonperturbative Aspects of Quantum Field Theory:
  Proceedings of the 35.~Internationale Universit\"atswochen f\"ur Kern- und
  Teilchenphysik, Schladming, Austria, March 2--9, 1996}, Vol. 479 of Lecture
  Notes in Physics, Springer, Berlin, Germany, 1997, pp. 311--362.
\newblock \href {http://arxiv.org/abs/hep-ph/9606222}
  {\path{arXiv:hep-ph/9606222}}.

\bibitem{Pich:1998xt}
A.~Pich, {Effective Field Theory}, in: R.~Gupta, A.~Morel, E.~de~Rafael,
  F.~David (Eds.), Probing the Standard Model of Particle Interactions, North
  Holland, Amsterdam, The Netherlands, 1999, pp. 949--1049.
\newblock \href {http://arxiv.org/abs/hep-ph/9806303}
  {\path{arXiv:hep-ph/9806303}}.

\bibitem{Burgess:1998ku}
C.~P. Burgess, {Goldstone and pseudo-Goldstone bosons in nuclear, particle and
  condensed-matter physics}, Phys. Rep. 330 (2000) 193--261.
\newblock \href {http://arxiv.org/abs/hep-th/9808176}
  {\path{arXiv:hep-th/9808176}}, \href
  {http://dx.doi.org/10.1016/S0370-1573(99)00111-8}
  {\path{doi:10.1016/S0370-1573(99)00111-8}}.

\bibitem{Scherer:2002tk}
S.~Scherer, {Introduction to chiral perturbation theory}, Adv. Nucl. Phys. 27
  (2003) 277--538.
\newblock \href {http://arxiv.org/abs/hep-ph/0210398}
  {\path{arXiv:hep-ph/0210398}}.

\bibitem{Bijnens:2006zp}
J.~Bijnens, {Chiral perturbation theory beyond one loop}, Prog. Part. Nucl.
  Phys. 58 (2007) 521--586.
\newblock \href {http://arxiv.org/abs/hep-ph/0604043}
  {\path{arXiv:hep-ph/0604043}}, \href
  {http://dx.doi.org/10.1016/j.ppnp.2006.08.002}
  {\path{doi:10.1016/j.ppnp.2006.08.002}}.

\bibitem{Scherer:2009bt}
S.~Scherer, {Chiral perturbation theory: Introduction and recent results in the
  one-nucleon sector}, Prog. Part. Nucl. Phys. 64 (2010) 1--60.
\newblock \href {http://arxiv.org/abs/0908.3425} {\path{arXiv:0908.3425}},
  \href {http://dx.doi.org/10.1016/j.ppnp.2009.08.002}
  {\path{doi:10.1016/j.ppnp.2009.08.002}}.

\bibitem{Fabri:1966fp}
E.~Fabri, L.~E. Picasso, Quantum field theory and approximate symmetries, Phys.
  Rev. Lett. 16 (1966) 408--410.
\newblock \href {http://dx.doi.org/10.1103/PhysRevLett.16.408.2}
  {\path{doi:10.1103/PhysRevLett.16.408.2}}.

\bibitem{Lange:1966zz}
R.~V. Lange, {Nonrelativistic Theorem Analogous to the Goldstone Theorem},
  Phys. Rev. 146 (1966) 301--303.
\newblock \href {http://dx.doi.org/10.1103/PhysRev.146.301}
  {\path{doi:10.1103/PhysRev.146.301}}.

\bibitem{Weinberg:1996v2}
S.~Weinberg, The Quantum Theory of Fields, Vol.~II, Cambridge University Press,
  Cambridge, UK, 1996.

\bibitem{Yang:1962zz}
C.~N. Yang, {Concept of Off-Diagonal Long-Range Order and the Quantum Phases of
  Liquid He and of Superconductors}, Rev. Mod. Phys. 34 (1962) 694--704.
\newblock \href {http://dx.doi.org/10.1103/RevModPhys.34.694}
  {\path{doi:10.1103/RevModPhys.34.694}}.

\bibitem{Miransky:1993mi}
V.~A. Miransky, Dynamical Symmetry Breaking in Quantum Field Theories, World
  Scientific, Singapore, 1993.

\bibitem{Barut:1977ba}
A.~O. Barut, R.~Raczka, Theory of Group Representations and Applications,
  Polish Scientific Publishers, Warszawa, Poland, 1977.

\bibitem{Araki:1963aw}
H.~Araki, E.~J. Woods, Representations of the canonical commutation relations
  describing a nonrelativistic infinite free bose gas, J. Math. Phys. 4 (1963)
  637--662.
\newblock \href {http://dx.doi.org/10.1063/1.1704002}
  {\path{doi:10.1063/1.1704002}}.

\bibitem{Nambu:1961tp}
Y.~Nambu, G.~Jona-Lasinio, {Dynamical Model of Elementary Particles Based on an
  Analogy with Superconductivity. I}, Phys. Rev. 122 (1961) 345--358.
\newblock \href {http://dx.doi.org/10.1103/PhysRev.122.345}
  {\path{doi:10.1103/PhysRev.122.345}}.

\bibitem{Nambu:1961fr}
Y.~Nambu, G.~Jona-Lasinio, {Dynamical Model of Elementary Particles Based on an
  Analogy with Superconductivity. II}, Phys. Rev. 124 (1961) 246--254.
\newblock \href {http://dx.doi.org/10.1103/PhysRev.124.246}
  {\path{doi:10.1103/PhysRev.124.246}}.

\bibitem{Michel:1971th}
L.~Michel, L.~A. Radicati, {Properties of the Breaking of Hadronic Internal
  Symmetry}, Ann. Phys. 66 (1971) 758--783.
\newblock \href {http://dx.doi.org/10.1016/0003-4916(71)90079-0}
  {\path{doi:10.1016/0003-4916(71)90079-0}}.

\bibitem{Michel:1980pc}
L.~Michel, {Symmetry defects and broken symmetry. Configurations Hidden
  Symmetry}, Rev. Mod. Phys. 52 (1980) 617--651.
\newblock \href {http://dx.doi.org/10.1103/RevModPhys.52.617}
  {\path{doi:10.1103/RevModPhys.52.617}}.

\bibitem{Kim:1981xu}
J.~S. Kim, {General Method for Analyzing Higgs Potentials}, Nucl. Phys. B196
  (1982) 285--300.
\newblock \href {http://dx.doi.org/10.1016/0550-3213(82)90040-2}
  {\path{doi:10.1016/0550-3213(82)90040-2}}.

\bibitem{Vollhardt:1990vw}
D.~Vollhardt, P.~W{\"o}lfle, The Superfluid Phases of Helium 3, Taylor and
  Francis, London, UK, 1990.

\bibitem{Slansky:1981yr}
R.~Slansky, {Group Theory for Unified Model Building}, Phys. Rep. 79 (1981)
  1--128.
\newblock \href {http://dx.doi.org/10.1016/0370-1573(81)90092-2}
  {\path{doi:10.1016/0370-1573(81)90092-2}}.

\bibitem{Meljanac:1985br}
S.~Meljanac, {Origin of Counter-Examples to Michel's Conjecture}, Phys. Lett.
  B168 (1986) 371--375.
\newblock \href {http://dx.doi.org/10.1016/0370-2693(86)91646-1}
  {\path{doi:10.1016/0370-2693(86)91646-1}}.

\bibitem{Abud:1983id}
M.~Abud, G.~Sartori, {The Geometry of Spontaneous Symmetry Breaking}, Ann.
  Phys. 150 (1983) 307--372.
\newblock \href {http://dx.doi.org/10.1016/0003-4916(83)90017-9}
  {\path{doi:10.1016/0003-4916(83)90017-9}}.

\bibitem{Iida:2000ha}
K.~Iida, G.~Baym, {The superfluid phases of quark matter: Ginzburg--Landau
  theory and color neutrality}, Phys. Rev. D63 (2001) 074018:1--074018:19.
\newblock \href {http://arxiv.org/abs/hep-ph/0011229}
  {\path{arXiv:hep-ph/0011229}}, \href
  {http://dx.doi.org/10.1103/PhysRevD.63.074018}
  {\path{doi:10.1103/PhysRevD.63.074018}}.

\bibitem{Brauner:2008ma}
T.~Brauner, {Helical ordering in the ground state of spin-one color
  superconductors as a consequence of parity violation}, Phys. Rev. D78 (2008)
  125027:1--125027:19.
\newblock \href {http://arxiv.org/abs/0810.3481} {\path{arXiv:0810.3481}},
  \href {http://dx.doi.org/10.1103/PhysRevD.78.125027}
  {\path{doi:10.1103/PhysRevD.78.125027}}.

\bibitem{Bailin:1983bm}
D.~Bailin, A.~Love, {Superfluidity and Superconductivity in Relativistic
  Fermion Systems}, Phys. Rep. 107 (1984) 325--385.
\newblock \href {http://dx.doi.org/10.1016/0370-1573(84)90145-5}
  {\path{doi:10.1016/0370-1573(84)90145-5}}.

\bibitem{Schmitt:2004et}
A.~Schmitt, {Ground state in a spin-one color superconductor}, Phys. Rev. D71
  (2005) 054016:1--054016:28.
\newblock \href {http://arxiv.org/abs/nucl-th/0412033}
  {\path{arXiv:nucl-th/0412033}}, \href
  {http://dx.doi.org/10.1103/PhysRevD.71.054016}
  {\path{doi:10.1103/PhysRevD.71.054016}}.

\bibitem{Mermin:1974me}
N.~D. Mermin, $d$-wave pairing near the transition temperature, Phys. Rev. A9
  (1974) 868--872.
\newblock \href {http://dx.doi.org/10.1103/PhysRevA.9.868}
  {\path{doi:10.1103/PhysRevA.9.868}}.

\bibitem{Alford:2007xm}
M.~G. Alford, A.~Schmitt, K.~Rajagopal, T.~{Sch\"afer}, {Color
  superconductivity in dense quark matter}, Rev. Mod. Phys. 80 (2008)
  1455--1515.
\newblock \href {http://arxiv.org/abs/0709.4635} {\path{arXiv:0709.4635}},
  \href {http://dx.doi.org/10.1103/RevModPhys.80.1455}
  {\path{doi:10.1103/RevModPhys.80.1455}}.

\bibitem{Goldstone:1961eq}
J.~Goldstone, {Field Theories with Superconductor Solutions}, Nuovo Cim. 19
  (1961) 154--164.
\newblock \href {http://dx.doi.org/10.1007/BF02812722}
  {\path{doi:10.1007/BF02812722}}.

\bibitem{Goldstone:1962es}
J.~Goldstone, A.~Salam, S.~Weinberg, {Broken Symmetries}, Phys. Rev. 127 (1962)
  965--970.
\newblock \href {http://dx.doi.org/10.1103/PhysRev.127.965}
  {\path{doi:10.1103/PhysRev.127.965}}.

\bibitem{Low:2001bw}
I.~Low, A.~V. Manohar, {Spontaneously Broken Spacetime Symmetries and
  Goldstone's Theorem}, Phys. Rev. Lett. 88 (2002) 101602:1--101602:4.
\newblock \href {http://arxiv.org/abs/hep-th/0110285}
  {\path{arXiv:hep-th/0110285}}, \href
  {http://dx.doi.org/10.1103/PhysRevLett.88.101602}
  {\path{doi:10.1103/PhysRevLett.88.101602}}.

\bibitem{Boulware:1962zz}
D.~G. Boulware, W.~Gilbert, {Connection between Gauge Invariance and Mass},
  Phys. Rev. 126 (1962) 1563--1567.
\newblock \href {http://dx.doi.org/10.1103/PhysRev.126.1563}
  {\path{doi:10.1103/PhysRev.126.1563}}.

\bibitem{Englert:1964et}
F.~Englert, R.~Brout, {Broken Symmetry and the Mass of Gauge Vector Mesons},
  Phys. Rev. Lett. 13 (1964) 321--322.
\newblock \href {http://dx.doi.org/10.1103/PhysRevLett.13.321}
  {\path{doi:10.1103/PhysRevLett.13.321}}.

\bibitem{Guralnik:1964eu}
G.~S. Guralnik, C.~R. Hagen, T.~W.~B. Kibble, {Global Conservation Laws and
  Massless Particles}, Phys. Rev. Lett. 13 (1964) 585--587.
\newblock \href {http://dx.doi.org/10.1103/PhysRevLett.13.585}
  {\path{doi:10.1103/PhysRevLett.13.585}}.

\bibitem{Elitzur:1975im}
S.~Elitzur, {Impossibility of Spontaneously Breaking Local Symmetries}, Phys.
  Rev. D12 (1975) 3978--3982.
\newblock \href {http://dx.doi.org/10.1103/PhysRevD.12.3978}
  {\path{doi:10.1103/PhysRevD.12.3978}}.

\bibitem{Frishman:1966fk}
Y.~Frishman, A.~Katz, Corollaries of the goldstone theorem, Phys. Rev. Lett. 16
  (1966) 370--371.
\newblock \href {http://dx.doi.org/10.1103/PhysRevLett.16.370}
  {\path{doi:10.1103/PhysRevLett.16.370}}.

\bibitem{Sachdev:1996sa}
S.~Sachdev, T.~Senthil, Zero temperature phase transitions in quantum
  heisenberg ferromagnets, Ann. Phys. 251 (1996) 76--122.

\bibitem{Ho:1998ho}
T.-L. Ho, Spinor bose condensates in optical traps, Phys. Rev. Lett. 81 (1998)
  742--745.

\bibitem{Ohmi:1998om}
T.~Ohmi, K.~Machida, Bose--einstein condensation with internal degrees of
  freedom in alkali atom gases, J. Phys. Soc. Jpn. 67 (1998) 1822--1825.

\bibitem{Uchino:2009ya}
S.~Uchino, M.~Kobayashi, M.~Ueda, {Bogoliubov Theory and Lee--Huang--Yang
  Correction in Spin-1 and Spin-2 Bose--Einstein Condensates in the Presence of
  the Quadratic Zeeman Effect}ArXiv:0912.0355 [cond-mat.quant-gas].

\bibitem{Honerkamp:2003hh}
C.~Honerkamp, W.~Hofstetter, Ultracold fermions and the su($n$) hubbard model,
  Phys. Rev. Lett. 92 (2004) 170403:1--170403:4.
\newblock \href {http://arxiv.org/abs/cond-mat/0309374}
  {\path{arXiv:cond-mat/0309374}}.

\bibitem{Honerkamp:2004ho}
C.~Honerkamp, W.~Hofstetter, Bcs pairing in fermi systems with $n$ different
  hyperfine states, Phys. Rev. B70 (2004) 094521:1--094521:10.
\newblock \href {http://arxiv.org/abs/cond-mat/0403166}
  {\path{arXiv:cond-mat/0403166}}.

\bibitem{He:2006ne}
L.~He, M.~Jin, P.~Zhuang, {Superfluidity in a three-flavor Fermi gas with SU(3)
  gauge symmetry}, Phys. Rev. A74 (2006) 033604:1--033604:8.
\newblock \href {http://arxiv.org/abs/cond-mat/0604580}
  {\path{arXiv:cond-mat/0604580}}, \href
  {http://dx.doi.org/10.1103/PhysRevA.74.033604}
  {\path{doi:10.1103/PhysRevA.74.033604}}.

\bibitem{Schafer:2001bq}
T.~{Sch\"afer}, D.~T. Son, M.~A. Stephanov, D.~Toublan, J.~J.~M. Verbaarschot,
  {Kaon condensation and Goldstone's theorem}, Phys. Lett. B522 (2001) 67--75.
\newblock \href {http://arxiv.org/abs/hep-ph/0108210}
  {\path{arXiv:hep-ph/0108210}}, \href
  {http://dx.doi.org/10.1016/S0370-2693(01)01265-5}
  {\path{doi:10.1016/S0370-2693(01)01265-5}}.

\bibitem{Miransky:2001tw}
V.~A. Miransky, I.~A. Shovkovy, {Spontaneous Symmetry Breaking with Abnormal
  Number of Nambu--Goldstone Bosons and Kaon Condensate}, Phys. Rev. Lett. 88
  (2002) 111601:1--111601:4.
\newblock \href {http://arxiv.org/abs/hep-ph/0108178}
  {\path{arXiv:hep-ph/0108178}}, \href
  {http://dx.doi.org/10.1103/PhysRevLett.88.111601}
  {\path{doi:10.1103/PhysRevLett.88.111601}}.

\bibitem{Andersen:2006ys}
J.~O. Andersen, {Pion and kaon condensation at finite temperature and density},
  Phys. Rev. D75 (2007) 065011:1--065011:11.
\newblock \href {http://arxiv.org/abs/hep-ph/0609020}
  {\path{arXiv:hep-ph/0609020}}, \href
  {http://dx.doi.org/10.1103/PhysRevD.75.065011}
  {\path{doi:10.1103/PhysRevD.75.065011}}.

\bibitem{Beraudo:2004zr}
A.~Beraudo, A.~De~Pace, M.~Martini, A.~Molinari, {Spontaneous symmetry breaking
  and response functions}, Ann. Phys. 317 (2005) 444--473.
\newblock \href {http://arxiv.org/abs/nucl-th/0409039}
  {\path{arXiv:nucl-th/0409039}}, \href
  {http://dx.doi.org/10.1016/j.aop.2004.12.002}
  {\path{doi:10.1016/j.aop.2004.12.002}}.

\bibitem{Blaschke:2004cs}
D.~Blaschke, D.~Ebert, K.~G. Klimenko, M.~K. Volkov, V.~L. Yudichev, {Abnormal
  number of Nambu--Goldstone bosons in the color-asymmetric dense color
  superconducting phase of a Nambu--Jona-Lasinio-type model}, Phys. Rev. D70
  (2004) 014006:1--014006:11.
\newblock \href {http://arxiv.org/abs/hep-ph/0403151}
  {\path{arXiv:hep-ph/0403151}}, \href
  {http://dx.doi.org/10.1103/PhysRevD.70.014006}
  {\path{doi:10.1103/PhysRevD.70.014006}}.

\bibitem{Buballa:2002wy}
M.~Buballa, J.~{Ho\v{s}ek}, M.~Oertel, {Anisotropic admixture in
  color-superconducting quark matter}, Phys. Rev. Lett. 90 (2003)
  182002:1--182002:4.
\newblock \href {http://arxiv.org/abs/hep-ph/0204275}
  {\path{arXiv:hep-ph/0204275}}, \href
  {http://dx.doi.org/10.1103/PhysRevLett.90.182002}
  {\path{doi:10.1103/PhysRevLett.90.182002}}.

\bibitem{Brauner:2005di}
T.~Brauner, {Goldstone boson counting in linear sigma models with chemical
  potential}, Phys. Rev. D72 (2005) 076002:1--076002:10.
\newblock \href {http://arxiv.org/abs/hep-ph/0508011}
  {\path{arXiv:hep-ph/0508011}}, \href
  {http://dx.doi.org/10.1103/PhysRevD.72.076002}
  {\path{doi:10.1103/PhysRevD.72.076002}}.

\bibitem{Brauner:2007uw}
T.~Brauner, {Goldstone bosons in presence of charge density}, Phys. Rev. D75
  (2007) 105014:1--105014:13.
\newblock \href {http://arxiv.org/abs/hep-ph/0701110}
  {\path{arXiv:hep-ph/0701110}}, \href
  {http://dx.doi.org/10.1103/PhysRevD.75.105014}
  {\path{doi:10.1103/PhysRevD.75.105014}}.

\bibitem{Kapusta:2006kg}
J.~I. Kapusta, C.~Gale, Finite-Temperature Field Theory: Principles and
  Applications, Cambridge University Press, Cambridge, UK, 2006.

\bibitem{Georgi:1982jb}
H.~Georgi, Lie Algebras in Particle Physics, Frontiers in Physics, Perseus
  Books, Reading, MA, USA, 1999.

\bibitem{Brauner:2006xm}
T.~Brauner, {Spontaneous symmetry breaking in the linear sigma model at finite
  chemical potential: One-loop corrections}, Phys. Rev. D74 (2006)
  085010:1--085010:13.
\newblock \href {http://arxiv.org/abs/hep-ph/0607102}
  {\path{arXiv:hep-ph/0607102}}, \href
  {http://dx.doi.org/10.1103/PhysRevD.74.085010}
  {\path{doi:10.1103/PhysRevD.74.085010}}.

\bibitem{Weinberg:1978kz}
S.~Weinberg, {Phenomenological Lagrangians}, Physica A96 (1979) 327--340.

\bibitem{Coleman:1969sm}
S.~R. Coleman, J.~Wess, B.~Zumino, {Structure of Phenomenological Lagrangians.
  I}, Phys. Rev. 177 (1969) 2239--2247.
\newblock \href {http://dx.doi.org/10.1103/PhysRev.177.2239}
  {\path{doi:10.1103/PhysRev.177.2239}}.

\bibitem{Callan:1969sn}
C.~G. Callan, S.~R. Coleman, J.~Wess, B.~Zumino, {Structure of Phenomenological
  Lagrangians. II}, Phys. Rev. 177 (1969) 2247--2250.
\newblock \href {http://dx.doi.org/10.1103/PhysRev.177.2247}
  {\path{doi:10.1103/PhysRev.177.2247}}.

\bibitem{Leutwyler:1993iq}
H.~Leutwyler, {On the Foundations of Chiral Perturbation Theory}, Ann. Phys.
  235 (1994) 165--203.
\newblock \href {http://arxiv.org/abs/hep-ph/9311274}
  {\path{arXiv:hep-ph/9311274}}, \href
  {http://dx.doi.org/10.1006/aphy.1994.1094}
  {\path{doi:10.1006/aphy.1994.1094}}.

\bibitem{Gasser:1983yg}
J.~Gasser, H.~Leutwyler, {Chiral Perturbation Theory to One Loop}, Ann. Phys.
  158 (1984) 142--210.
\newblock \href {http://dx.doi.org/10.1016/0003-4916(84)90242-2}
  {\path{doi:10.1016/0003-4916(84)90242-2}}.

\bibitem{Gasser:1984gg}
J.~Gasser, H.~Leutwyler, {Chiral Perturbation Theory: Expansions in the Mass of
  the Strange Quark}, Nucl. Phys. B250 (1985) 465--516.
\newblock \href {http://dx.doi.org/10.1016/0550-3213(85)90492-4}
  {\path{doi:10.1016/0550-3213(85)90492-4}}.

\bibitem{Roman:1999ro}
J.~M. Roman, J.~Soto, {Effective Field Theory Approach to Ferromagnets and
  Antiferromagnets in Crystalline Solids}, Int. J. Mod. Phys. B13 (1999)
  755--789.
\newblock \href {http://arxiv.org/abs/cond-mat/9709298}
  {\path{arXiv:cond-mat/9709298}}, \href
  {http://dx.doi.org/10.1142/S0217979299000655}
  {\path{doi:10.1142/S0217979299000655}}.

\bibitem{Hofmann:1998pp}
C.~P. Hofmann, {Spin-wave scattering in the effective Lagrangian perspective},
  Phys. Rev. B60 (1999) 388--405.
\newblock \href {http://arxiv.org/abs/cond-mat/9805277}
  {\path{arXiv:cond-mat/9805277}}, \href
  {http://dx.doi.org/10.1103/PhysRevB.60.388}
  {\path{doi:10.1103/PhysRevB.60.388}}.

\bibitem{Hofmann:2001ck}
C.~P. Hofmann, {Spontaneous magnetization of the O($3$) ferromagnet at low
  temperatures}, Phys. Rev. B65 (2002) 094430:1--094430:12.
\newblock \href {http://arxiv.org/abs/cond-mat/0106492}
  {\path{arXiv:cond-mat/0106492}}, \href
  {http://dx.doi.org/10.1103/PhysRevB.65.094430}
  {\path{doi:10.1103/PhysRevB.65.094430}}.

\bibitem{Hofmann:1997qm}
C.~P. Hofmann, {Effective analysis of the O($N$) antiferromagnet:
  Low-temperature expansion of the order parameter}, Phys. Rev. B60 (1999)
  406--413.
\newblock \href {http://arxiv.org/abs/hep-ph/9706418}
  {\path{arXiv:hep-ph/9706418}}, \href
  {http://dx.doi.org/10.1103/PhysRevB.60.406}
  {\path{doi:10.1103/PhysRevB.60.406}}.

\bibitem{Kampfer:2005ba}
F.~{K\"ampfer}, M.~Moser, U.~J. Wiese, {Systematic low-energy effective theory
  for magnons and charge carriers in an antiferromagnet}, Nucl. Phys. B729
  (2005) 317--360.
\newblock \href {http://arxiv.org/abs/cond-mat/0506324}
  {\path{arXiv:cond-mat/0506324}}, \href
  {http://dx.doi.org/10.1016/j.nuclphysb.2005.09.004}
  {\path{doi:10.1016/j.nuclphysb.2005.09.004}}.

\bibitem{Brugger:2006dz}
C.~{Br\"ugger}, F.~{K\"ampfer}, M.~Moser, M.~Pepe, U.-J. Wiese, {Two-hole bound
  states from a systematic low-energy effective field theory for magnons and
  holes in an antiferromagnet}, Phys. Rev. B74 (2006) 224432.
\newblock \href {http://arxiv.org/abs/cond-mat/0606766}
  {\path{arXiv:cond-mat/0606766}}, \href
  {http://dx.doi.org/10.1103/PhysRevB.74.224432}
  {\path{doi:10.1103/PhysRevB.74.224432}}.

\bibitem{Hofmann:2009ru}
C.~P. Hofmann, {Thermodynamics of O($N$) antiferromagnets in $2+1$ dimensions},
  Phys. Rev. B81 (2010) 014416:1--014416:17.
\newblock \href {http://arxiv.org/abs/0909.5239} {\path{arXiv:0909.5239}}.

\bibitem{Andersen:2003qj}
J.~O. Andersen, {Theory of the weakly interacting Bose gas}, Rev. Mod. Phys. 76
  (2004) 599--639.
\newblock \href {http://arxiv.org/abs/cond-mat/0305138}
  {\path{arXiv:cond-mat/0305138}}, \href
  {http://dx.doi.org/10.1103/RevModPhys.76.599}
  {\path{doi:10.1103/RevModPhys.76.599}}.

\bibitem{Andersen:2002nd}
J.~O. Andersen, {Effective Field Theory for Goldstone Bosons in Nonrelativistic
  Superfluids}ArXiv:cond-mat/0209243.

\bibitem{Yukalov:2008yu}
V.~I. Yukalov, Representative statistical ensembles for bose systems with
  broken gauge symmetry, Ann. Phys. 323 (2008) 461--499.
\newblock \href {http://arxiv.org/abs/arXiv:0801.3168 [cond-mat.stat-mech]}
  {\path{arXiv:arXiv:0801.3168 [cond-mat.stat-mech]}}.

\bibitem{Son:2005rv}
D.~T. Son, M.~Wingate, {General coordinate invariance and conformal invariance
  in nonrelativistic physics: Unitary Fermi gas}, Ann. Phys. 321 (2006)
  197--224.
\newblock \href {http://arxiv.org/abs/cond-mat/0509786}
  {\path{arXiv:cond-mat/0509786}}, \href
  {http://dx.doi.org/10.1016/j.aop.2005.11.001}
  {\path{doi:10.1016/j.aop.2005.11.001}}.

\bibitem{Son:2002zn}
D.~T. Son, {Low-Energy Quantum Effective Action for Relativistic
  Superfluids}ArXiv:hep-ph/0204199.

\bibitem{Manuel:2004iv}
C.~Manuel, A.~Dobado, F.~J. Llanes-Estrada, {Shear viscosity in a CFL quark
  star}, J. High Energy Phys. 09 (2005) 076:1--076:22.
\newblock \href {http://arxiv.org/abs/hep-ph/0406058}
  {\path{arXiv:hep-ph/0406058}}.

\bibitem{Weinberg:1972fn}
S.~Weinberg, {Approximate Symmetries and Pseudo-Goldstone Bosons}, Phys. Rev.
  Lett. 29 (1972) 1698--1701.
\newblock \href {http://dx.doi.org/10.1103/PhysRevLett.29.1698}
  {\path{doi:10.1103/PhysRevLett.29.1698}}.

\bibitem{Coleman:1973jx}
S.~R. Coleman, E.~J. Weinberg, {Radiative Corrections as the Origin of
  Spontaneous Symmetry Breaking}, Phys. Rev. D7 (1973) 1888--1910.
\newblock \href {http://dx.doi.org/10.1103/PhysRevD.7.1888}
  {\path{doi:10.1103/PhysRevD.7.1888}}.

\bibitem{Volovik:1982vo}
G.~E. Volovik, M.~V. Khazan, {Dynamics of the $A$-phase of ${}^3$He at low
  pressures}, Sov. Phys. JETP 55 (1982) 867--871.

\bibitem{Volovik:1983vo}
G.~E. Volovik, M.~V. Khazan, {A classification of the collective modes of the
  order parameter in superfluid ${}^3$He}, Sov. Phys. JETP 58 (1983) 551--555.

\bibitem{Muther:2002ej}
H.~{M\"uther}, A.~Sedrakian, {Breaking rotational symmetry in two-flavor color
  superconductors}, Phys. Rev. D67 (2003) 085024:1--085024:6.
\newblock \href {http://arxiv.org/abs/hep-ph/0212317}
  {\path{arXiv:hep-ph/0212317}}, \href
  {http://dx.doi.org/10.1103/PhysRevD.67.085024}
  {\path{doi:10.1103/PhysRevD.67.085024}}.

\bibitem{Sannino:2001fd}
F.~Sannino, W.~{Sch\"afer}, {Relativistic massive vector condensation}, Phys.
  Lett. B527 (2002) 142--148.
\newblock \href {http://arxiv.org/abs/hep-ph/0111098}
  {\path{arXiv:hep-ph/0111098}}, \href
  {http://dx.doi.org/10.1016/S0370-2693(01)01521-0}
  {\path{doi:10.1016/S0370-2693(01)01521-0}}.

\bibitem{Gusynin:2003yu}
V.~P. Gusynin, V.~A. Miransky, I.~A. Shovkovy, {Spontaneous rotational symmetry
  breaking and roton like excitations in gauged $\sigma$-model at finite
  density}, Phys. Lett. B581 (2004) 82--92.
\newblock \href {http://arxiv.org/abs/hep-ph/0311025}
  {\path{arXiv:hep-ph/0311025}}, \href
  {http://dx.doi.org/10.1016/j.physletb.2003.11.042}
  {\path{doi:10.1016/j.physletb.2003.11.042}}.

\bibitem{Leutwyler:1996er}
H.~Leutwyler, {Phonons as Goldstone bosons}, Helv. Phys. Acta 70 (1997)
  275--286.
\newblock \href {http://arxiv.org/abs/hep-ph/9609466}
  {\path{arXiv:hep-ph/9609466}}.

\bibitem{Casalbuoni:2003wh}
R.~Casalbuoni, G.~Nardulli, {Inhomogeneous superconductivity in condensed
  matter and QCD}, Rev. Mod. Phys. 76 (2004) 263--320.
\newblock \href {http://arxiv.org/abs/hep-ph/0305069}
  {\path{arXiv:hep-ph/0305069}}, \href
  {http://dx.doi.org/10.1103/RevModPhys.76.263}
  {\path{doi:10.1103/RevModPhys.76.263}}.

\bibitem{Dzyaloshinsky:1958dz}
I.~Dzyaloshinsky, A thermodynamic theory of ``weak'' ferromagnetism of
  antiferromagnetics, J. Phys. Chem. Solids 4 (1958) 241--255.

\bibitem{Moriya:1960mo}
T.~Moriya, Anisotropic superexchange interaction and weak ferromagnetism, Phys.
  Rev. 120 (1960) 91--98.

\end{thebibliography}
\bibliographystyle{elsarticle-num}

\end{document}